\documentclass[preprint,showpacs,preprintnumbers,amsmath,amssymb, superscriptaddress]{revtex4}

\usepackage{latexsym}
\usepackage{amssymb}
\usepackage{epsfig,amsmath,graphics}
\usepackage{epstopdf}
\usepackage{verbatim}
\usepackage[hypertex]{hyperref}
\usepackage{graphicx}
\usepackage{subfigure}

\newcommand{\OO}{\mathcal{O}}

\newcommand{\GeV}{\text{GeV}}
\newcommand{\TeV}{\text{TeV}}

\newcommand{\half}{\frac{1}{2}}

\newcommand{\D}{^2\text{H}}

\newcommand{\He}{\text{He}}
\newcommand{\Li}{\text{Li}}
\newcommand{\Be}{\text{Be}}
\newcommand{\Hy}{\text{H}}
\newcommand{\s}{\text{s}}

\newcommand{\cm}{\text{cm}}
\newcommand{\sr}{\text{sr}}
\newcommand{\kpc}{\text{kpc}}
\newcommand\ee{\end{equation}}
\newcommand\be{\begin{equation}}

\newcommand{\LiSeven}{^{7}\text{Li}}
\newcommand{\LiSix}{^{6}\text{Li}}
\newcommand{\FiveBar}{\bar{5}}
\newcommand{\TenBar}{\bar{10}}

\newcommand{\FiveBarDagger}{\bar{5}^{\dagger}}
\newcommand{\TenDagger}{10^{\dagger}}

\newcommand{\HuDagger}{H_{u}^{\dagger}}

\newcommand{\chiDagger}{\chi^{\dagger}}
\newcommand{\chiBar}{\bar{\chi}}
\newcommand{\chiBarDagger}{\bar{\chi}^{\dagger}}

\newcommand{\OrderOne}{\mathcal{O}\left(1\right)}
\newcommand{\HBr}{Br(H)}
\newcommand{\Mchi}{M_{\chi}}
\newcommand{\Mgut}{M_\text{GUT}}
\newcommand{\gut}{\text{GUT}}

\newcommand{\mugut}{\left(\frac{\mu}{M_{\text{GUT}}}\right)}

\def\r{\right)}
\def\l{\left(}

\begin{document}


\title{Astrophysical Probes of Unification}

\author{Asimina Arvanitaki}
\affiliation{Berkeley Center for Theoretical Physics, University of California, Berkeley, CA, 94720}
\affiliation{Theoretical Physics Group, Lawrence Berkeley National Laboratory, Berkeley, CA, 94720}

\author{Savas Dimopoulos}
\affiliation{Department of Physics, Stanford University, Stanford, California 94305}

\author{Sergei Dubovsky}
\affiliation{Department of Physics, Stanford University, Stanford, California 94305}
\affiliation{ Institute for Nuclear Research of the Russian Academy of Sciences,
        60th October Anniversary Prospect, 7a, 117312 Moscow, Russia}

\author{Peter W. Graham}
\affiliation{Department of Physics, Stanford University, Stanford, California 94305}

\author{Roni Harnik}
\affiliation{Department of Physics, Stanford University, Stanford, California 94305}

\author{Surjeet Rajendran}
\affiliation{SLAC National Accelerator Laboratory, Stanford University, Menlo Park, California 94025}
\affiliation{Department of Physics, Stanford University, Stanford, California 94305}

\date{\today}

\begin{abstract}
Traditional ideas for testing unification involve searching for the decay of the proton and its branching modes. We point out that several astrophysical experiments are now reaching sensitivities that allow them to explore supersymmetric unified theories. In these theories the electroweak-mass dark matter particle can decay, just like the proton, through dimension six operators with lifetime $\sim 10^{26} $ sec. Interestingly, this timescale is now being investigated in several experiments including ATIC, PAMELA, HESS, and Fermi. Positive evidence for such decays may be opening our first direct window to physics at the supersymmetric unification scale of $M_{GUT} \sim10^{16}$ GeV, as well as the TeV scale.  Moreover, in the same supersymmetric unified theories, dimension five operators can lead a weak-scale superparticle to decay with a lifetime of $\sim100$ sec. Such decays are recorded by a change in the primordial light element abundances and may well explain the present discord between the measured Li abundances and standard big bang nucleosynthesis, opening another window to unification. These theories make concrete predictions for the spectrum and signatures at the LHC as well as Fermi.
\end{abstract}

\maketitle

\tableofcontents



\section{Long Lifetimes from the Unification Scale and Astrophysical Signals}
One of the most interesting lessons of our times is the evidence for a new fundamental scale in nature, the grand unification (GUT) scale near $M_\text{GUT} \sim 10^{16} ~\GeV$, at which, in the presence of superparticles, the gauge forces unify \cite{Dimopoulos:1981yj, Dimopoulos:1981zb}.  This is a precise, quantitative (few percent) concordance between theory and experiment and one of the compelling indications for physics beyond the Standard Model.  Together with neutrino masses \cite{GellMann:1976pg, Yanagida:1980xy}, it provides independent evidence for new physics near $10^{16} ~\GeV$, significantly below the Planck mass of $M_\text{pl} \approx 10^{19} ~\GeV$.  The LHC may considerably strengthen the evidence for grand unification if it discovers superparticles.  Furthermore, future proton decay experiments may provide direct evidence for physics at $\Mgut$.  In this paper we consider frameworks in which GUT scale physics is probed by cosmological and astrophysical observations.

In grand unified theories the proton can decay because the global baryon-number symmetry of the low energy Standard Model is broken by GUT scale physics.  Indeed, only local symmetries can guarantee that a particle remains exactly stable since global symmetries are generically broken in fundamental theories.  Just as the proton is long-lived but may ultimately decay, other particles, for example the dark matter, may decay with long lifetimes.  If a TeV mass dark matter particle decays via GUT suppressed dimension 6 operators, its lifetime would be
\begin{equation}
\label{Eqn: dim 6 lifetime}
\tau \sim 8 \pi \frac{\Mgut^4}{m^5} = 3 \times 10^{27} ~\s \left( \frac{\TeV}{m} \right)^5 \left( \frac{\Mgut}{2 \times 10^{16} ~\GeV} \right)^4
\end{equation}
Similarly a long-lived particle decaying through dimension 5 GUT suppressed operators has a lifetime
\begin{equation}
\tau \sim 8 \pi \frac{\Mgut^2}{m^3} = 7 ~\s \left( \frac{\TeV}{m} \right)^3 \left( \frac{\Mgut}{2 \times 10^{16} ~\GeV} \right)^2
\end{equation}
Both of these timescales have potentially observable consequences.  The dimension 6 decays cause a small fraction of the dark matter to decay today, producing potentially observable high energy cosmic rays.  The dimension 5 decays happen during Big Bang Nucleosynthesis (BBN) and can leave their imprint on the light element abundances.
There is, of course, uncertainty in these predictions for the lifetimes because the physics at the GUT scale is not known.

If the dark matter decays through dimension 6 GUT suppressed operators with a lifetime as in Eqn. \ref{Eqn: dim 6 lifetime} it can produce high energy photons, electrons and positrons, antiprotons, or neutrinos.  Interestingly, the lifetime of order $10^{27} ~\s$ leads to fluxes in the range that is being explored by a variety of current experiments such as HESS, MAGIC, VERITAS, WHIPPLE, EGRET, WMAP, HEAT, PAMELA, ATIC, PPB-BETS, SuperK, AMANDA, Frejus, and upcoming experiments such as the Fermi (GLAST) gamma ray space telescope the Planck satellite, and IceCube, as shown in Table \ref{Tab: astro limits}.  This is an intriguing coincidence, presented in section 2, that may allow these experiments to probe physics at the GUT scale, much as the decay of the proton and a study of its branching ratios would. Possible hints for excesses in some of these experiments may have already started us on such an  exciting path.

\begin{table}
\begin{center}
\begin{math}
\begin{footnotesize}
\begin{array}{|c|c|c|c|c|c|}
\hline
 & \text{Extragalactic } \gamma \text{-rays} & \text{Galactic } \gamma \text{'s} & \text{antiprotons} & \text{positrons} & \text{neutrinos} \\
\text{Decay} & & & & & \text{Super-K} \\
 \text{channel} & \text{EGRET} & \text{HESS} & \text{PAMELA} & \text{PAMELA} & \text{AMANDA, Frejus}\\
\hline
q \overline{q} & 4 \times 10^{25} ~\s & - & 10^{27} ~\s & - & - \\
\hline
e^+ e^- & 8 \times 10^{22} ~\s & 2 \times 10^{22} ~\s ~ \sqrt{\frac{m_\psi}{\TeV}} \left( \text{K} \right) & 10^{24} ~\s & 2 \times 10^{25} ~\s \left( \frac{\TeV}{m_\psi} \right) & 3 \times 10^{21} ~\s ~ \left( \frac{m_\psi}{\TeV} \right)\\
\hline
\mu^+ \mu^- & 8 \times 10^{22} ~\s & 2 \times 10^{22} ~\s ~ \sqrt{\frac{m_\psi}{\TeV}} \left( \text{K} \right) & 10^{24} ~\s & 2 \times 10^{25} ~\s \left( \frac{\TeV}{m_\psi} \right) & 3 \times 10^{24} ~\s ~ \left( \frac{m_\psi}{\TeV} \right) \\
\hline
\tau^+ \tau^- & 10^{25} ~\s & 10^{22} ~\s ~ \sqrt{\frac{m_\psi}{\TeV}} \left( \text{K} \right) & 10^{24} ~\s & 10^{25} ~\s \left( \frac{\TeV}{m_\psi} \right) & 3 \times 10^{24} ~\s ~ \left( \frac{m_\psi}{\TeV} \right) \\
\hline
W W & 3 \times 10^{25} ~\s & -  & 3 \times 10^{26} ~\s  & 4 \times 10^{25} ~\s & 8 \times 10^{23} ~\s ~ \left( \frac{m_\psi}{\TeV} \right)  \\
\hline
 & 9 \times 10^{24} ~\s \left( m_\psi = 100 ~ \GeV \right) & 2 \times 10^{24} ~\s ~ \sqrt{\frac{m_\psi}{\TeV}} \left( \text{K} \right) & &  & \\
\gamma \gamma & 2 \times 10^{22} ~\s \left( m_\psi = 800 ~ \GeV \right) &  & 2 \times 10^{25} ~\s & 8 \times 10^{23} ~\s \left( \frac{\TeV}{m_\psi} \right) & - \\
 & 4 \times 10^{23} ~\s \left( m_\psi = 3200 ~ \GeV \right) & 5 \times 10^{25} ~\s ~ \sqrt{\frac{m_\psi}{\TeV}} \left( \text{NFW} \right) & &  & \\
\hline
\nu \overline{\nu} & 8 \times 10^{22} ~\s & - & 10^{24} ~\s & 10^{23} ~\s & 10^{25} ~\s ~ \left( \frac{m_\psi}{\TeV} \right) \\
\hline
\end{array}
\end{footnotesize}
\end{math}
\caption[Astrophysical Limits on Decaying Dark Matter]{\label{Tab: astro limits} The lower limit on the lifetime of a dark matter particle with mass in the range $10 ~\GeV \lesssim m_\psi \lesssim 10 ~\TeV$, decaying to the products listed in the left column.  The experiment and the observed particle being used to set the limit are listed in the top row.  HESS limits only apply for $m_\psi > 400 ~\GeV$ and are shown for two choices of halo profiles: the Kravtsov (K) and the NFW.  PAMELA limits are most accurate in the range $100 ~\GeV \lesssim m_\psi \lesssim 1 ~\TeV$.  All the limits are only approximate.  Generally conservative assumptions were made and there are many details and caveats as described in Section \ref{Sec: Astro Limits}.}
\end{center}
\end{table}

GUT scale physics can also manifest itself in astrophysical observations by leaving its imprint on the abundances of light elements created during BBN.  For example neutrons from the decay of a heavy particle create hot tracks in the surrounding plasma in which additional nucleosynthesis occurs.  In particular, these energetic neutrons impinge on nuclei and energize them, causing a cascade of reactions.  This most strongly affects the abundances of the rare elements produced during BBN, especially $^6\Li$ and $^7\Li$ and possibly $^9\Be$.

In fact, measurements of both isotopes of $\Li$ suggest a discrepancy from the predictions of standard BBN (sBBN).  The observed $\LiSeven$ abundance of $\frac{^7\Li}{\text{H}} \sim (1 - 2) \times 10^{-10}$ is a factor of several below the sBBN prediction of $\frac{^7\Li}{\text{H}} \approx ( 5.2 \pm 0.7 ) \times 10^{-10}$ \cite{Cyburt:2008kw}.  In contrast, observations indicate a primordial $\LiSix$ abundance over an order of magnitude above the sBBN prediction.  The Lithium abundances are measured in a sample of low-metallicity stars.  The $\Li$ isotopic ratio in all these stars is similar to that in the lowest metallicity star in the sample: $\frac{^6\Li}{^7\Li} = 0.046 \pm 0.022$ \cite{Asplund:2005yt}.  This implies a primordial $\LiSix$ abundance in the range $\frac{^6\Li}{\Hy} \approx (2 - 10) \times 10^{-12}$, while sBBN predicts $\frac{^6\Li}{\Hy} \approx 10^{-14}$ \cite{Jedamzik:2007cp, VangioniFlam:1998gq}.  The apparent presence of a Spite plateau in the abundances of both $\LiSix$ and $\LiSeven$ as a function of stellar temperature and metallicity is an indication that the measured abundances are indeed primordial.  Of course, though there is no proven astrophysical solution, either or both of these anomalies could be due to astrophysics and not new particle physics.  Nevertheless, the Li problems are suggestive of new physics because a new source of energy deposition during BBN naturally tends to destroy $\LiSeven$ and produce $\LiSix$, a nontrivial qualitative condition that many single astrophysical solutions do not satisfy.  Further, energy deposition, for example due either to decays or annihilations, during BBN most significantly affects the Li abundances, not the other light element abundances, making the Li isotopes the most sensitive probes of new physics during BBN.  Finally, a long-lived particle decaying with a $\sim 1000$ s lifetime can naturally destroy the correct amount of $\LiSeven$ and produce the correct amount of $\LiSix$ without significantly altering the abundances of the other light elements $\D$, $^3\He$, and $^4\He$.

In this paper we explore astrophysical signals of GUT scale physics in the framework of supersymmetric unified theories (often referred to as SUSY GUTs or supersymmetric standard models (SSM) \cite{Dimopoulos:1981zb}), as manifested by particle decays via dimension 5 and 6 GUT suppressed operators.  In order to preserve the success of gauge coupling unification, we work with the minimal particle content of the MSSM with additional singlets or complete SU(5) multiplets.  The effects of GUT physics on a low energy experiment are generally suppressed. However, these effects can accumulate and lead to vastly different physics over long times depending on the details of the higher dimension operators generated by GUT physics. For example, particles that would have been stable in the absence of GUT scale physics can decay with very different lifetimes and decay modes depending on the particular GUT physics. Conversely, such decays are sensitive diagnostics of the physics at the GUT scale. The details of these decays depend both upon the physics at the GUT scale and the low energy MSSM spectrum of the theory. So, astrophysical observations of such decays, in conjunction with independent measurements of the low energy MSSM spectrum at the LHC, would open a window to the GUT scale - just as proton decay would.  The role of the water and photomultipliers that register a proton decay event  are now replaced by the universe and either the modified Li abundance, or the excess cosmic rays that may be detected in todays plethora of experiments.

To characterize the varieties of GUT physics that can give rise to decays of would-be-stable particles
we enumerate the possible dimension 5 and 6 operators, each one of which defines a separate general class of theories that breaks a selection rule or conservation law that would have stabilized the particle in question. This approach has the advantage of being far more general than a concrete theory and encompasses all that can be known from the limited low-energy physics experiments available to us. In other words, all measurable consequences depend on the form of the operator and not on the detailed microphysics at the unification scale (e.g. the GUT mass particles) that give rise to it.

The presence of supersymmetry, together with gauge symmetries and Poincare invariance, and the simplicity of the near-MSSM particle content greatly reduces the number of possible higher dimension operators of dimension 5 and 6 and allows for the methodic enumeration of the operators and the decays they cause. This is what we do in Sections \ref{Sec: Dim 6 decays}, \ref{LithiumIntroduction}, and \ref{LithiumPAMELA}.  We work in an SU(5) framework because any of the SU(5) invariant operators we consider can be embedded into invariant operators of any larger GUT gauge group so long as it contains SU(5).  Of course, such operators may in general contain several SU(5) operators so the detailed conclusions can be affected.  As an example, we study a model that is made particularly simple in an SO(10) GUT in Section \ref{LithiumPAMELA}.

The dimension 6 operators of Section \ref{Sec: Dim 6 decays} have many potentially observable astrophysical signals at experiments searching for gamma rays, positrons, antiprotons or neutrinos.  We separate them into R-conserving and R-breaking classes and in each case we also build UV models which give rise to these operators. The dimension 5 operators of Section \ref{LithiumIntroduction} have potentially observable effects on BBN, may solve the Lithium problems, and give dramatic out-of-time decays in the LHC detectors. Section \ref{LithiumPAMELA} presents frameworks and simple theories that have both dimension 5 and 6 operators destabilizing particles to lifetimes of both 100 sec and $10^{27}$ sec to explain Li and lead to astrophysical signatures today in PAMELA/ATIC and other observatories.

In Section \ref{Sec: Astro signals} we similarly look at the consequences for Fermi/Glast and PAMELA/ATIC.  In Section \ref{Sec: LHC signals} we outline LHC-observable consequences of some of these scenarios (operators) that could lead to their laboratory confirmation.

Finally, there are two more classes of theories that fit into the elegant framework of supersymmetric unification. The first of these are SUSY theories that solve the strong CP problem with an axion. The other is the Split SUSY framework. Both these frameworks provide particles that can have lifetimes long enough to account for the primordial lithium discrepancies without additional inputs, and are included in Section \ref{LithiumIntroduction}.
 



\section{Astrophysical Limits on Decaying Dark Matter}
\label{Sec: Astro Limits}

In this section we describe the existing astrophysical limits on decaying dark matter, as summarized in Table \ref{Tab: astro limits}.  Note that the limits on dark matter decaying into many different final states (e.g. photons, leptons, quarks, or neutrinos) are similar even though they arise from different experiments.  These different observations are all sensitive to lifetimes in the range given by a dimension 6 decay operator, as in Eqn. \eqref{Eqn: dim 6 lifetime}.  This can be understood, at least for the satellite and balloon experiments, because these all generally have similar acceptances of $\sim (1 ~\text{m}^2) (1~ \text{yr}) (1 ~\sr) \approx 3 \times 10^{11} ~ \cm^2 ~\s ~ \sr$.  For comparison, the number of incident particles from decaying dark matter is $\sim \int^{10 ~\kpc} \frac{d^3r}{r^2}  (0.3 ~\frac{\GeV}{m_\psi ~\cm^3}) (10^{-28} ~\s^{-1}) \approx 10^{-10} ~\cm^{-2} ~\s^{-1} ~\sr^{-1}$, where these could be photons, positrons or antiprotons for example, depending on what is produced in the decay.  This implies such experiments observe $\sim \left( 3 \times 10^{11} ~ \cm^2 ~\s ~ \sr \right) \times \left( 10^{-10} ~\cm^{-2} ~\s^{-1} ~\sr^{-1} \right) \approx 30$ events which is in the right range to observe dark matter decaying with a dimension 6 decay lifetime.

Not only are there limits from astrophysical observations, there may be indications of dark matter decaying or annihilating from recent experiments.  PAMELA \cite{Adriani:2008zr} has observed a rise in the positron fraction of cosmic rays around 50 GeV.  ATIC \cite{ATIC} and PPB-BETS \cite{PPB-BETS} have observed a bump in the spectrum of electrons plus positrons with a peak around 500 GeV.  Finally, there are claims that WMAP has observed a `haze' which could be synchrotron radiation from high energy electrons and positrons near the galactic center.  This is consistent with dark matter annihilating \cite{Hooper:2007kb} and possibly also with dark matter decaying \cite{Ishiwata:2008qy}.

We will call the decaying dark matter particle $\psi$ with mass $m_\psi$.

\subsection{Diffuse Gamma-Ray Background From EGRET}

Observations of the diffuse gamma-ray background by EGRET have been used to set limits on particles decaying either into $q \overline{q} $ or $\gamma \gamma$ \cite{Kribs:1996ac}, as reproduced in Table \ref{Tab: astro limits}.  We adapted these limits for the other decay modes shown in the Table.  We took the age of the universe to be $t_0 = 13.72 \pm 0.12 \text{~Gyr} \approx 4.3 \times 10^{17} ~\s$ and the abundance of dark matter to be $\Omega_\text{DM} h^2 \approx 0.11$ \cite{Komatsu:2008hk}.  For the $e^+ e^-$, $\mu^+ \mu^-$, and $\nu\nu$ decay modes the limit comes from assuming that these produce a hard $W$ or $Z$ from final state radiation $\sim 10^{-2}$ of the time.  The limit on the lifetime is given conservatively as $3 \times 10^{-3}$ times the limit on the $W^+ W^-$ decay mode, since only one gauge boson is radiated and its energy is slightly below $\frac{m_\psi}{2}$.  For the $\tau^+ \tau^-$ decay mode, the strongest limit comes from considering the hadronic branching fraction of the $\tau$.  The $\tau$ decays into leptons, $e \overline{\nu}_e \nu_\tau$ and $\mu \overline{\nu}_\mu \nu_\tau$, 30\% of the time.  The rest of the decay modes have several hadrons and one $\nu_\tau$ which carries away at most one-half of the energy \cite{Amsler:2008zz}.  Thus we estimate the hadronic fraction of the energy from the decay as $\frac{1}{2} \times 0.7 \approx 0.4$.  The limit on $q \overline{q}$ is relatively insensitive to the mass of the decaying particle in our range of interest $100 ~\GeV \lesssim m_\psi \lesssim 10 ~\TeV$.  We take this to imply that it depends only on the total energy produced in the decay and not as much on the shape of the spectrum, giving a limit on the decay width into $\tau$'s which is a factor of 0.4 of that into $q \overline{q}$.  To set a limit on decays into $W^+ W^-$, the ratio between the photon yield from $W^+ W^-$ and $q \overline{q}$ is approximated as $\frac{2}{3}$ from \cite{Cirelli:2008pk}.

\subsection{Galactic Gamma-Rays From HESS}

HESS observations of gamma rays above 200 GeV from the Galactic ridge \cite{Aharonian:2006au} can also be used to limit the partial decay rates of dark matter with mass $m_\psi > 400 ~\GeV$.  The limit on the flux of gamma rays comes from this HESS analysis in which the flux from an area near the galactic center ($-0.8^\circ < l < 0.8^\circ$ and $0.8^\circ < b < 1.5^\circ$) was taken as background and subtracted from the flux in the galactic center region ($-0.8^\circ < l < 0.8^\circ$ and $-0.3^\circ < b < 0.3^\circ$) and the resulting flux reported.  Our limit on the decay mode $\psi \to \gamma \gamma$ is found by taking a similar difference in the flux from decays and setting this equal to the observed flux.  Because we are considering decays to $\gamma \gamma$ they give a line in the gamma-ray spectrum whose intensity need only be compared to the observed flux in one energy bin.  This is similar to the analysis in \cite{Mack:2008wu} and, as a check, their limit on a dark matter annihilation cross section agrees with our quoted limit on the lifetime.

The photon flux from decays is given by
\begin{equation}
\label{Eqn: Decay flux}
\Phi_\text{decay} =  \frac{\Gamma N_\gamma}{4 \pi m_\psi} \int_{\Delta \Omega} \rho \, dr d\Omega
\end{equation}
where the integral is taken over a line of sight from the earth within a solid angle $\Delta \Omega$, $r$ is the distance from the earth, $m_\psi$ is the mass of the dark matter and $\Gamma$ is its decay rate, and $N_\gamma$ is the number of photons from the  decay which we will set equal to 2 for our limits.  The density of dark matter is taken as
\begin{equation}
\label{Eqn:halo profile}
\rho(s) = \frac{\rho_0}{ \left( \frac{s}{r_s} \right)^\gamma  \left( 1 +  \left( \frac{s}{r_s} \right)^\alpha \right)^\frac{\beta - \gamma}{\alpha} }
\end{equation}
where $s$ is the radial coordinate from the galactic center.  We use the Kravtsov profile \cite{Kravtsov:1997dp} with $\left( \alpha, \beta, \gamma \right) = \left( 2, 3, 0.4 \right)$, $r_s = 10 ~\text{kpc}$, and $\rho_0$ is fixed by $\rho(8.5 ~\text{kpc}) = 0.37 ~\frac{\GeV}{\text{cm}^3}$ for our limits.  This gives conservative limits since the flux from the galactic center is much less than in the commonly-used NFW profile \cite{Navarro:1995iw} with $\left( \alpha, \beta, \gamma \right) = \left( 1, 3, 1 \right)$, $r_s = 20 ~\text{kpc}$, and $\rho(8.5 ~\text{kpc}) = 0.3 ~\frac{\GeV}{\text{cm}^3}$.  There is an even more sharply peaked profile, the Moore profile \cite{Moore:1997sg}, which is defined by Eqn. \eqref{Eqn:halo profile} with $\left( \alpha, \beta, \gamma \right) = \left( 1.5, 3, 1.5 \right)$, $r_s = 28 ~\text{kpc}$, and $\rho(8.5 ~\text{kpc}) = 0.27 ~\frac{\GeV}{\text{cm}^3}$.  There is also the very conservative Burkert profile \cite{Burkert:1995yz, Salucci:2000ps, Salucci:2007tm}
\begin{equation}
\rho(s) = \frac{\rho_0 }{ \left( 1+ \frac{s}{r_s} \right) \left( 1+ \frac{s^2}{r^2_s} \right) }
\end{equation}
where $\rho_0 = 0.839 ~\frac{\GeV}{\text{cm}^3}$ and $r_s = 11.7 ~\text{kpc}$.
We sometimes translate limits on the annihilation cross section, $\sigma v$, into limits on the decay rate using the flux from annihilations
\begin{equation}
\label{Eqn: Annihilation flux}
\Phi_\text{annihilation} = \int \frac{\rho^2 \sigma v N_\gamma}{8 \pi m^2_\psi} dr d\Omega.
\end{equation}

A conservative limit on decays to $\gamma \gamma$ is calculated using the Kravtsov profile.  If the NFW profile is used instead, the limit is stronger $\tau > 5 \times 10^{25} ~\s ~ \sqrt{\frac{m_\psi}{1 ~\TeV}}$.  If we had used the Burkert profile the limit would have been much weaker $\tau > 3 \times 10^{22} ~\s ~ \sqrt{\frac{m_\psi}{1 ~\TeV}}$.  In this case the central profile is so flat that there is essentially no difference between the galactic center signal and the flux from the nearby region used for background subtraction, making the limit from our procedure very weak.  However there would then presumably be a much better limit from just comparing the actual flux at the center (without background subtraction) to what was observed.  Thus we believe that the quoted limit in Table \ref{Tab: astro limits} is conservative.

The decay widths to $e^+ e^-$, $\mu^+ \mu^-$, and $\tau^+ \tau^-$ can be limited from the HESS observations.  These light leptons will bremsstrahlung relatively hard photons with a spectrum that can be estimated as (see for example \cite{Birkedal:2005ep})
\begin{equation}
\label{Eqn: FSR spectrum}
\frac{d\Gamma_{ll\gamma}}{dx} \approx \frac{\alpha}{\pi} \left( \frac{1+ \left( 1-x \right)^2 }{x} \right)  \text{log}\left( \frac{m^2_\psi \left( 1-x \right) }{m^2_l} \right) \Gamma_{ll}
\end{equation}
where $\Gamma_{ll}$ is the decay width of $\psi \to ll$, $\Gamma_{ll\gamma}$ is the decay width of $\psi \to ll\gamma$, $\alpha$ is the fine structure constant, $x = \frac{2 E_\gamma}{m_\psi}$, and $m_l$ is the mass of lepton $l$.  Ignoring the logarthmic dependence on $m_\psi$ we estimate this as giving $10^{-2}$ photons per decay for the light leptons and $\half 10^{-2}$ for $\tau$'s with energy high enough to count in the `edge' feature in the final state radiation spectrum.  We then scale the limits from decay to $\gamma \gamma$ by those factors because the edge is assumed to be visible as the line from $\gamma \gamma$.  Clearly, a more realistic analysis would include a better determination of the observability of the edge feature.

We do not place limits on decays to $WW$ or $\nu \nu$ because the spectrum of photons produced by final state radiation does not have a large hard component.  These can produce many softer photons but these are better limited by lower energy gamma ray observations such as EGRET and are counted in the first column.  Although the $q\overline{q}$ decay mode may have a large FSR component, this will still give a bound worse than the HESS bound on the $\gamma \gamma$ mode.  Additionally the $q\overline{q}$ mode produces $\pi^0$'s which decay to photons but these are at low energies so HESS cannot place good limits on them.  So the $q\overline{q}$ mode is also better limited by the EGRET observations.

\subsection{Neutrino Limits From SuperK, AMANDA, and Frejus}

To find the limits on dark matter decays from astrophysical neutrino observations we start by finding the limits on decays directly into two neutrinos using \cite{Yuksel:2007ac}.  The given limit on annihilation cross section into $\nu\nu$ is almost independent of the dark matter mass $m_\psi$ in the range $10 ~\GeV < m_\psi < 1 ~\TeV$ so we simply take it to be a constant.  This can be converted to a limit on the decay rate by comparing the neutrino flux from annihilations at the limit to the neutrino flux from decays using Eqns. \eqref{Eqn: Decay flux} and \eqref{Eqn: Annihilation flux}.  This gives the limit on the decay rate to $\nu \nu$.  This limit comes from considering the signal from a $30^\circ$ half-angle cone around the galactic center (so called "Halo Angular" in \cite{Yuksel:2007ac}).  The limit from using the signal from the full sky is only a factor of 3 worse.  Though the signal comes from the direction of the galactic center, the cone is wide enough that the given limit is essentially independent of whether the Kravtsov, NFW, or Burkert profile is used.

The limit on the decay rate to $e^+ e^-$ is set by the minimum branching ratio for the electrons to bremsstrahlung a hard $W$ or $Z$, which is $\sim 10^{-2}$.  So we conservatively take the limit to be $3 \times 10^{-3}$ times the limit from neutrino observations on decays into $W^+W^-$ because only one gauge boson is radiated and its energy is slightly less than $\frac{m_\psi}{2}$.  The limits on $\mu^+ \mu^-$ come from the fact that muons always decay as $\mu \to e \overline{\nu}_e \nu_\mu$ and the $\nu_\mu$ tends to carry away almost half the energy.  Further all produced neutrinos will oscillate a large number of times over these galactic distances before reaching the Earth so all neutrinos count equally for detection (and \cite{Yuksel:2007ac} already assumed for their limit that all three neutrino species are produced equally).  So we ignore the soft $\overline{\nu}_e$ and give the limit as $\frac{1}{4} \tau_{\nu\nu}$ where $\tau_{\nu\nu}$ is the limit on the lifetime into $\nu\nu$.  The factor $\frac{1}{4}$ comes from assuming the $\nu_\mu$ has half the energy of the $\mu$ and the neutrino background scales as $E^{-2}$.  Equivalently, the bound on $\tau_{\nu\nu}$ scales with $m_\psi$ (because the bound on annihilations from \cite{Yuksel:2007ac} is constant in $m_\psi$) so a decay to $\mu^+ \mu^-$ is like a decay to $\nu\nu$ but with a decaying particle of half the mass and half the dark matter density.  Similarly, the limit on $\tau^+ \tau^-$ of $\frac{1}{4} \tau_{\nu\nu}$ is set by assuming that the $\nu_\tau$, which is always produced in $\tau$ decay, generally carries away about half the energy of the $\tau$.  Most $\tau$ decays are two- or three-body so we expect this to be a good approximation \cite{Amsler:2008zz}.

The $W$ decays one-third of the time to $l\nu$ \cite{Amsler:2008zz} and the neutrino has about half the energy so the decay to $W^+W^-$ is limited to $\frac{1}{12} \tau_{\nu\nu}$.  Similarly the $Z$ decays 20\% of the time to $\nu\nu$ \cite{Amsler:2008zz} and since there are two neutrinos, each of which carries away about half the energy, the limit on decays to $ZZ$ is slightly stronger: $\frac{2}{5} \frac{1}{4} \tau_{\nu\nu}$.

We do not limit decays to $\gamma \gamma$ or $q \overline{q}$ because we expect these to be better limited by direct gamma ray and antiproton observations.

\subsection{Positrons and Antiprotons from PAMELA}

  
We translate recently published limits from PAMELA on the annihilation cross sections of dark matter into the various final states into limits on the decay rate.  This can be done because a dark matter particle of mass $m_\psi$ decaying into one of the given final states (e.g. $q\overline{q}$) yields exactly the same spectrum of products as two dark matter particles of mass $\half m_\psi$ annihilating into the same final state.  To translate the limit on annihilation cross section we set $\Phi_\text{decay} = \Phi_\text{annihilation}$ from Eqns \eqref{Eqn: Decay flux} and \eqref{Eqn: Annihilation flux} but we must use $\half m_\psi$ instead of $m_\psi$ in Eqn. \eqref{Eqn: Annihilation flux} for $\Phi_\text{annihilation}$.  Also, we integrate over the entire sky, $\Delta \Omega = 4 \pi$, but only over a local sphere out to a radius $r_\text{max} = 5 ~\text{kpc}$.  This is a crude model for the fact that antiprotons and positrons do not propagate simply like gamma rays do.  We can ignore the subtleties of this propagation because we are not computing the actual flux observed, just the ratio between the flux from decays and from annihilations.  We then just take the simple model that these particles only arrive at earth from a distance of $r_\text{max} \sim \OO(5 ~\text{kpc})$.  Using these assumptions, the lifetime of a decaying dark matter particle with mass $m_\psi$ that corresponds to the annihilation rate of a dark matter particle with mass $\half m_\psi$ and cross section $\sigma v$ is:
\begin{equation}
\label{Eqn: annihilations to decays}
\tau = 4 \times 10^{28} \s \left( \frac{m_\psi}{\TeV} \right) \left( \frac{3 \times 10^{-26} \frac{\text{cm}^3}{\s} }{\sigma v} \right).
\end{equation}
It turns out that this is almost independent of halo profile and the size, $r_\text{max}$, of the local sphere used to define it. 
We expect limits from ATIC to be similar to our limits from PAMELA because roughly the same signal that fits ATIC will fit PAMELA (see for example Section \ref{Sec: Astro signals}).
 
The limits from positrons in the $e^+ e^-$, $\mu^+ \mu^-$, $\tau^+ \tau^-$, and $WW$ channels are translated from the annihilation cross section limits from \cite{Cholis:2008hb}.  Really we use the largest annihilation cross section which could explain the observed PAMELA positron excess \cite{Adriani:2008zr} given propagation uncertainties (model B of  \cite{Cholis:2008hb}) and translate this into a decay rate.  This means that lifetimes around and up to an order of magnitude greater than those given in Table \ref{Tab: astro limits} are the best fit lifetimes for explaining the PAMELA positron excess.  This is most true for the lepton channels, while decays to $WW$ do not seem to fit the shape of the positron spectrum very well (see e.g. \cite{Cholis:2008hb, Cirelli:2008pk}).  Note that these lifetimes are in qualitative agreement with those found in \cite{Ibarra:2008jk}.  Note that we do not place a limit on the $q\overline{q}$ channel because this is better limited by the PAMELA antiproton measurement.  The limit on $\nu \nu$ comes from assuming the usual $3 \times 10^{-3}$ factor times the $WW$ limit from one of the neutrinos producing a $W$ from final state radiation.  It is possible that a stronger limit could be set by considering soft W bremsstrahlung from the neutrino, turning the neutrino into a hard positron and the limit would then come from that positron and not from the decay of the W.  We do not attempt to estimate this.

The limit from positrons on the $\gamma \gamma$ decay channel arises when positrons are produced through an off-shell photon.  The relative branching ratio of this decay is well known from $\pi^0$ decay \cite{Pich:1983zk}
\begin{equation}
\label{Eqn: off-shell photon to fermions}
\frac{\Gamma_{\psi \to \gamma f^+ f^-}}{\Gamma_{\psi \to \gamma \gamma}} = \frac{4\alpha Q^2}{3 \pi} \left( \text{ln} \left( \frac{m_\psi}{m_f} \right) - \frac{7}{4}  \right)
\end{equation}
where $f$ is a fermion of charge $Q$, lighter than $\psi$.  Counting the production of muons as well (since they always decay to electrons), a conservative estimate for this branching fraction is 4\% in our range of masses $100 ~\GeV \lesssim m_\psi \lesssim 1 ~\TeV$.  We assume the produced positron to have an energy around $\sim \frac{m_\psi}{3}$.  The scaling of the limit on the $e^+e^-$ channel with $m_\psi$ comes from the scaling of the number density of dark matter.  Thus the energy of the produced positron is not very relevant in our range of $m_\psi$ so the limit on the $\gamma \gamma$ decay channel is just 4\% of the limit on $e^+e^-$.

The limits from antiprotons on the $q\overline{q}$ and $WW$ channels come from comparing the antiproton fluxes computed in \cite{Cirelli:2008pk} with data from PAMELA  \cite{Boezio:2008mp, Adriani:2008zq}, finding the limit this gives on annihilation cross section, and converting this into a limit on the decay rate using Eqn. \eqref{Eqn: annihilations to decays}.  These are almost exactly the same limits as would be derived by translating the limits on annihilation cross section from \cite{Donato:2008jk} into limits on decay rate.  The limits on the $e^+ e^-$, $\mu^+ \mu^-$, $\tau^+ \tau^-$, and $\nu \nu$ channels come from hard W bremsstrahlung from the leptons producing antiprotons and we take this to be $3 \times 10^{-3}$ of the limit on $WW$ (the same factor as used above).  Note that these antiproton limits are most applicable in the range $100 ~\GeV \lesssim m_\psi \lesssim 1~\TeV$ and are essentially independent of mass in that range (equivalently the annihilation cross section limits scale linearly with $m_\psi$).

The limit from antiprotons on the $\gamma \gamma$ decay channel comes when one of the photons is off-shell and produces a $q\overline{q}$ pair.  Using Eqn. \eqref{Eqn: off-shell photon to fermions} and summing the contributions from the four light quarks (using their current masses) with a factor of 3 for number of colors, we estimate a branching ratio $B(\psi \to \gamma q \overline{q}) \sim 4\%$ at $m_\psi = 100 ~\GeV$ (rising to $\sim 6 \%$ at $m_\psi = 1 ~\TeV$).  Note that this is in rough agreement with the one-photon rate found in \cite{Covi:2001nw}.  If we conservatively assume that the quark pair has energy $=\half m_\psi$ then we find that the limit is $0.02$ of the limit on the $q\overline{q}$ decay channel.


\subsubsection{Explaining the Electron/Positron Excess}
\label{sec:explaining-pamela}

In the previous section we set limits on the decay rate of a dark matter particle using several experiments including PAMELA.  The limits from positron observations were less stringent than they would have been had PAMELA not seen an excess.  In this section we consider what is necessary to explain the PAMELA/ATIC positron excesses.  A more detailed analysis of individual models is presented in Section \ref{Sec: Astro signals}.

Due to the large positron signal and hard spectrum detected by PAMELA the best fit to the data is achieved with a decaying (or annihilating) particle with direct channel to leptons.  Further, the lack of a signal in antiprotons disfavors channels with a large hadronic branching fraction such as $q\overline{q}$ or $WW$.  Thus the PAMELA excess is fit well by a dark matter particle that decays to $e^+e^-$, $\mu^+\mu^-$, or $\tau^+\tau^-$ with the lighter two leptons providing the best fit \cite{Cholis:2008hb, Cirelli:2008pk}.  As in the previous section, we translate the cross sections given in \cite{Cholis:2008hb} using Eqn. \eqref{Eqn: annihilations to decays} to find the range of lifetimes which best fit the PAMELA positron data given the propagation uncertainties.  We find the best fit for masses in the range $100 ~\GeV \lesssim m_\psi \lesssim 1 ~\TeV$ is
\begin{equation}
\label{Eqn: pamela lifetimes}
2 \times 10^{25} ~\s \left( \frac{\TeV}{m_\psi} \right) \lesssim \tau \lesssim 8 \times 10^{26} ~\s \left( \frac{\TeV}{m_\psi} \right)
\end{equation}
The range comes from the uncertainties in the propagation of positrons in our galaxy and we estimate it by using the maximum and minimum propagation models from \cite{Cholis:2008hb} (models B and C).  It is difficult to fit just the PAMELA positron excess (not even considering antiprotons) with decays to $WW$ unless the dark matter is light $m_\psi \lesssim 300 ~\GeV$ or very heavy $m_\psi \gtrsim 4 ~\TeV$ \cite{Cirelli:2008pk}.  We will not consider these mass ranges because we also wish to fit the ATIC spectrum which requires masses in the range $1 ~\TeV \lesssim m_\psi \lesssim 2 ~\TeV$.  Note that while this paper was in preparation \cite{Nardi:2008ix} appeared which generally agrees with the results presented here.

So for explaining the PAMELA/ATIC excesses, we are most interested in a model in which dark matter decay releases between 1 TeV and 2 TeV of  energy, dominantly decaying to hard leptons with a lifetime given in Eqn. \eqref{Eqn: pamela lifetimes}.  Such a model must also avoid the limits on other subdominant decay modes to hadrons which are given in Table \ref{Tab: astro limits}.  For example a $1.6 ~\TeV$ dark matter particle produces an endpoint of the electron/positron spectrum around $800 ~\GeV$, in agreement with ATIC.  Such a particle must have a lifetime to decay to leptons in the range $10^{25} ~\s \lesssim \tau \lesssim 5 \times 10^{26} ~\s$ and a lifetime to decay to $q\overline{q}$ longer than $\tau \gtrsim 10^{27} ~\s$.

\section{Dark Matter Decays by Dimension 6 Operators}
\label{Sec: Dim 6 decays}

The decays  of a particle with dark matter abundance into the standard model are constrained by many astrophysical observations. These limits (see section \ref{Sec: Astro Limits}) on the decay lifetimes are in the range $10^{23}$ to $10^{26}$ seconds. Current experiments like PAMELA, ATIC, Fermi {\it etc.} probe even longer lifetimes. These lifetimes are in the range expected for the decays of a TeV mass particle through dimension 6 GUT suppressed operators. In this section, we present a general operator analysis of such dimension 6 operators. 

A dimension 6 operator generated by integrating out a particle of mass $M$ scales as $M^{-4}$. This strong dependence on $M$ implies a wide range of possible lifetimes from small variations in $M$. The decay lifetime is also a strong function of the phase space available for the decay, with the lifetime scaling up rapidly with the number of final state particles produced in the decay. Since we wish to explore a wide class of possible operators, we will consider decays with different numbers of particles in their final state. Motivated by the decay lifetime $\sim 10^{26}$ s being probed by experiments, we exploit the strong dependence of the lifetime on the scale $M$ to appropriately lower $M$ to counter the suppression from multi body phase space factors and yield a lifetime  $\sim 10^{26}$ s. In Table \ref{rough-lifetimes}, we present the scale $M$ required to yield this lifetime for scenarios with different numbers of final state particles. The scale $M$ varies from the putative scale where the gauge couplings meet $\sim 2 \times 10^{16}$ GeV for a two body decay to the right handed neutrino mass scale $\sim 10^{14}$ GeV for a five body decay. We note that the scale $10^{14}$ GeV also emerges as the KK scale in the Horava-Witten scenario. In the rest of the paper, we will loosely refer to these scales as $M_{GUT}$. 




\begin{table}[h]
\begin{center}
\begin{tabular}{|c|c|}
\hline
Number of Final & Scale $M$ (GeV) \\
State Particles & \\
\hline
2  &  $10^{16}$ \\
3  &  $3 \times10^{15}$\\
4  &  $5 \times 10^{14}$\\
5 &  $10^{14}$\\
\hline
\end{tabular}
\end{center}
\caption[The No. of final state particles and the scale $M$]{\label{rough-lifetimes} A rough estimate of the scale $M$ that suppresses the dimension 6 operator mediating the decay of a TeV mass particle in order to get a lifetime $\sim 10^{26}$ s for decays with various numbers of particles in the final state. Phase space is accounted for approximately using~\cite{Jackson}. Lifetimes scale as~$ M^4$. Specific decays may have other suppression or enhancement factors as discussed in the text.}
\end{table}



The observations of PAMELA/ATIC can be explained through the decays of a TeV mass particle with dark matter abundance if its lifetime $\sim 10^{26}$ s (see section \ref{Sec: Astro Limits}). PAMELA, in particular, observes an excess in the lepton channel and constrains the hadronic channels. In our operator analysis, we highlight operators that can fit the PAMELA/ATIC data. However, we also include operators that dominantly produce other final states like photons, neutrinos and hadrons in our survey. While these operators will not explain the PAMELA data, they provide new signals for upcoming experiments like Fermi. For concreteness, we consider $SU(5)$ GUT models. We classify the dimension 6 operators into two categories:  R-parity conserving operators and R-breaking ones.  In the R-conserving case, we will add singlet superfield(s) to the MSSM and consider decays from the MSSM to the singlet sector and vice versa. The singlets may be representatives of a more complicated sector (see section \ref{LithiumPAMELA}).  In the R-breaking part we will consider the decay of the MSSM LSP into standard model particles.  As a preview of our results, a partial list of operators and their associated final states are summarized in Table~\ref{DM-operators}.


\begin{table}[h]
\begin{center}
\begin{footnotesize}
\begin{tabular}{|c|c|c|c|c|}
\hline
Operator in SU(5)& Operator in MSSM& Final State & Lifetime (sec) & Mass Scale (GeV) \\
& & & ($M_{GUT} \sim10^{16}$ GeV) & (lifetime $\sim 10^{26}$ sec) \\
\hline
R-parity conserving& ~&~&~&   \\
\hline
$S^\dagger S 10^\dagger 10$ &$S^\dagger S Q^\dagger Q,~S^\dagger S U^\dagger U,~S^\dagger S  E^\dagger E$ & leptons & $10^{26}$& $10^{16}$   \\
$S^\dagger S H_{u(d)}^\dagger  H_{u(d)}$&$S^\dagger S H_{u(d)}^\dagger  H_{u(d)}$ &quarks& $10^{26}$& $10^{16}$\\
$S^\dagger 10_f \bar5_f^\dagger 10_f$& $S^\dagger Q L^\dagger U, S^\dagger U D^\dagger E, S^\dagger Q D^\dagger Q$&quarks and leptons& $5 \times 10^{28}$&$10^{15}$ \\
$S^\dagger \bar5_f H_u^\dagger 10_f$& $S^\dagger L H_u^\dagger E,~S^\dagger D H_u^\dagger Q$&leptons& $ 10^{26}$&$10^{16}$ \\
$S^2 \mathcal{W}_\alpha \mathcal{W^\alpha}$ &$S^2 \mathcal{W}_{EM} \mathcal{W}_{EM},~S^2 \mathcal{W}_{Z} \mathcal{W}_{Z}$ &  $\gamma$ (line) & $ 10^{26}$& $10^{16}$\\
\hline
Hard R violating& &~&~&\\
\hline
$\bar5_f(\Sigma \bar5_f)\bar5_f(\Sigma \bar5_f)\bar5_f$&$DDDLL$&quarks and leptons&$10^{37}$& $10^{13}$\\
\hline
Soft R violating& &~&~&\\
\hline
$\mathcal{L} \ni \frac{m_{SUSY}^4}{M_{GUT}^2} H_u \tilde{\bar5}_{f}$ &$\frac{m_{SUSY}^4}{M_{GUT}^2} H_u \tilde{\ell}$&quarks& $4\times 10^{30}$& $7 \times 10^{14}$\\
$\mathcal{L} \ni \frac{m_{SUSY}^3}{M_{GUT}^2} \tilde{H}_u\bar5_{f}$& $\frac{m_{SUSY}^3}{M_{GUT}^2} \tilde{H}_u \ell $&leptons & $6\times 10^{32} $&$10^{14}$ \\
$\mathcal{L} \ni \frac{m_{SUSY}}{M_{GUT}^2}H_d \tilde{W}  \partial\hspace{-2mm}/\hspace{.3mm} \bar 5_f^\dagger $&$\frac{m_{SUSY}}{M_{GUT}^2}H_d \tilde{W} \partial\hspace{-2mm}/\hspace{.3mm} \ell^\dagger$ &$\gamma+\nu$ & $2\times 10^{32}$&$10^{14}$\\
\hline
\end{tabular}
\end{footnotesize}
\end{center}
\caption[Dimension 6 Operator Catalog]{\label{DM-operators}A partial list of dimension 6 GUT suppressed decay operators. For each operator, we list its most probable MSSM final state. The lifetime column gives the shortest lifetime that this operator can yield when the scale suppressing the operator is $\sim 10^{16}$ GeV. In the mass scale column, we list the highest possible scale that can suppress the operator in order for it to yield a lifetime $\sim 10^{26}$ seconds. Assumptions (see text) about the low energy MSSM spectrum were made in order to derive these results. All the operators are in superfield notation except for the soft R violating operators.}
\end{table}

\subsection{ $R$-parity Conserving Operators}
\label{RParityConserving}

The lifetimes of the decays of MSSM particles to the MSSM LSP are far shorter than the dimension 6 GUT suppressed lifetime  $\sim 10^{26}$ s currently probed by experiments. In a R-parity conserving theory, the MSSM cannot lead to decays with such long lifetimes since the MSSM LSP, being the lightest particle that carries $R$ parity, is stable \cite{Dimopoulos:1981zb}. These decays require the introduction of new, TeV scale multiplets. Limits from dark matter direct detection \cite{DirectDetection, Beck:1993sb} and heavy element searches \cite{HeavyElement} greatly constrain the possible standard model representations of the new particle species. In this paper, we will add a new singlet chiral superfield $S$  in addition to the MSSM\footnote{Adding new states that are charged under the standard model gauge group, such as a $\l 5, \FiveBar \r$, is possible. However, evading limits from direct detection requires some additional model building}. The singlet may emerge naturally as one of the light moduli of string theory or  it could be a representative of another sector of the theory (see section \ref{LithiumPAMELA}). 




Decays between the singlet sector and the MSSM can happen through the dimension 6 GUT suppressed operators in Table \ref{DM-operators} only if there are no other faster decay modes between the two sectors. Such decay modes might be allowed if there are lower dimensional operators between the singlet sector and the MSSM. In subsection  \ref{RConservingSingletConserving}, we consider models where lower dimensional operators are forbidden by the imposition of a singlet parity under which $S$ has parity -1. This parity could be softly broken if the scalar component of the singlet $\tilde{s}$ develops a TeV scale vev $\langle \tilde s \rangle$.  In subsection \ref{Sec: S-number violating}, we consider models without singlet parity that require additional model building to ensure the absence of dangerous lower dimensional operators between the singlets and the MSSM.

For the rest of the paper, we will adopt the following $SU(5)$ conventions: $S$ will refer to a $SU(5)$ singlet, $\l 5, \FiveBar\r$ to a fundamental and an antifundamental of $SU(5)$ and $\l 10, \TenBar\r$ to the antisymmetric tensor of $SU(5)$. The subscript $f$ identifies standard model fields, $\mathcal{W}_\alpha$ denotes standard model gauge fields and $H_u$ and $H_d$ are standard model higgs fields. The subscript GUT refers to a field with a GUT scale mass. $\tilde l$, $\tilde e$, $\tilde \nu$, $\tilde q$, $\tilde u$ and $\tilde d$ refer to sleptons and squarks. $\tilde{H}_u$ and $\tilde{H}_d$ refer to higgsinos and $\tilde{W}$ refers to a wino. 

\subsubsection{Models with Singlet Parity}
\label{RConservingSingletConserving}

The R parity conserving operators in Table \ref{DM-operators} that also conserve singlet parity are

\begin{equation}
\label{Eqn:SingletParity}
\frac{S^{\dagger} S \FiveBarDagger_f \FiveBar_f}{M_{GUT}^2} \text{, } \frac{S^{\dagger} S \TenDagger_f 10_f}{M_{GUT}^2} \text{, } \frac{S^{\dagger} S H_{u\l d \r}^{\dagger} H_{u\l d \r}}{M_{GUT}^2} \text{ and } \frac{S^2 \mathcal{W}_{\alpha} \mathcal{W}^{\alpha}}{M_{GUT}^2}
\end{equation}
The decay topologies of these operators are determined by the low energy spectrum of the theory. SUSY breaking will split $m_s$ and $m_{\tilde{s}}$, the singlet fermion and scalar masses respectively. It is conceivable that SUSY breaking soft masses may make $m_{\tilde{s}}^2$ negative leading to a TeV scale vev $\langle \tilde s \rangle$ for $\tilde s$. In this case, additional interactions in the singlet sector will be required to stabilize the vev at the TeV scale. A decay mode with a singlet vev will typically dominate over modes without a vev since the former have fewer particles in their final state leading to smaller phase space suppression (see discussion in sub section \ref{LeptonicDecays}). 

Upon SUSY breaking, there are two distinct decay topologies: 

\begin{enumerate}
\item A  component of the singlet is heavier than the MSSM LSP and this component can then decay to it. The relic abundance of the singlet can be generated if the singlet is a part of a more complicated sector, for example, through the decays of heavier standard model multiplets (see section \ref{LithiumPAMELA}). 

\item Alternatively, the MSSM LSP is heavier than the singlets. In this case, the MSSM LSP will decay to the singlets. 
\end{enumerate}

We now divide our discussion further based on the final state particles produced in the decay. Motivated by PAMELA/ATIC and Fermi, we will be particularly interested in operators that produce leptons and photons. \newline


\paragraph{{\bf Leptonic decays} \newline \newline}
\label{LeptonicDecays}

The R and singlet parity conserving operators in Table \ref{DM-operators} that contain leptonic final states are

\begin{equation}
\frac{S^\dagger S \bar 5_f^\dagger \bar 5_f}{M_{GUT}^2} \qquad \mbox{and} \qquad\frac{S^\dagger S 10_f^\dagger 10_f}{M_{GUT}^2}
\label{SdS10d10}
\end{equation}

These operators can be generated by integrating out a GUT scale $U(1)_{B-L}$ gauge boson under which both the MSSM and the $S$ fields are charged. At low energies the first of these operators will lead to couplings of the form

\begin{equation}
\frac{ \tilde s }{M_{GUT}^2}\tilde l^* \, s^\dagger \partial\hspace{-2mm}/\hspace{.3mm}l  \qquad \mbox{and} \qquad
\frac{ \tilde s }{M_{GUT}^2} \tilde d^* \, s^\dagger \partial\hspace{-2mm}/\hspace{.3mm}d
\label{sds10d10comp0}
\end{equation}

while the second operator will lead to similar operators involving $u$, $q$ and $e$. Here we have suppressed all flavor indices.

We first consider the case when the singlet scalar develops a TeV scale vev $\langle \tilde{s} \rangle$. In this case, the operators in (\ref{sds10d10comp0}) mediate the decay of the singlets to the MSSM LSP or vice versa. With a vev insertion, the operators in (\ref{sds10d10comp0}) yield: 

\begin{equation}
\frac{\langle \tilde s \rangle}{M_{GUT}^2}\tilde l^* \, s^\dagger \partial\hspace{-2mm}/\hspace{.3mm}l  \qquad \mbox{and} \qquad
\frac{\langle \tilde s \rangle}{M_{GUT}^2} \tilde d^* \, s^\dagger \partial\hspace{-2mm}/\hspace{.3mm}d
\label{sds10d10comp1}
\end{equation}

\begin{figure}
\begin{center}
  \includegraphics[height=4.5cm]{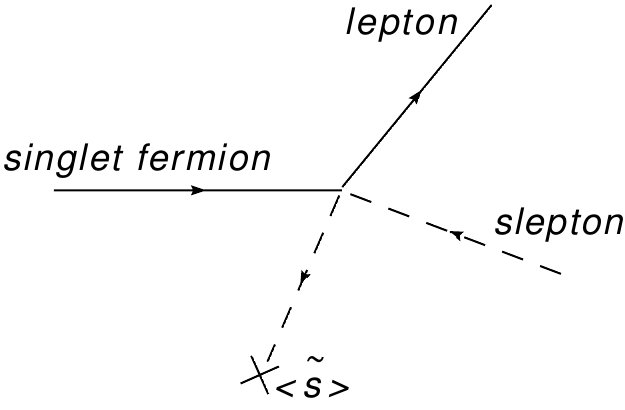}
   \caption[$s \to l, \tilde{l}$]{A singlet fermion decaying to a lepton, slepton pair with a singlet scalar vev $\langle \tilde s \rangle$ insertion. }\label{fig-singletsleptonlepton}
   \end{center}
\end{figure}

When the singlet fermion is heavier than the MSSM LSP, these interactions mediate its decay into  MSSM states. In particular, when the singlet fermion is heavier than a slepton, it can decay to a slepton, lepton pair. The component operators of ~(\ref{sds10d10comp1})  produce the two body final state (see figure \ref{fig-singletsleptonlepton}), 

\begin{equation}
s \to l^\pm\,\tilde l^\mp 
\end{equation}
with a lifetime of
\begin{equation}
\tau_{s\to l^\pm \tilde l^\mp}\sim 2 \times 10^{26} 
\left(\frac{1\mbox{ TeV}}{\Delta m}\right)^3
\left(\frac{1\mbox{ TeV}}{\langle \tilde s \rangle}\right)^2 
\left(\frac{M_{GUT}}{10^{16}\mbox{ GeV}}\right)^4 
\mbox{ sec}
\end{equation}
per lepton generation, where $\Delta m=m_s-m_{\tilde l}$. 

\begin{figure}
\begin{center}
  \includegraphics[height=4.5cm]{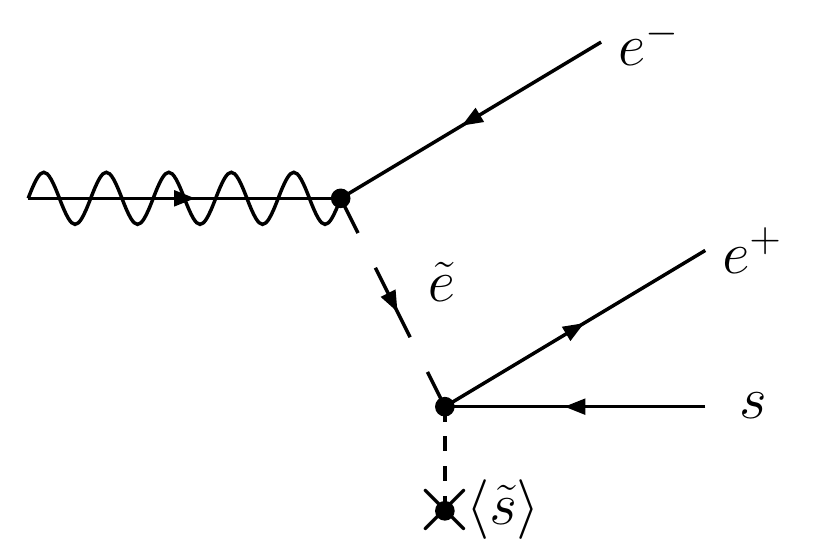}
   \caption[$LSP \to s, l^+, l^-$]{MSSM neutralino decaying to a singlet LSP and an $e^+e^-$ pair. The decay is dominantly into leptons because sleptons are typically lighter than squarks.}\label{fig-DMdecay}
   \end{center}
\end{figure}

When the MSSM LSP is heavier than the singlet fermion, it will decay to the singlet fermion and a lepton, anti lepton pair through the operators of equation ~(\ref{SdS10d10}). These decays (see Figure~\ref{fig-DMdecay}) are of the form

\begin{equation}
LSP \to l^++l^-+s 
\label{LSP2S}
\end{equation}
The lifetime for this three body decay is
\begin{equation}
\tau_{LSP\to sl^+l^-}\sim 10^{26} \left(\frac{1\mbox{ TeV}}{m_\chi}\right)^3\left(\frac{1\mbox{ TeV}}{\langle \tilde s \rangle}\right)^2 \left(\frac{M_{GUT}}{10^{15}\mbox{ GeV}}\right)^4 \left(\frac{R_l}{0.5}\right)^4 \mbox{ sec}
\end{equation}
where $R_l$ is the ratio of the LSP mass to the slepton mass and we assumed that $m_\chi\gg m_s$\,; if this is not the case $m_\chi$ should be replaced by the available energy in the decay $\Delta m=m_\chi-m_s$. Similarly,  when the singlet fermion is heavier than the MSSM LSP but lighter than the slepton, the singlet fermion will decay to the MSSM LSP through a 3 body decay mediated by an off-shell slepton. Note that the decay lifetime $\sim 10^{26}$ s when the mass scale suppressing the decay is $M_{GUT} \sim 10^{15} \text{ GeV}$. In a $U(1)_{B-L}$ UV completion of these operators, this scale is the vev of the broken $B-L$ gauge symmetry. The $B-L$ symmetry must be broken slightly below the GUT scale ({\it i.e.} the putative scale where the gauge couplings meet) in order for the three body decays mediated by this gauge sector to have lifetimes of interest to this paper. 

The decays discussed above can also produce quarks in their final states. The decay rate is a strong function of the phase space available for the decay and is hence a strong function of the squark and slepton masses. As discussed in section  \ref{Dim6LHCSignals}, the hadronic branching fraction of these decays can be suppressed if the squarks are slightly heavier than the sleptons. Since squarks are generically heavier than sleptons due to RG running, the suppressed hadronic branching fraction observed by PAMELA is a generic feature of these operators. An interesting possibility emerges when the spectrum allows for the decay of the singlet fermion to an on-shell slepton, lepton pair. This decay produces a primary source of monoenergetic hot leptons. However, the subsequent decay of the slepton to the MSSM LSP will also produce a lepton whose energy is cut off by the slepton and LSP mass difference. With two sources of injection, this decay could explain the secondary "bump" seen by ATIC in addition to the primary bump (see sections \ref{Dim6LHCSignals}  and \ref{Sec: Astro signals}).

The scalar singlet $\tilde{s}$  can decay when the singlet gets a vev. In the presence of such a vev, $\tilde{s}$ can decay to a pair of scalars ({\it i.e.} sleptons and squarks) or fermions ({\it i.e.} leptons and quarks). The decays of $\tilde{s}$ to a pair of sleptons is also mediated by the operators in (\ref{SdS10d10}) which in components yield terms of the form 

\begin{figure}
\begin{center}
  \includegraphics[height=4.5cm]{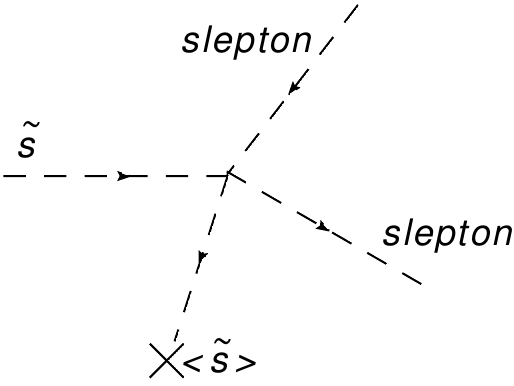}
   \caption[$\tilde s \to l^+ l^-$]{A singlet scalar $\tilde s$ decaying to a slepton pair with a singlet scalar vev $\langle \tilde s \rangle$ insertion. }\label{fig-singletleptonlepton}
   \end{center}
\end{figure}

\begin{equation}
\frac{\langle \tilde s \rangle}{M_{GUT}^2}\tilde s^* \, \partial \tilde{l}^* \partial\hspace{.3mm}\tilde{l}  \qquad \mbox{and} \qquad
\frac{\langle \tilde s \rangle}{M_{GUT}^2} \tilde s^* \, \partial \tilde{d}^* \partial\hspace{.3mm}\tilde{d}
\label{sds10d10compsleptons}
\end{equation}
as well as terms involving $\tilde{q}$, $\tilde{u}$, and $\tilde{e}$. 

$\tilde{s}$ will now decay directly to two sleptons (see figure \ref{fig-singletleptonlepton}) or two squarks, with a lifetime
\begin{equation}
\tau_{\tilde s\to \tilde{l}^\pm \tilde{l}^\mp}\sim 2 \times 10^{26} 
\left(\frac{1\mbox{ TeV}}{m_{\tilde s}}\right)^3
\left(\frac{1\mbox{ TeV}}{\langle \tilde s \rangle}\right)^2 
\left(\frac{M_{GUT}}{10^{16}\mbox{ GeV}}\right)^4 
\mbox{ sec}
\label{2sfermions}
\end{equation}
per generation (and per representation). The hadronic branching fraction of this decay is generically suppressed since squarks are expected to be heavier than sleptons (see  section \ref{Dim6LHCSignals}).  $\tilde{s}$ can also decay to a pair of fermions. These decays are mediated by the operators 

\begin{equation}
\frac{\langle \tilde s \rangle}{M_{GUT}^2}\tilde s^* \, l^\dagger \partial\hspace{-2mm}/\hspace{.3mm}l  \qquad \mbox{and} \qquad
\frac{\langle \tilde s \rangle}{M_{GUT}^2} \tilde s^* \, d^\dagger \partial\hspace{-2mm}/\hspace{.3mm}d
\label{sds10d10comp}
\end{equation}
which can also be extracted from (\ref{SdS10d10}). However, the decays of a scalar to a pair of fermions (of mass $m_f$) are helicity suppressed by $\l\frac{m_f}{m_{\tilde{s}}}\r^2$ \footnote{We thank Hitoshi Murayama for relevant comments.}. These decays will predominantly produce the most massive fermion pair that is allowed by phase space. 

Owing to this helicity suppression, the hadronic branching fraction from the decays of $\tilde{s}$ can be smaller than 0.1 and accommodate the PAMELA anti-proton constraint  if one of the following conditions are satisfied by the scalar singlet mass $m_{\tilde{s}}$, the slepton mass $m_{\tilde{l}}$ and the top quark mass $m_t$. 

\begin{itemize}
\item  $m_{\tilde{s}} \gg m_{\tilde{l}}$, $m_{\tilde{s}} \gg m_t$: In this case, the spectrum allows both sleptons and the top quark to be produced on shell. Hadrons are produced in this process from the decays of the top quark. The branching fraction for top production is  $\l\frac{m_t}{m_{\tilde{s}}}\r^2$ which is smaller than 0.1 if  $m_{\tilde{s}} \gtrapprox 3 m_t$. 
\item $m_{\tilde{s}} > 2 m_{\tilde{l}}$ and $m_{\tilde{s}} < 2 m_t$: In this case, sleptons are produced on shell. Hadrons can be produced in this process either through direct production of the $b$ quark or through off-shell tops. The branching fractions of these hadronic production channels are smaller than 0.1 since the direct production of  $b$ is suppressed by $\l\frac{m_b}{m_{\tilde{s}}}\r^2$ and the decays mediated by off-shell tops are suppressed by additional phase space factors. 
\end{itemize}

When the scalar singlet does not get a vev, $S$ parity is conserved. Decays between the singlets and the MSSM must either involve the decay of the heavier component of the singlet to its lighter partner and the MSSM or the decay of the MSSM to the two singlet components. We first consider the case when one of the singlet components is heavier than the MSSM LSP. Without loss of generality, we assume this component to be the scalar singlet $\tilde s$. $\tilde{s}$ can decay to its fermionic partner and a lepton-slepton pair through a 3 body decay mediated by the operators in (\ref{sds10d10comp0}) if the slepton is light enough to permit the decay. The lifetime for this decay mode is 

\begin{equation}
\tau_{\tilde s\to s\tilde l^\pm l^\mp}\sim  10^{26} 
\left(\frac{1000\mbox{ GeV}}{\Delta m}\right)^5
\left(\frac{M_{GUT}}{10^{15}\mbox{ GeV}}\right)^4 
\mbox{ sec}\,,
\end{equation}

The 3 body decay of $\tilde s$ to the singlet fermion and slepton-lepton pair may be kinematically forbidden if the slepton is heavy. $\tilde s$ then decays to the singlet fermion and the MSSM through a four body decay: 
\begin{equation}
\tilde{s}\to l^++l^-+s +LSP \,.
\label{LSP2SS}
\end{equation}
The lifetime in this case is 
\begin{equation}
\tau_{\tilde{s} \to LSP  s l^\pm l^\mp}\sim 10^{26} 
\left(\frac{1\mbox{ TeV}}{\Delta m}\right)^5
\left(\frac{M_{GUT}}{3 \times 10^{14}\mbox{ GeV}}\right)^4 
\left(\frac{R_l}{0.5}\right)^4 
\mbox{ sec}\,,
\end{equation}

If the MSSM neutralino is heavier than the singlets, then its decays through the operators  in (\ref{sds10d10comp0}) are also four body decays similar to the decay discussed above. The lifetime from this decay is $\sim 10^{26}$ s when $M_{GUT} \sim 3 \times 10^{14} \text{ GeV}$, which is roughly the scale of the right-handed neutrino  in a see-saw scenario. In fact, if this decay is mediated by a $U(1)_{B-L}$ gauge boson, the scale $M_{GUT}$ that suppresses this decay is the vev that breaks the $U(1)_{B-L}$ gauge symmetry which is roughly the mass of the right handed neutrino. 


The decays discussed in this section involve decays between the singlet sector and the MSSM LSP. In order for these dimension 6 GUT suppressed operators to be involved in the decays between these sectors, it is essential that there are no other faster decay modes available in the model. One such mode can be provided by a light gravitino. If the gravitino is the MSSM LSP, then the superparticles of the MSSM will rapidly decay to the gravitino with lifetimes $\sim \frac{ \text{TeV}^5}{F^2}$. If one component of the singlet is heavier than the gravitino, then that component will decay to its superpartner and the gravitino with a lifetime $\sim \frac{\Delta m^5}{F^2}$ where $\Delta m$ is the phase space available for this decay. Since we are interested in the decays of TeV mass particles, this scenario is relevant only when the gravitino mass $\l \frac{F}{M_{pl}} \r$ is less than a TeV {\it i.e.} $F \lessapprox \l10^{11} \text{ GeV}\r^2$. When $F \lessapprox  \l10^{11} \text{ GeV}\r^2$, the above decays occur with lifetimes $\sim 10^6$ s which are far too rapid. 

Another possibility is for the gravitino to be the MSSM LSP and be heavier than the singlets. In this case, the gravitino will decay to the singlet sector with a decay rate $\sim \l\frac{F^3}{M_{pl}^5}\r $ yielding a lifetime $\sim 10^6 \text{ s} \l \frac{\l10^{11} \text{ GeV}\r^2}{F} \r^3$ which is also far too rapid.  The dimension 6 GUT suppressed operators discussed in this section lead to astrophysically interesting decays only when the gravitino is not the MSSM LSP and has a mass larger than $\sim$ TeV. This forces the primordial SUSY breaking scale $F \gtrapprox \l10^{11} \text{ GeV}\r^2$, making the gravitino heavier and forcing it off-shell in the decays mediated by it. Integrating out the gravitino, operators of the form $ \l \frac{m_{\tilde{s}}^2}{F} \r^2  \l \frac{M_{pl}}{F}\r \l \tilde{s}^*   s \tilde{\bar{5}}_f^{*}  \FiveBar_f \r$ are generated.  The masses $m_{\tilde{s}}$ in this operator are the singlet and soft SUSY breaking masses $\sim 1$ TeV. The decays mediated by this operator have lifetimes $\gtrapprox 10^{37}$ s for $F \gtrapprox \l 10^{11} \text{ GeV} \r^2$ and will not compete with dimension 6 GUT operators discussed in this section. 

The constraints on the gravitino mass can be evaded if there are more singlets in the theory. For example, if the dark sector contains flavor, standard model particles may be emitted during ``flavor-changing'' decays in the dark sector.  Consider for example an $SU(6)$ extension of the GUT group on an orbifold. The chiral matter fields and their decomposition to $SU(5)$ representations are 
\begin{equation}
\bar 6_i = \bar 5_i + S_i  \qquad 15_i= 10_i + \hat 5_i\,.
\end{equation}
where the index $i$ represents flavor. We break the $SU(6)$ symmetry by projecting out the light modes of the 5-plet $\hat 5$ by orbifold boundary conditions at the GUT scale. In the absence of the 5-plet, the singlet superfields $S$ do not have yukawa couplings that connect them to the MSSM fields, eliminating direct decay modes between the singlets and the MSSM.

Once the heavy off-diagonal $SU(6)$ gauge multiplets are integrated out,  operators of the form
\begin{equation}
\frac{1}{M_{GUT}^2}S_i^\dagger S_j L_j^\dagger L_i \qquad \mbox{or}\qquad  
\frac{1}{M_{GUT}^2}S_i^\dagger S_j D_j^\dagger D_i
\end{equation}
are generated. These operators may lead to dark-flavor changing decays of $s_i\to s_j + l_i +l_j$ or $s_i\to s_j + d_i +d_j$ if there are mass splittings amongst the singlets. These splittings can arise due to explicit $SU(6)$ breaking terms (which may be present on the brane which breaks this symmetry). Soft SUSY breaking will also contribute to mass splittings in the singlet scalar sector. These splittings can cause the decay of a scalar singlet $\tilde{s}_i$ to a singlet fermion $s_j$ and a lepton $l_i$, slepton $\tilde l_j$ pair. In this case, the lepton and slepton emitted in this process will belong to different families. The hadronic branching fraction of these operators relative to the leptonic channel depends strongly on the masses of the various broken $SU(6)$ gauge bosons. This branching fraction is suppressed if the $SU(6)$ gauge boson that connects the singlets and the leptons is lighter than the boson that connects the singlets and the quarks.  

The decays mediated by these operators are immune to the effects of the gravitino since these decays explicitly require off-diagonal gauge bosons. A relic abundance of the singlets can again be generated through non-thermal processes as discussed in section \ref{LithiumPAMELA}.

\paragraph{{\bf Decays to Higgses} \newline }
\label{HiggsDecays}

The R and singlet parity conserving operators in Table \ref{DM-operators} that contain higgs final states are

\begin{equation}
\frac{S^\dagger S  H_u^\dagger \bar H_u}{M_{GUT}^2}
\qquad
\frac{S^\dagger S  H_d^\dagger H_d}{M_{GUT}^2}\,.
\label{SdSHdH}
\end{equation}
These operators are very similar to the operators discussed in subsection \ref{LeptonicDecays}. They can also be generated by integrating out a GUT scale $U(1)_{B-L}$ gauge sector and the topologies of the decays mediated by these operators are also similar to the decay topologies of the operators discussed in that subsection. However, since these operators involve final state higgses, they will always yield an $\OrderOne$ hadronic branching fraction. 

In any particular UV completion, the leptonic operators discussed in subsection \ref{LeptonicDecays} and the higgs operators presented in this section may be simultaneously present. The hadronic branching fraction of the decays between the singlets and the MSSM is a strong function of the phase space available for the various decay modes. If the sleptons are lighter than the squarks and the higgsinos, the decays will predominantly proceed via the leptonic channels and hence these UV completions will also be compatible with the constraints on the hadronic channel imposed by PAMELA.

\paragraph{{\bf Decays to Gauge Bosons} \newline}

The only operator in Table \ref{DM-operators} that contains gauge boson final states is 

\begin{equation}
\frac{\mathcal{W}^\alpha \mathcal{W}_\alpha  S^2}{M_{GUT}^2} 
\label{s2w2}
\end{equation}

This operator may be generated by integrating out a heavy axion-like or dilaton field that couples linearly to both $\mathcal{W}^2$ and to $S^2$. For example, at the GUT scale we may write a superpotential
\begin{equation}
W= \Mgut\overline{10}_{\gut}10_{\gut} + \Mgut \bar X_{\gut} X_{\gut} 
+X_{\gut} \overline{10}_{\gut}10_{\gut}   + X_{\gut}S^2\,.
\end{equation}
where $\Mgut$  is a GUT scale mass. Integrating out the heavy $10_{\gut}$s will lead to a one loop coupling of the form $X_{\gut}\mathcal{W}^\alpha\mathcal{W}_\alpha/\Mgut$ (see subsection \ref{Axinos}). Using this effective operator and integrating out $X_{\gut}$ leads to the operator
\begin{equation}
\frac{\alpha}{4\pi \Mgut^2} S^2  \mathcal{W}^\alpha \mathcal{W}_\alpha 
\end{equation}

The decay topology of this operator is similar to the topologies already discussed in subsection \ref{LeptonicDecays} but leads to new final states. For example, when the scalar singlet $\tilde s$ develops a TeV scale vev, this operator can lead to decays between the MSSM LSP and the singlet fermion that result in the direct production of a monochromatic photon. The lifetime for this decay is 

\begin{equation}
\tau_{s\to \gamma + LSP} \sim 3 \times 10^{29}
\left(\frac{\langle \tilde s \rangle}{1\,\mathrm{TeV}}\right)^2
\left(\frac{\Delta m}{1\,\mathrm{TeV}}\right)^3
\left(\frac{3 \times 10^{15}\,\mathrm{GeV}}{M_{GUT}}\right)^4\, \mathrm{sec} 
\label{photondecay}
\end{equation}
with $\Delta m = m_s-m_{LSP}$.

 This signal is noteworthy since Fermi will have an enhanced sensitivity to a photon line up to a TeV. Direct decays to monochromatic photons is also a qualitatively different feature permitted for decaying dark matter.  The production of monochromatic photons from dark matter annihilations is loop suppressed and hence annihilations always lead to bigger signals in other standard model channels before yielding signals in the photon channel. However, direct decays of dark matter to monochromatic photons can happen independently of its decays to other channels. 

The operator discussed in this section will also induce decays involving $Z$s or decays to a chargino and a $W$ boson. The relative rates compared to the photon decay will be set by the decay topology, the bino vs. wino composition of dark matter, as well as the relative size of the $U(1)_Y$ vs. $SU(2)_L$ couplings. Hadronic decays to a gluino and a gluon are also possible if they are kinematically allowed.


\subsubsection{$S$-Number Violating Operators}
\label{Sec: S-number violating}

A different class of operators are Kahler terms that break the global $S$ number, for example
\begin{equation}
\label{Eqn: S-number violating ops}
S^\dagger\, 10_f H_u^\dagger \bar 5_f \qquad  
S^\dagger\, 10_f H_d^\dagger 10_f \qquad
S^\dagger\, 10_f \bar 5_f^\dagger 10_f
\end{equation}
The first two operators involve an R-even $S$ and allow the lightest neutralino of the MSSM to decay to the fermion component of $S$.  If the neutralino has a significant Higgsino component the decay proceeds just through the single dimension 6 vertex and so is three-body.  In this case the first operator produces both leptons and hadrons and the second produces only hadrons.  In this case, it is possible to make the first operator produce mostly leptons if SU(5) breaking effects in the UV completion make the coefficient of the $S^\dagger H_u^\dagger L E$ component dominate over that of the $S^\dagger H_u^\dagger Q D$.  If the neutralino is mostly gaugino the decay proceeds through an off-shell Higgsino, squark, or slepton and is four-body.  In this case, the first operator could produce both hadrons and leptons.  Here, the lepton-only channel will dominate if a slepton is lighter than all the Higgsinos and squarks.  Again, the second produces only hadrons, as in Table \ref{DM-operators}.

The third operator in Eqn. \eqref{Eqn: S-number violating ops} has an R-odd $S$, so the neutralino will decay to the scalar component of S.  This generally produces both leptons and hadrons
\begin{equation}
N_0\to \tilde{s}+  l^\pm + 2 \mbox{ jets}
\qquad
N_0\to \tilde{s}+  \nu + 2 \mbox{ jets} 
\qquad \mbox{or} \qquad
N_0 \to \tilde{s} + 3 \mbox{ jets}
\end{equation}
If the right handed sleptons are the lightest sleptons the channel with charged leptons can dominate.

The decays from the MSSM to S mediated by the operators in Eqn. \eqref{Eqn: S-number violating ops} are either three or four body and their rates are as shown in Table \ref{DM-operators}, probing scales as shown in Table \ref{rough-lifetimes}.

Of course, these operators also allow the $S$ to decay to MSSM particles, if $S$ has a primordial abundance.  A primordial abundance of $S$ could be generated, for example, by decays at around 1000 s by dimension 5 operators, as happens in the model in Section \ref{SU5U1B-L}.  The second and third operators in Eqn. \eqref{Eqn: S-number violating ops} will always produce hadrons in the decay of $S$, but the first operator could produce mostly leptons if, for example, the lepton component dominates due to SU(5) breaking effects as described above.  In a theory with this operator $S^\dagger\, 10_f H_u^\dagger \bar 5_f$, if the LSP of the MSSM has a large Higgsino component then the decay of the fermion S will be three-body to a Higgsino and two leptons.  In such a scenario, the scalar $\tilde{s}$ could have 
two-body decays to lepton or slepton pairs. Also possible are
three-body decays to a Higgs and two leptons or to a Higgsino, a slepton, and a lepton with the slepton then decaying to the LSP and a lepton.  Even if the $\tilde{s}$ is lighter than the LSP it will still decay just to a Higgs and two leptons or just to two leptons if the channel where the Higgs goes to its VEV dominates.  The details of the decay depend on the mass spectrum of the MSSM particles.  Such a decay could yield an interesting electron/positron spectrum.  Cosmic ray observations could then give important evidence for the mass spectrum of the MSSM, as discussed in Section \ref{Sec: Astro signals}.

\begin{figure}
\begin{center}
\includegraphics[width=3.0 in]{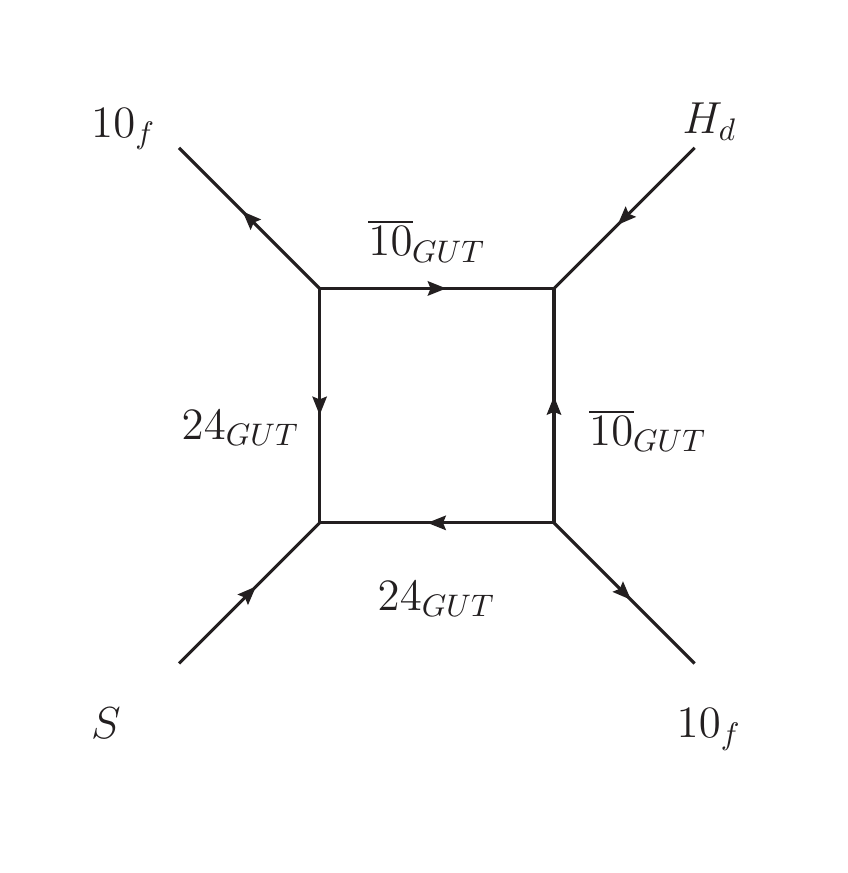}
\caption{ \label{Fig: S-violating box diagram} A way to generate the operator $S^\dagger\, 10_f H_d^\dagger 10_f$.}
\end{center}
\end{figure}

One example of a UV theory which can generate the operator $S^\dagger\, 10_f H_d^\dagger 10_f$ is shown in Fig. \ref{Fig: S-violating box diagram}.  We have added two new fields at the GUT scale, $\overline{10}_ \gut$ and $24_ \gut$ and their conjugate fields, $10_\gut$ and $24_\gut^c$, in order to give them vector-like GUT masses.  The superpotential is taken to be
\begin{equation}
W = S 24_\gut 24_\gut + 10_f 24_\gut \overline{10}_\gut + H_d \overline{10}_\gut \overline{10}_\gut
\end{equation}
This preserves a `heavy parity' under which the GUT scale fields $\overline{10}_\gut$ and $24_\gut$ and their conjugates are odd and everything else is even.  This ensures no mixing happens between the new GUT scale fields and the MSSM fields.  Further this preserves a PQ symmetry with charges $Q(H_u) = Q(H_d) = Q(24_\gut) = 2$, $Q(10_f) = Q(\overline{5}_f) = Q(\overline{10}_\gut) = -1$ and $Q(S) = -4$.  These symmetries and R-parity forbid all dangerous operators of dimension lower than 6 that would cause a faster decay, except for $S H_u H_d$ which is a superpotential term and so will not be generated if it does not exist at tree level.

Similar box diagram ways exist to generate the other $S$-number violating operators.  A combination of PQ symmetry and R parity forbids dangerous operators of dimension 5 or lower for the three operators in  Eqn. \eqref{Eqn: S-number violating ops} except for $S H_u H_d$ in the case of the first two operators and $S \overline{5}_f H_u$ for the third operator.  These are in the superpotential and so will not be generated if not there at tree level.

Another possible UV model to generate these operators is to expand the GUT gauge group beyond SU(5) and integrate out the heavy (GUT scale) gauge bosons.  For  example, the operator $S^\dagger\, 10_f \bar 5_f^\dagger 10_f$ may be generated by integrating out an $SO(10)$ gauge boson at the GUT scale. If this is the case, the field $S$ is a right handed neutrino and some model building would be required to assure the decay of the neutralino does not happen by dimension five operators mediated by Yukawa couplings. This may happen if the lightest right handed sneutrino has no Yukawa coupling, suggesting one of the neutrinos would be completely massless.  Generating one of the other operators involving a Higgs would require an even larger group than SO(10).  We do not consider such models further.

\subsection{R-parity Breaking Operators}
The minimal extension of the SSM allowing the dark matter to decay without introducing any new light particles arises if R-parity is broken. R-parity is a symmetry imposed  to forbid renormalizable superpotential operators,
\begin{eqnarray}
\label{RMSSM}
UDD,~QDL, ~ LLE ,~\text{and}~ H_u L,
\end{eqnarray}
that would otherwise cause very rapid proton decay. As a byproduct, it stabilizes the Lightest Supersymmetric Particle (LSP) which, if neutral, makes an excellent dark matter candidate. Consequently, dark matter decay may indicate that R-parity is broken.


In this section, we will connect the smallness of R-breaking to the hierarchy between the weak and the GUT scales. If R-parity violating effects are mediated through GUT scale particles, their effects can be suppressed by the GUT scale. But the SUSY non-renormalization theorem is not enough to protect the theory from dimension 4 or dimension 5 R-breaking operators, which would lead to too rapid LSP decay; for example, if R-parity is broken and there is no additional symmetry replacing it, kinetic mixings, such as $H_d^\dagger L$, are allowed and are not suppressed by the high scale.

\begin{figure}[htbp]
\begin{center}
\includegraphics[width=3in]{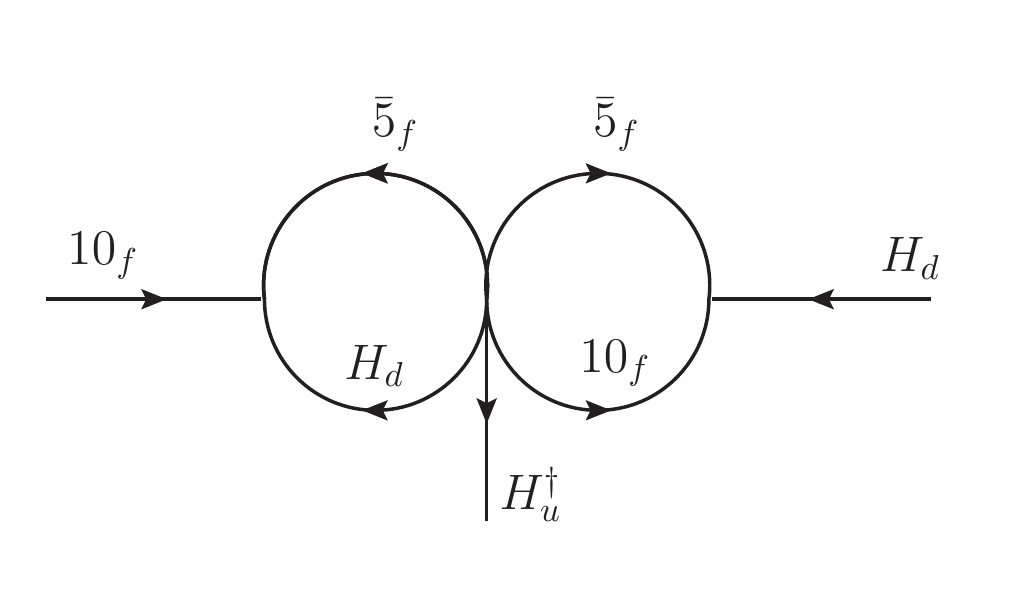}
\caption[R-breaking induced Kahler term]{${\cal D}_\alpha^210_fH_u^\dagger H_d$ generated by   $H_u H_d \bar5_{f} \bar5_{f} 10_{f}$ and the MSSM Yukawas}
\label{closedlegs1}
\end{center}
\end{figure}

To illustrate this point consider a dimension 6 operator $H_u H_d \bar5_{f} \bar5_{f} 10_{f}$.
In the presence of the MSSM Yukawa interactions, there is no symmetry that forbids the dimension 5 operator $10_fH_u^\dagger H_d$, and  prevents it from being generated in a UV completed theory, as can be seen from the existence of diagram \ref{closedlegs1}.
Note, that in this example there is no problem at the effective field theory level because the diagram in Fig. ~\ref{closedlegs1}  cannot generate $10_fH_u^\dagger H_d$. If one starts with an effective field theory involving only dimension 6 operators, the low energy loops will never generate dimension 5 terms because Lorentz symmetry  leads to cancellation of the linear divergencies.  So the diagram in  Fig. ~\ref{closedlegs1}
by itself gives rise only to  dimension 6 operators like ${\cal D}_\alpha^210_fH_u^\dagger H_d$, but not to 
dimension 5 terms. Still the existence of such a diagram is a signal that it will be challenging to UV complete such a theory without generating the $10_fH_u^\dagger H_d$ term as well.

\begin{figure}[tbp]
\begin{center}
\includegraphics[width=3in]{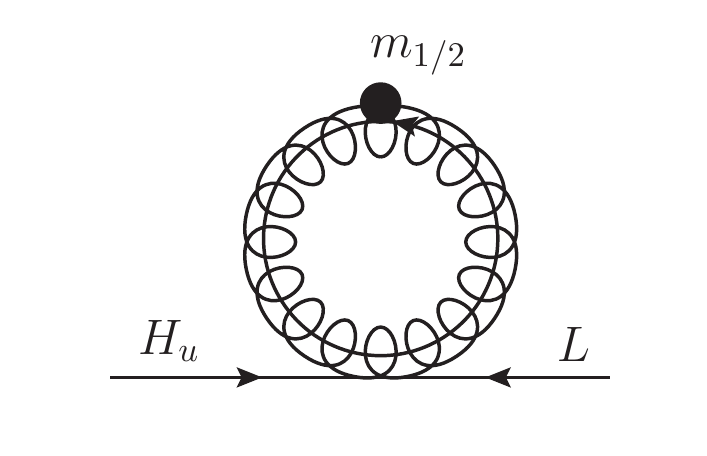}
\caption[$L H_u$ generated from a dimension 6 operator and gaugino masses]{Gaugino loop generating $L H_u$ from ${1\over M_{GUT}^2} H_u L W_\alpha W^\alpha$ after SUSY breaking} 
\label{gauginoloop}
\end{center}
\end{figure}

In fact, there are also examples of operators for which the problem arises directly at the effective field theory level.
 For example, consider the superpotential dimension 6 operator, ${1\over M_{GUT}^2} H_u L W_\alpha W^\alpha$.
 By closing the gaugino legs (see Fig.~\ref{gauginoloop}) it gives rise to the R-parity breaking
 $H_uL$ term. This is a superpotential term, so one may think it is not generated. Indeed, this loop 
 is zero in the SUSY limit. 
 However, in the presence of a SUSY breaking gaugino mass $m_{1/2}$ the loop is non-zero and quadratically divergent, so it gives rise to the term $m_{1/2}H_uL$. The presence of such a quadratic divergence 
 does not contradict the lore that the quadratic divergencies are cancelled in softly broken SUSY, because 
 at the end of the day we obtained a mass term of order the SUSY breaking scale $m_{1/2}$.

These problems lead us to two possible ways of consistently implementing R-parity breaking implying dimension 6 dark matter decays. The first possibility is to replace R-parity by another discrete symmetry that forbids the dimension 4 terms in (\ref{RMSSM}) and also dimension 5 operators that give rise to the dark matter decays, but allows dimension 6 decays.
An alternative proposal is that the R-parity is violated by a tiny amount, that is not put in by hand, but related to the 
SUSY breaking scale. Let us illustrate each of these options in more detail.

\subsubsection{Hard R-parity Breaking}

R-parity is not the only discrete symmetry that forbids the dangerous lepton and baryon number violating operators
(\ref{RMSSM}). Alternative discrete symmetries  may arise from broken gauge symmetries and insure the longevity of the proton. Heavy fields may have couplings that preserve these symmetries but not R-parity \cite{LawrenceZ3, Ibanez:1991pr}. As a result, these symmetries allow for the LSP to decay at the non-renormalizable level. One such operator, 
\begin{eqnarray}
\frac{DDDLL}{M_{GUT}^2}
\end{eqnarray}
arises when there is GUT scale antisymmetric representation of SU(5), while the fundamental theory obeys a $Z_3$ symmetry (see Table \ref{Z3}). This operator is generated by the combination of the couplings

\begin{eqnarray}
W \supset 10_{GUT} \bar5_{f} \bar5_{f} + \bar 10_{GUT} \bar 10_{GUT} \bar5_{f} + \Sigma 10_{GUT}\bar{10}_{GUT},
\end{eqnarray}
that violates R-parity. $\Sigma$ is the adjoint that breaks SU(5) down to the SM gauge group. It is essential that it splits the colored and the electroweak parts of  $10_{GUT}$, otherwise R-parity violation would come through $(\bar 5_{f})^5$ which is zero.

\begin{table}[htdp]
\begin{center}
\begin{tabular}{|c|c|}
\hline
Particle & $Z_3$ charge\\
\hline
$10_{f}$ &  $e^{i \frac{4 \pi}{3}}$\\
$\bar5_{f}$ & 1\\
$H_u$ &  $e^{i \frac{4 \pi}{3}}$\\
$H_d$ & $ e^{i \frac{2 \pi}{3}}$\\
$10_{GUT}$ & 1\\
\hline

\end{tabular}
\end{center}
\label{Z3}
\caption[$Z_3$ symmetry charge assignments replacing R-parity]{The charges of the SM fields and the GUT scale fields under a $Z_3$ discrete that substitutes R-parity}
\end{table}%

\begin{figure}[htbp]
\begin{center}
\includegraphics[width=3.5in]{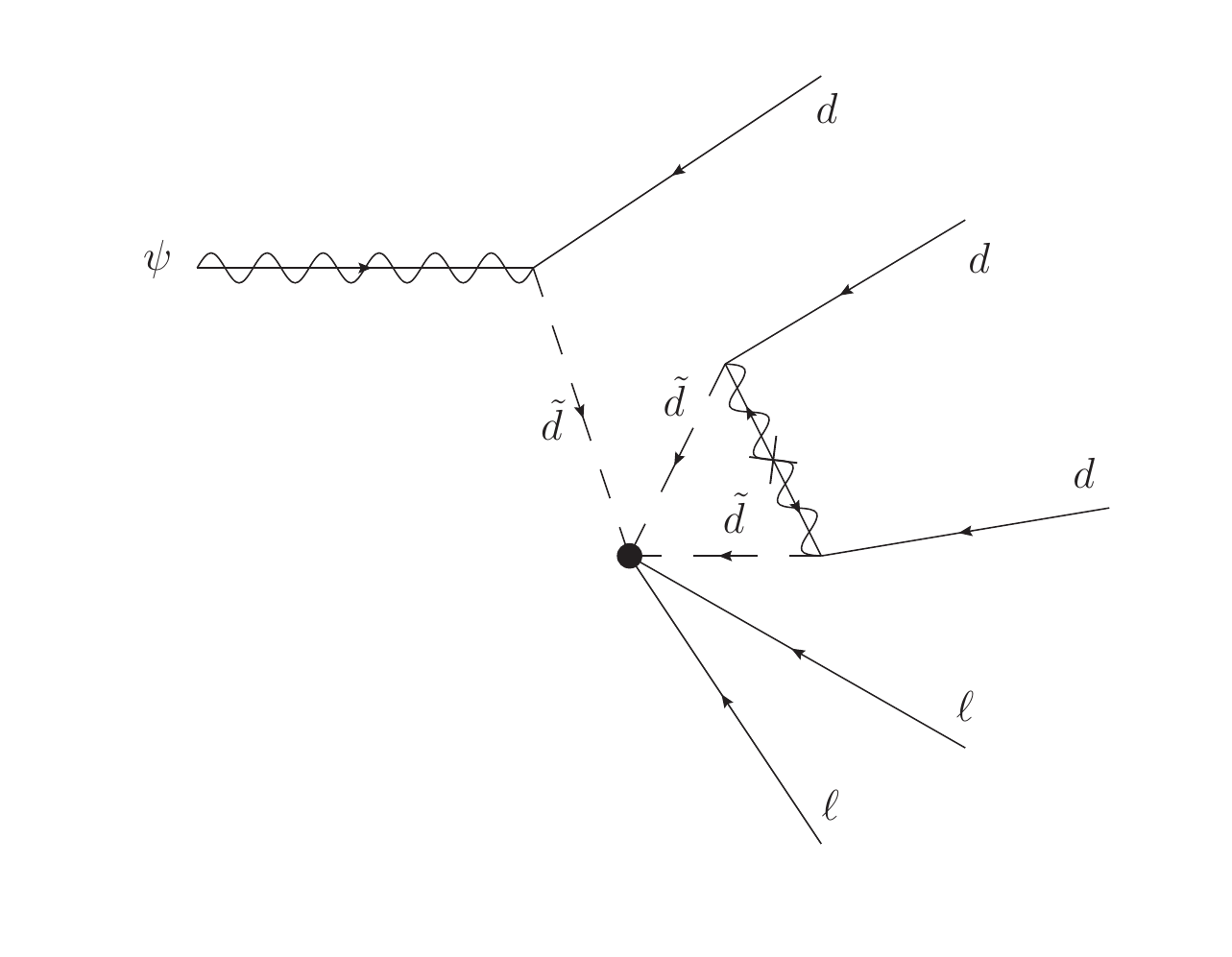}
\caption[LSP decay with hard R-parity breaking]{LSP decay in a theory with a $Z_3$ symmetry substituting R-parity}
\label{dddll}
\end{center}
\end{figure}

The $Z_3$ symmetry also insures that there are no kinetic mixings between light and heavy fields that introduce rapid DM decay. The LSP decays to 5 SM fermions through a sparticle loop(see Fig. \ref{dddll})
\begin{eqnarray}
\chi \rightarrow 3~\text{jets}+ 2 \ell
\end{eqnarray}
The rate of decay is of order,
\begin{eqnarray}
\Gamma_{\psi}\sim 10^{-14} \frac{m_\psi^5}{M_{GUT}^4}\sim(3 \times10^{26} sec)^{-1} \big( \frac{m_\psi}{1~\text{TeV}} \big)^5 \big( \frac{m_{GUT}}{10^{13}~ \text{GeV}}\big)^4\;,
\end{eqnarray}
where we took into account the five body phase space and loop suppression.

This example  illustrates two generic features of LSP decays in the presence of $Z_N$ symmetries replacing R-parity. Typically the allowed operators have a large number of legs, such that the decay rates are significantly suppressed  
by final state phase space. Such operators also tend to involve  quarks, resulting in order one hadronic branching fractions.

\subsubsection{Soft R-parity Breaking}

Even if R-parity is violated only through GUT fields, in the absence of a symmetry, kinetic mixings can still generate order one R-parity violating effects. This problem could be avoided if R parity is broken only through SUSY breaking effects involving GUT fields. These effects will be communicated to the MSSM through the GUT fields  resulting in  suppressions $\sim \l\frac{m_{SUSY}}{\Mgut}\r^2$.
Take, for example, two pairs  of  heavy $5\oplus \bar5$ and two heavy singlets, $S_1$ and $S_2$. The interactions between the GUT and MSSM fields are:
\begin{gather}
W \supset S_1 \bar5_{{GUT}_1} H_u + S_2 5_{{GUT}_2} \bar5_f + \nonumber \\
M_{GUT} S_1^2 + M_{GUT} S_2^2 + \
M_{GUT}\bar5_{{GUT}_1} 5_{{GUT}_1}+  M_{GUT}\bar5_{{GUT}_2} 5_{{GUT}_2}.
\end{gather}
The important point is that  the R-parities of these heavy fields are not fully defined with this superpotential-- $S_1,5_{{GUT}_1} $ have equal R-parity and $S_2,5_{{GUT}_2} $ have opposite R-parity. To break R-parity 
one needs two soft terms for heavy fields, for instance:
\begin{eqnarray}
m_{SUSY}^2 \tilde{S}_1 \tilde{S}_2~ \text{and} ~m_{SUSY}^2 \tilde{\bar5}_{{GUT}_1} \tilde{5}_{{GUT}_2}
\end{eqnarray}
Consequently, the R-parity breaking coefficients in the MSSM sector are proportional to the product of the two soft masses and are always suppressed by at least $M_{GUT}^2$ . Indeed, the loop of heavy fields generates the R-breaking $B\mu$-term
\[
\frac{m_{SUSY}^4}{M_{GUT}^2}h_u \tilde{\bar5}_f\;,
\]
which is effectively dimension 6.


Even though this scenario works at the spurion level, it is hard to implement in a full theory of SUSY breaking. First, we need to sequester the source of R-parity violation  from the MSSM fields but not from the GUT fields. In a toy model with one extra dimension, the MSSM and R-breaking fields are located on different branes, while the GUT fields are free to propagate in the bulk. Then R-parity breaking is communicated to the MSSM fields only through loops of GUT particles. Care has to be taken in order to suppress the effects of gravity that propagates everywhere and may communicate unsuppressed R-violation to the MSSM. This is ensured if the soft terms for the heavy fields are generated by a gauge mediation mechanism. However, as discussed in sub section \ref{RConservingSingletConserving}, the MSSM LSP can decay through these dimension 6 GUT suppressed operators only if the LSP is not the gravitino. This requirement forces the $F$-term responsible for the MSSM soft masses to be much larger than the one responsible for R-breaking. Consequently, we need two very different scales of SUSY breaking, one for the MSSM sector and another for the GUT sector. It is not clear if such a SUSY breaking mechanism can be successfully embedded into a UV completion.

\section{The Primordial Lithium Problems and Dimension 5 decays}
\label{LithiumIntroduction}

Recent observations \cite{SpiteandCo} of the $\LiSeven$/H and $\LiSix$/H ratio in metal-poor halo stars suggest a discrepancy between the standard big bang nucleosynthesis (BBN) and observationally inferred primordial light element abundances. As pointed out in \cite{Jedamzik2008, Jedamzik:2004er},  the decay of a particle $\chi$  with a lifetime $\tau \sim 100 - 1000 \text{ s}$ could explain this anomaly if the energy density $\Omega_{\chi} h^2$ of $\chi$ and the hadronic branching fraction $\HBr$ are such that  $\Omega_{\chi} h^2 \HBr \sim 10^{-4}$. The required density of $\chi$ at the time of its decay is similar to the expected relic density of a particle of mass $\Mchi \sim  100 \text{ GeV}$ with electroweak interactions.  The decay of such a particle can explain the $\LiSeven$ and $\LiSix$ abundances if its lifetime $\tau \sim 100 - 1000 \text{ s}$. Previous work \cite{Savas1987} has concentrated on obtaining this lifetime within the context of the MSSM either through the decays of the gravitino to the LSP or the decays of the NLSP to the gravitino depending upon the low energy SUSY spectrum. Due to the small production cross-section of the gravitino, the former  scenario is difficult to test at collider experiments and either requires non-thermal mechanisms or tuning of the reheat temperature of the Universe to furnish the relic abundance of the gravitino. The latter scenario involving the decay of the NLSP to the gravitino has been discussed extensively in the literature \cite{JedamzikFengetc}. The desire to generate a dark matter abundance of gravitinos from the decay of the NLSP while simultaneously allowing for the required hadronic branching fraction forces the NLSP to be the stau with a mass  $\sim 1.5 \text{ TeV}$ in the generic parameter space of the theory. The LHC reach for such long lived charged staus is $\sim 500 \text{ GeV}$ \cite{KraanSkands} making it difficult to test this scenario at the LHC.  Other previous work has suggested that late decays can affect nucleosynthesis \cite{Khlopov:1987bh, Khlopov:1999rs}.

In this paper, we point out that supersymmetric GUT theories provide a natural home to another wide class of models that could explain the Lithium anomalies.  Dimension 5 operators suppressed by the GUT scale are a generic feature of supersymmetric GUT theories. If $\chi$ decays through such an operator, then its decay rate $\Gamma \sim \left(\frac{\Mchi^3}{M_{GUT}^2}\right)$ is naturally $\sim \left(100 \text{ s}\right)^{-1}$. 


In the following, we perform a general operator analysis of the possible dimension 5 GUT suppressed operators that can solve the primordial Lithium abundance problem.  For concreteness, we consider $SU(5)$ grand unification. A solution to the primordial Lithium problem involving dimension 5 GUT suppressed operators requires the introduction of a  new TeV scale particle species $\chi$. We restrict ourselves to models that involve the addition of complete, vector-like $SU(5)$ multiplets in order to retain the success of gauge coupling unification in the MSSM.  We will only consider operators that preserve R-parity, automatically ensuring the existence of a stable dark matter candidate.  The operators are classified on the basis of the $SU(5)$ representation of $\chi$. In each case, we discuss the relic abundance and hadronic branching fraction of the decaying species that solves the primordial Lithium problem and the corresponding LHC signatures.

\subsection{Operator Classification}
\label{Classification}

We illustrate the possible dimension 5 GUT suppressed operators  in Table \ref{Tab:LithiumOperators}. The operators are classified on the basis of the $SU(5)$ representation of $\chi$.
  The R-parity of $\chi$ is  indicated by subscripts $e$ (R-even) and $o$ (R-odd) and chosen to make the operators in  Table \ref{Tab:LithiumOperators}  R-invariant. We restrict ourselves to the cases when $\chi$ is a singlet, a fundamental $(5,\bar{5})$ or
  an antisymmetric tensor $(10,\bar{10})$ of the  $SU(5)$ group. As illustrated in the second and third column of 
  Table \ref{Tab:LithiumOperators},  it is straightforward to construct a number of dimension five operators 
  inducing the decay of $\chi$  or LSP decay, depending which one is lighter,
  in all these cases.
   The hadronic branching fraction for the corresponding decays is typically quite significant, as the operators involve either Higgs fields or quarks. 
     
 Note that fundamental or antisymmetric representations of $SU(5)$ are not good Dark Matter candidates; a stable fundamental
would imply dirac DM, which is excluded by direct detection searches \cite{DirectDetection, Beck:1993sb}, while an $SU(5)$ antisymmetric does not have a neutral component. Only when $\chi$ is a singlet is the LSP phenomenologically allowed to be heavier than $\chi$ and decay to it. In this case, $\chi$ naturally inherits the LSP thermal relic abundance. If this singlet is heavier than the LSP, one needs to introduce TeV scale interactions that would give $\chi$ a thermal calculable abundance, or rely on non-thermal production mechanisms such as tuning of the reheat temperature, in order to yield the required $\chi$ abundance. Due to this difference between $\chi$'s that do and do not carry  SM charges, we have only included the quadratic in $\chi$ operator in the of a singlet $\chi$.   This operator allows for the LSP decay into $\chi$'s, but doesn't make the lightest component of $\chi$ unstable.

    
A small subtlety in all the cases
is  that we can also construct relevant and marginal gauge invariant operators
involving $\chi$ and the MSSM fields.  If present, they
would mediate the decay of $\chi$ without GUT scale suppression. These dangerous operators are collected in the last
column of Table \ref{Tab:LithiumOperators}. In particular, the  kinetic mixing terms such as $\chiBarDagger_o \FiveBar_f$
(here $\chiBar_o$ is an R-odd antifundamental GUT multiplet; the effect of this term is equivalent to the mixing between  
$\chiBar_o$  and $\FiveBar_f$ in  the MSSM Yukawa  interactions) are not protected by SUSY non-renormalization theorems
and will be inevitably generated unless forbidden by some additional symmetries.  However, it is relatively straightforward to impose Peccei--Quinn symmetries  that either forbid these operators or make them GUT suppressed as well.

Let us illustrate how this works in several concrete examples. If $\chi$ is an R-even singlet, the only dangerous operator is $\chi_e H_uH_d$. We can forbid it by imposing a discrete $Z_2$ symmetry,
$\chi\to-\chi$. With this symmetry, the only allowed dimension five operator is $\chi_e^2H_uH_d$. As a result, $\chi$
is the dark matter, and the lithium problem is solved by the decay of the LSP into pairs of
$\chi$ particles. 

Other superpotential dimension 5 terms for an R-even singlet are of the form $\chi W_i$, where
$W_i$ are Yukawa terms present in the MSSM Lagrangian. In this case, we need to use a symmetry other than $Z_2$ under which $\chi$
is neutral. The Higgs fields carry charges $Q_u$ and $Q_d$ such that $Q_u+Q_d\neq0$, and the charges of the MSSM matter fields  are determined by requiring that the MSSM Yukawa terms do not violate the symmetry.  Then the MSSM $\mu$ term is forbidden by this symmetry, however, it can be generated 
if the symmetry is spontaneously broken by the TeV vev of the field $S$ with charge $(-Q_u-Q_d)$ coupled to Higgses
through the term $S H_uH_d$. The new massless Goldstone boson will not appear if the actual symmetry of the action
is just a discrete
subgroup of this continuous symmetry, so that one can make all components of $S$ massive. The lowest dimension operator involving $\chi$, $S$ and MSSM Higgses is $S\chi H_uH_d$. When $S$ develops a vev
it gives rise to the marginal operator $\chi H_uH_d$ mediating $\chi$ decay, whose coefficient is naturally suppressed by the ratio of the soft PQ breaking scale, $\mu$, to the scale where the dimension five operator is generated, $M_{GUT}$.

It is straightforward to generalize these arguments to other cases as well. For instance, for the fundamental
$\chi$ to avoid the kinetic mixings with Higgses or matter fields one can impose a discrete symmetry which is a subgroup of the continuous PQ symmetry with the following charge assignments for the MSSM fields
\[
Q_{10_f}=1\text{, }Q_{H_d}=-Q_{H_u}=2\text{, }Q_{\FiveBar_f}=-3
\]
 The charges are chosen in such a way that the MSSM superpotential is invariant. If the charge of $\chi$
 is equal to $Q_\chi=8$ the dimension 5 operator $\chi_e H_u \FiveBar_f \FiveBar_f$ is allowed, while all other operators mediating $\chi$ decay are forbidden. In fact, one can check that this is the only possible charge assignment 
 that avoids kinetic mixing and allows $\chi$ to decay without requiring soft breaking of the PQ symmetry.
 There are more possibilities if the Higgs charges are not opposite so that the PQ symmetry
 is softly broken by the $\mu$ term.
 Finally, let us use this example to illustrate that there is no problem to go beyond the effective theory
 analysis and construct renormalizable models generating the required operators at the GUT scale. Namely,
 let us consider the following renormalizable superpotential
 \[
 W=10_{\gut}\FiveBar_f\FiveBar_f+\TenBar_{\gut}H_u\chi_e+\Mgut\TenBar_{\gut}10_{\gut}\;.
 \]
 Here $(10_{\gut},\TenBar_{\gut})$ is a pair of R-even GUT scale fields with the PQ charge $Q_{10}=6$.
 By integrating out the heavy fields one obtains the dimension five operator $\chi_e H_u \FiveBar_f \FiveBar_f$
 at low energies.
 
In addition to decays involving the chiral supermultiplets of the MSSM, singlet $\chi$s can also have decays involving the gauge supermultiplets $W_{\alpha}$ through operators of the form $\chi W_{\alpha} W^{\alpha}$. These operators can be generated through an axion-like mechanism where $\chi$ is the goldstone boson of some global symmetry broken at the GUT scale. 


\begin{table}[ t]
\centering
\begin{tabular}{|l||l||l||l|}
\hline
$\chi \; SU(5)$ Rep. & Superpotential Terms  & Kahler terms &  Soft PQ breaking\\
\hline
Singlet & 
$\chi_e 10_f 10_f H_u \text{, } \chi_e 10_f \FiveBar_f H_d \text{, } $   
&
$\chi_e \TenDagger_f 10_f \text{, } 
$   
 $ \chi_e \HuDagger H_{u} \text{, }
 $ 
 &
  $ \mugut  \chi_e H_u H_d$\\
   &$ \chi_{e,o}^2 H_u H_d \text{, } \chi_o 10_f \FiveBar_f \FiveBar_f \text{, }$  & $\chi_o \FiveBarDagger_fH_d$&$\mugut\chi_o H_u\FiveBar_f$\\
   &$ \chi_e \mathcal{W}_{\alpha} \mathcal{W}^{\alpha} $ & & \\
   \hline
$\left(5, \FiveBar\right)$ &
 $\chi_e H_u \FiveBar_f \FiveBar_f\text{, } \chiBar_e H_u H_uH_d\text{, } $ 
 & $\chiBarDagger_e10_f10_f\text{, }$ 
 & $\mugut \left(\chiDagger_e H_u \text{, } \chiBarDagger_e H_d \text{, } \chiBarDagger_o \FiveBar_f\right)$ \\
  &
 $\chi_o10_f10_f10_f\text{, }\chi_o\FiveBar_fH_uH_d$
 &
$ \chi_o\TenDagger_fH_u$
 &  $\mu \l\frac{\mu}{\Mgut}\r\chi_o \FiveBar_f$\\
\hline
$\left(10, \TenBar\right)$& $\chi_e 10_f 10_f H_d \text{, } \chiBar_e 10_f \FiveBar_f H_u \text{, }$  &  $\chiBarDagger_e 10_f \FiveBarDagger_f \text{, } \chiBarDagger_e \FiveBar_f \FiveBar_f$  &$\mugut\l\chiDagger_o 10_f\text{, }\chi_e\FiveBar_f\FiveBar_f\r$\\
&$\chiBar_o \FiveBar_f \FiveBar_f \FiveBar_f \text{, } \chiBar_e\FiveBar_f \FiveBar_f H_d$ & 
$\chiBar_o H_u\FiveBarDagger_f$
&$\mu \l\frac{\mu}{\Mgut}\r\chiBar_o 10_f$ \\
\hline
\end{tabular}
\caption[Dimension 5 Operator Catalog]{\label{Tab:LithiumOperators}The possible dimension 5 GUT suppressed operators classified on the basis of their generation in the superpotential or through soft breaking of PQ symmetry or through kinetic mixing in the Kahler potential. The subscript f denotes standard model families, $\mathcal{W}_{\alpha}$ are  gauge fields and $H_u$, $H_d$ are the Higgs fields of the MSSM. The R-parity of $\chi$ is denoted by its subscripts $e$ and $o$ for even and odd parities respectively.  }
\end{table}

\subsection{Relic Abundance}

\begin{figure}
\begin{center}
\includegraphics[scale=1.0]{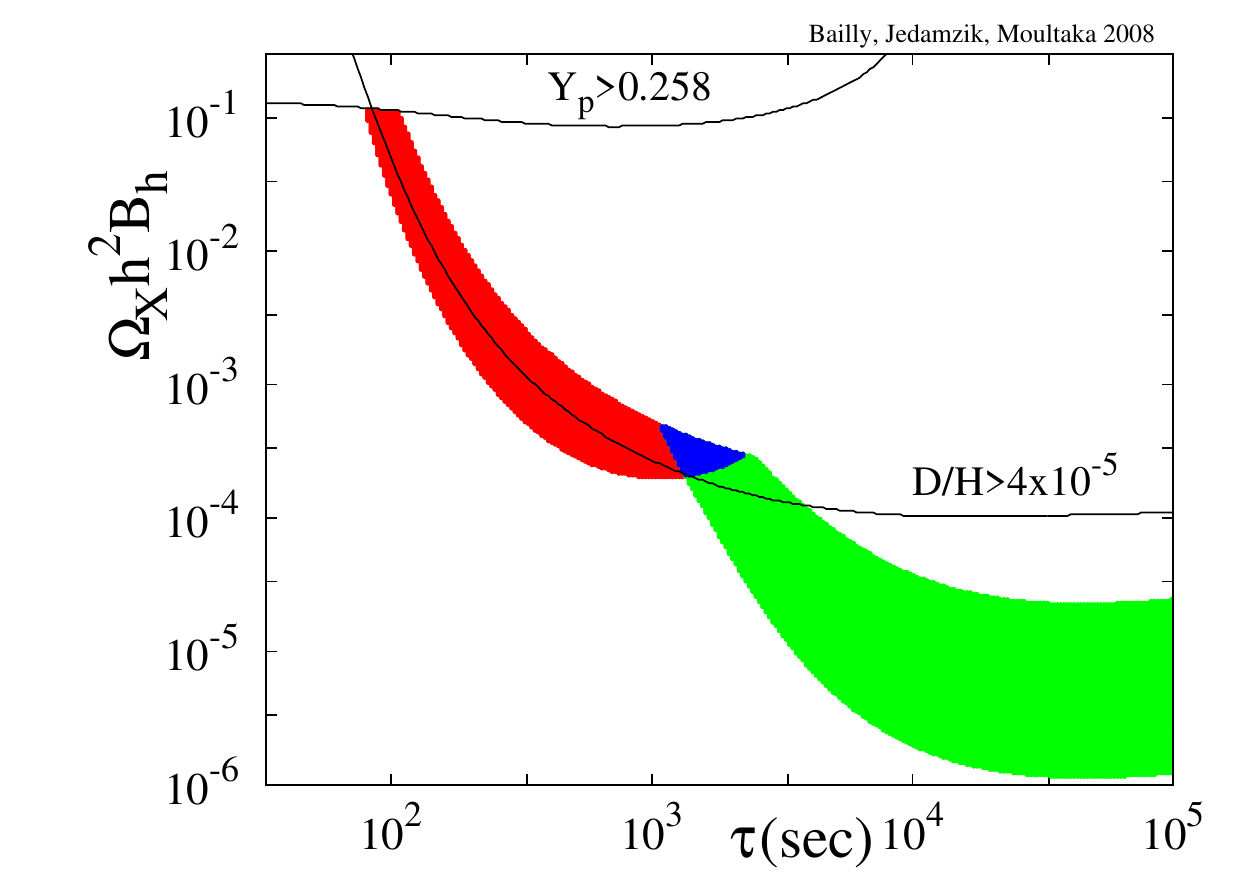}
\caption[Required Hadronic Energy Injection]{ \label{Fig:LithiumRelicAbundance} (Color online) This figure from \cite{Jedamzik2008} plots the energy density that must be injected into hadrons versus the decay lifetime in order to solve the primordial lithium problems. Decays in the red region solve the $\LiSeven$ problem and decays in the green region solve the $\LiSix$ problem.  }
\end{center}
\end{figure}

The energy density $\Omega_\chi h^2 \HBr$ that must be injected into hadrons is plotted against the lifetime $\tau$ of the decaying particle in Figure \ref{Fig:LithiumRelicAbundance}. The $\LiSeven$ problem can be solved when the lifetime $\tau \sim \text{100 s }  \left(\frac{0.1}{\Omega_\chi h^2 \HBr}\right)^{\frac{1}{3}}$. The operators described in section \ref{Classification} can cause decays between the MSSM and the $\chi$ sector with these lifetimes.  

\subsubsection{Electroweak Relics}
\label{ChiDecayToMSSM}
Let us first consider how the Lithium problems can be solved by the colorless components of  fundamental or antisymmetric representations of $SU(5)$. Standard model gauge interactions generate a thermal abundance of the electroweak multiplets in  $\left(5, \FiveBar\right)$ and $\left(10, \TenBar\right)$. We focus on the standard model operators that are extracted from the  $SU(5)$ invariant operators in  Table \ref{Tab:LithiumOperators} and contain the electroweak multiplets (the lepton doublet $L$ in  $\left(5, \FiveBar\right)$  and the right handed positron $E$ in $\left(10, \TenBar\right)$) from the $\chi$. These operators can be classified into three categories: operators that involve quarks or only contain higgses, operators that are purely  leptonic and operators that involve leptons and higgs doublets. This classification is presented in Table \ref{Tab:LowEnergyLithium}.

Operators in Table \ref{Tab:LithiumOperators} that contain higgs triplets when the $\chi$s are electroweak multiplets have not been included in Table \ref{Tab:LowEnergyLithium} since the higgs triplets are at the GUT scale and cannot cause a dimension 5 decay of the electroweak $\chi$. For example, with $\chi$ a  $\left(10, \TenBar\right)$, the only gauge invariant operator that can be extracted from  $\chiBar \FiveBar_f \FiveBar_f H_d$ in Table \ref{Tab:LithiumOperators}  is $\bar{E_\chi} D D H_d^T$ where $H_d^T$ is the color triplet higgs. This operator cannot cause a dimension 5 decay of the $E_\chi$ and is not listed in Table \ref{Tab:LowEnergyLithium}. There are also operators in Table \ref{Tab:LithiumOperators} from which standard model operators that belong to more than one category in Table \ref{Tab:LowEnergyLithium} can be extracted. For example, the operator $\chiBar 10_f \FiveBar_f H_u$ generates an operator containing quarks,  $\bar{E}_\chi Q_f D_f H_d$,  and an operator containing leptons and higgses, $\bar{E}_\chi E_f L_f H_d$. We include this operator in both categories in Table \ref{Tab:LowEnergyLithium}. A UV completion of this operator involves integrating out  GUT scale $SU(5)$ multiplets. $\OrderOne$ $SU(5)$ breaking effects at the GUT scale (like doublet-triplet splitting) can result in one of the operators(say, $\bar{E}_\chi Q_f D_f H_d$)  being suppressed relative to the other ($\bar{E}_\chi E_f L_f H_d$). 

The relic energy density $\Omega_\chi h^2$ of these electroweak multiplets is  $\sim 0.1 \left(\frac{M_\chi}{\text{TeV}}\right)^2$. The decays mediated by the operators  in Table \ref{Tab:LowEnergyLithium}  that involve quarks or only contain higgses  have  $\OrderOne$ hadronic branching fractions. These operators can solve the $\LiSeven$ problem if the $\chi$ lifetime is $\sim 100 \text{ s } \left(\frac{\text{TeV}}{M_\chi}\right)^{\frac{2}{3}}$ (figure \ref{Fig:LithiumRelicAbundance}). However, these decays cannot  solve the $\LiSix$ problem. A solution to the $\LiSix$ problem requires a hadronic energy density injection  $\Omega_\chi h^2 \HBr\lessapprox 10^{-4}$ around a 1000 s. Collider bounds on charged particles imply that $M_\chi > 100 \text{ GeV}$. Consequently, the relic energy density  $\Omega_\chi h^2$ is greater than $10^{-3}$. Due to the $\OrderOne$ hadronic branching fraction, the hadronic energy density injected is also greater than $10^{-3}$. This injection is too large and over produces $\LiSix$ (figure  \ref{Fig:LithiumRelicAbundance}).

\begin{figure}
\begin{center}
\includegraphics[scale=1.0]{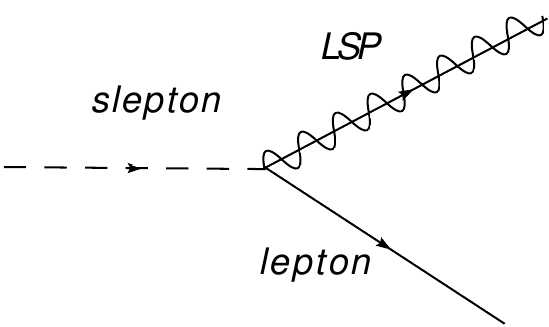}
\caption[$\tilde{l} \to l, LSP]{ \label{Fig:SleptonLeptonLSP} The decay of a slepton to a lepton and the LSP.  }
\end{center}
\end{figure}

\begin{figure}
\begin{center}
\includegraphics[scale=1.0]{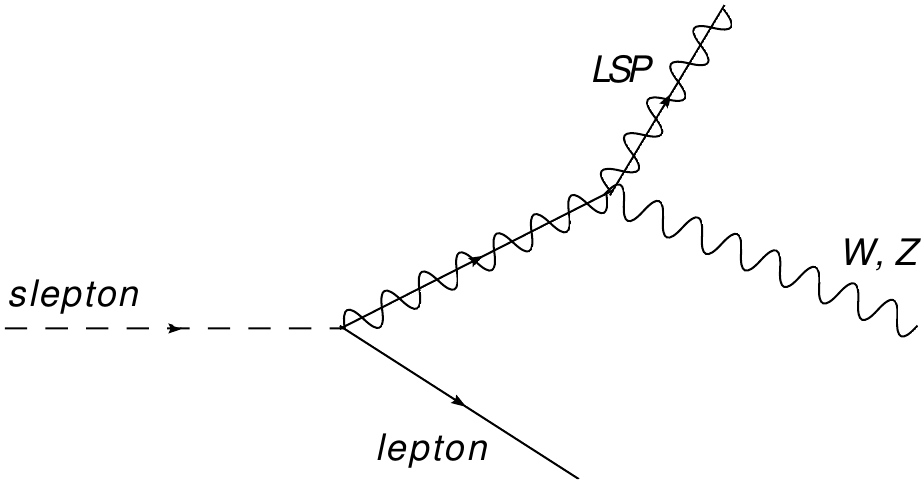}
\caption[$\tilde{l} \to l, W, LSP$]{ \label{Fig:SleptonLeptonWLSP} The decay of a slepton to a lepton, gauge boson and the LSP through an off-shell chargino.  }
\end{center}
\end{figure}

The MSSM products of the decays mediated by the purely leptonic operators in Table \ref{Tab:LowEnergyLithium} may or may not contain sleptons. For example, in the decay of the fermionic component $\chi$ino of the $\chi$, the operator $\chiBarDagger \FiveBar_f \FiveBar_f$ yields a lepton, slepton pair while the operator $\chiBarDagger \FiveBar_f$ does not produce any sleptons. The hadronic branching fraction of decays that involve sleptons depends upon the MSSM spectrum. An $\OrderOne$ hadronic branching fraction will be produced from the decay of the sleptons if the SUSY spectrum contains charginos or other neutralinos between the slepton and the LSP. These decays can solve the $\LiSeven$ problem but not the $\LiSix$ problem. When this is not the case, the slepton predominantly decays to its leptonic partner and the LSP (figure \ref{Fig:SleptonLeptonLSP}). The slepton decay can directly produce hadrons through off-shell charginos (figure \ref{Fig:SleptonLeptonWLSP}). But, these decays are suppressed by phase space factors and additional gauge couplings. The branching fraction for these processes is $\sim 10^{-4}$. The leptons produced in these decays could be $\tau$s. However, even though the $\tau$ has an $\OrderOne$ branching fraction into hadrons, the hadrons produced in this process are pions. A solution to the primordial lithium problem requires the injection of neutrons\cite{Jedamzik2008, Jedamzik:2004er} and hadronic energy injected in the form of pions is ineffective in achieving this goal. The dominant hadronic branching fraction in these leptonic decays is provided by final state radiation of $Z$ and $W$ bosons off the produced lepton doublets. These bosons decay to hadrons with an $\OrderOne$ branching fraction. Using the branching fraction for final state radiation of $Z$ and $W$ bosons from \cite{FengFSR}, we estimate that the relic abundance and hadronic branching fraction from the decays of a $600$ GeV - $1$ TeV $\chi$ satisfy the constraint $\Omega_{\chi} h^2 \HBr \sim 10^{-4}$. These decays can solve the $\LiSeven$ and $\LiSix$ problems if the lifetime $\sim$ 1000 s. 

\begin{figure}
\begin{center}
\includegraphics[scale=1.0]{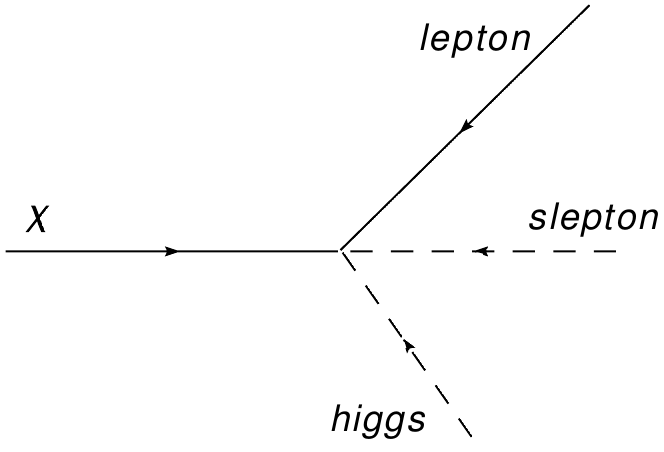}
\caption[$\chi \to \tilde{l}, l, H_{u\l d \r}$]{ \label{Fig:ChiSleptonLeptonHiggs} $\chi$ decay to a slepton,  lepton and higgs.  }
\end{center}
\end{figure}

\begin{figure}
\begin{center}
\includegraphics[scale=1.0]{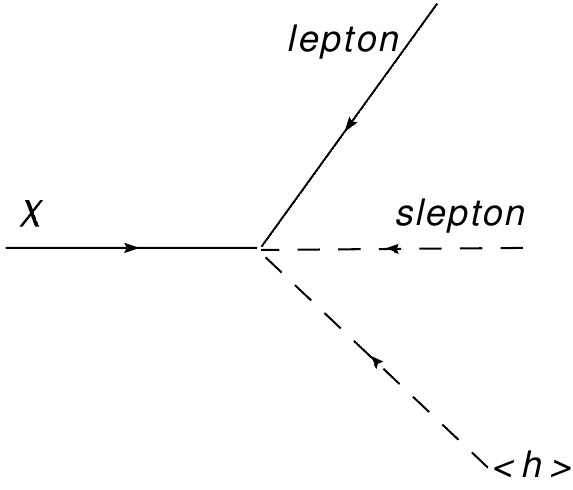}
\caption[$\chi \to \tilde{l}, l$]{ \label{Fig:ChiSleptonLeptonHiggsVev} $\chi$ decay to a slepton and lepton with a higgs vev $\langle h \rangle$ insertion.}
\end{center}
\end{figure}

The hadronic branching fraction of the decays mediated by the operators involving both leptons and higgses in Table \ref{Tab:LowEnergyLithium} is model dependent. Naively, these operators should have an $\OrderOne$ hadronic branching fraction since the higgs decays predominantly to $b$ quarks. However, the higgs operators in these fields can be replaced with the higgs vev $\langle h \rangle$, resulting in an effective purely leptonic decay mode. These leptonic decay modes produce hadrons through final state radiation of $Z$ and $W$ bosons with a branching fraction $\sim 10^{-2}$ as discussed above. The hadronic branching fraction of operators with both leptons and higgses is the ratio of the decay rate to processes involving the higgs and the rate to processes where the higgs is replaced by $\langle h \rangle$. For example, the operator $\chi H_u \FiveBar_f \FiveBar_f$ can cause the $\chi$ to decay to a higgs,  slepton and  lepton (figure \ref{Fig:ChiSleptonLeptonHiggs}) with a rate $\Gamma_h \sim \left(\frac{M_\chi^3}{192 \pi^3 M_{GUT}^2}\right)$. When the higgs is replaced by $\langle h \rangle$, this operator causes the $\chi$ to decay to a lepton, slepton pair (figure \ref{Fig:ChiSleptonLeptonHiggsVev}) with a rate $\Gamma_l \sim \left( \frac{\langle h \rangle ^2} { 8 \pi M_{GUT}^2}\right) M_\chi$. The decay to a higgs directly produces hadrons and the branching fraction for this decay mode is $\frac{\Gamma_h}{\Gamma_l} = \left(\frac{1}{24 \pi^2}\right)\left(\frac{M_\chi }{ \langle h \rangle}\right)^2 \sim 10^{-2} \left(\frac{M_\chi}{\text{500 GeV}}\right)^2$. With $M_\chi \sim \text{500 GeV}$, this hadronic branching fraction is sufficiently small to allow this decay to solve both the $\LiSeven$ and $\LiSix$ problems. However, we cannot replace the higgs field by $\langle h \rangle$ in every such operator in Table \ref{Tab:LowEnergyLithium} that has leptons and higgses. After electroweak symmetry breaking, the masses of the electrically charged $l^{+}_\chi$ and neutral $l^{0}_\chi$ components of the lepton doublet in  $\left(5, \FiveBar \right)$ are split. The charged fermion  $l^{f^{+}}_\chi$ is heavier than the neutral fermion $l^{f^{0}}_\chi$ by $\sim \alpha M_Z$ \cite{MinimalDarkMatter, ScottWells} while the masses of the corresponding scalar components are additionally split by $\sim \cos \l 2 \beta \r M_W^2$ \cite{SUSYPrimer}. When the fermion components are lighter than the scalars, all the components of $l_\chi$ rapidly decay to $l^{f^{0}}_\chi$. In this case, the lithium problems can be solved with  $\left(5_\chi, \FiveBar_\chi \right)$s only through the decays of $l^{f^{0}}_\chi$. With $l^{f^{0}}_\chi$, the operator $\chi \TenDagger_f H_u$ does not lead to any $SU(3) \times U(1)_{EM}$ invariant operators when $H_u$ is replaced by $\langle h \rangle$. The only decay mode for the $l^{f^{0}}_\chi$ that is permitted by this operator is a decay to the right handed positron and a charged higgs, resulting in an $\OrderOne$ hadronic branching fraction. As discussed earlier, the decays mediated by this operator cannot solve the $\LiSix$ problem but can address the $\LiSeven$ problem. 

The relic abundance of a singlet $\chi$ is model dependent. A thermal abundance of $\chi$ can be generated if the $\chi$ is coupled to new, low energy gauge interactions like a $U(1)_{B-L}$. It could also be produced through the decays of new TeV scale standard model multiplets or through a tuning of the reheat temperature of the Universe. The hadronic branching fraction of the singlet $\chi$ operators in Table \ref{Tab:LithiumOperators} is also model dependent. Operators like  $\chi 10_f 10_f H_u$ have an $\OrderOne$ hadronic branching fraction since $\chi Q_f U_f H_u$,  the only standard model operator that can be extracted from it, contains quark fields. However, an operator like $\chi 10_f \FiveBar_f H_d$ yields both $\chi Q_f D_f H_d$ and $\chi E_f L_f H_d$. A UV completion of this operator involves integrating out  GUT scale $SU(5)$ multiplets. $\OrderOne$ $SU(5)$ breaking effects at the GUT scale (like doublet-triplet splitting) can result in the operator $\chi Q_f D_f H_d$ being suppressed relative to $\chi E_f L_f H_d$. Due to these tunable model dependences, the decays of singlet $\chi$s can solve both $\LiSeven$ and $\LiSix$ problems as long as their relic abundance $\Omega_\chi h^2 \gtrapprox 10^{-4}$. 
 
 Yet another possibility available in the singlet case is that the  MSSM LSP is heavier than the R-odd 
 component of $\chi$.
 Then the LSP will decay to $\chi$ and
   the $\chi$ abundance will be close to the dark matter abundance today.
 In fact, if no other stable particles are added then $\chi$ itself will be a dark matter particle.
  This scenario shares many similarities with the scenario where the gravitino is the lightest R-odd particle, so that the LSP
  can decay. 
   If there is no additional mechanism for generating $\chi$ (such as the coupling to new low energy gauge interactions like $U(1)_{B-L}$) the decaying MSSM LSP should have rather high relic abundance, 
  $\Omega_{LSP}h^2\gtrsim 0.1$, depending on the mass ratio between the LSP and $\chi$. This makes it somewhat challenging
  to solve $\LiSix$ problem. This is achievable though if the LSP is a slepton coupled to the singlet through purely leptonic operators.

It is worth stressing that a generic property of all our models is the presence of several long-living particles with somewhat different 
lifetimes and masses. There are two sources for proliferation of different long-living species.
The first is related to R-parity. Indeed, as the $\chi$-parity $\chi\to-\chi$ is broken only by dimension
five operators, the lightest particle in the $\chi$ multiplet is always long-lived. However, if  SUSY breaking mass splitting between the lightest R-odd and R-even particles is smaller than the mass of the MSSM
LSP, both of these particles are metastable and will decay only through $\chi$ parity violating dimension five operators.
Their presence doesn't change much in our discussion. 
Another reason for the existence of several long-lived particles is that 
we are adding new fields in the complete GUT multiplets that contain also colored components. 
Let's now discuss their story.

\begin{table}[ t]
\centering
\begin{tabular}{|l||l||l||l|}
\hline
$\chi \; SU(5)$ Rep. & Quark and Higgs operators & Purely Leptonic &  Leptonic operators\\
& & Operators &  with higgs doublets \\
\hline
Singlet &$ \chi 10_f \FiveBar_f H_d \left(= \chi Q_f D_f H_d\right) \text{, } $ & $\chi 10_f \FiveBar_f H_d \left(=\chi E_f L_f H_d\right) \text{, }$ & $\mugut \chi H_u \FiveBar_f\text{, }$ \\
& $\chi 10_f 10_f H_u\text{, } \chi^2 H_u H_d\text{, } \chi 10_f \FiveBar_f \FiveBar_f,  $ & $\chi 10_f \FiveBar_f \FiveBar_f \text{, } \chi \TenDagger_f 10_f $ & $ \chi \FiveBarDagger_f H_d$\\
 & $\chi \TenDagger_f 10_f \text{, } \chi W_{\alpha} W^{\alpha} \text{, }  \chi H_u^{\dagger} H_u $ & & \\
\hline
$\left(5, \FiveBar\right)$ &
 $  \chiBarDagger 10_f 10_f (= \bar{l}^{\dagger}_\chi Q_f U_f) \text{, } \chiBar H_u H_uH_d\text{, }$ 
 & $ \mugut   \chiBarDagger \FiveBar_f\text{,  }$ 
 & $\chi \FiveBar_fH_uH_d \text{, }  \chi \TenDagger_fH_u, $\\
 &
 $ \chi 10_f 10_f 10_f$
 &$ \mu \mugut  \chi \FiveBar_f$
$$
 &$\mugut \l\chiDagger H_u \text{, } \chiBarDagger H_d\r$\\
\hline
$ \left(10, \TenBar\right)$& $\chiBar 10_f \FiveBar_f H_u \left(= \bar{E}_\chi Q_f D_f H_d\right)\text{, } $  &  $ \chiBarDagger \FiveBar_f \FiveBar_f\text{, }  \mu \mugut  \chiBar 10_f$  &$ \chiBar 10_f \FiveBar_f H_u \left(= \bar{E}_\chi E_f L_f H_d\right)\text{, } $\\
&$\chi 10_f 10_f H_d \text{, } \chiBar \FiveBar_f \FiveBar_f \FiveBar_f \text{, }  
\chiBarDagger 10_f \FiveBarDagger_f$ &
&$\mugut \l\chiDagger 10_f\text{, } \chi \FiveBar_f\FiveBar_f\r,$ \\
& & & $\chiBar H_u\FiveBarDagger_f $\\
\hline
\end{tabular}
\caption[MSSM Decay Channels of Dimension 5 Operators]{\label{Tab:LowEnergyLithium}The classifications of dimension 5 operators based on the standard model operators that can be extracted from them. These operators are generated using the electroweak multiplets in $\chi$. Operators that require the higgs triplet fields are ignored. The subscript f denotes standard model families and $H_u$, $H_d$ are the Higgs fields of the MSSM. The R-parity of $\chi$ is chosen in order to make these operators R-invariant. }
\end{table}

\subsubsection{Colored Relics}
For the fundamental $\chi$ we have at least one pair of long-lived quarks $\chi_{d,\bar{d}}$, where
$\chi_{d}$ has quantum numbers of the right-handed MSSM $d$-quarks (and, as before, if the mass splitting between R-even and R-odds $\chi$-quarks
is small enough, we have another pair of long-lived colored particles). For antisymmetric $\chi$ we have long-lived
vector-like $\chi$-quarks both with quantum numbers of the right-handed MSSM $u$-quark $\chi_{u,\bar u}$ and left-handed
$u$ and $d$ quarks, $\chi_{Q,\bar Q}$.

The
evolution history in the early Universe is significantly more involved  for colored particles \cite{ColoredLuty} and the corresponding relic abundances are border-line to be uncalculable. The point is that in general heavy long-lived
colored particles experience two epochs of annihilation as the Universe expands. First, they suffer from the conventional
perturbative annihilations at high temperatures before the QCD phase transition. The resulting relic abundance of the colored particles is at the level $\Omega_\chi h^2\sim  10^{-3}\l {M_\chi\over TeV}\r^2$. This is a very interesting number---as follows from Figure \ref{Fig:LithiumRelicAbundance}, if correct it would imply that a long-lived colored particle with a mass
in the subTeV range solves both Lithium problems.

However, colored particles experience a second stage of annihilations after the QCD phase transition, that can
significantly reduce the abundance. Indeed, after
the QCD phase transition they hadronize---get dressed by a soft QCD cloud of the size of order $\sim\Lambda_{QCD}^{-1}$.
It is plausible that when two slowly moving hadrons  involving heavy colored particles collide, a bound state containing two heavy  particles forms with geometric cross-section $\sim 30$~mbarn. This conclusion is somewhat counterintuitive---naively, one may think that the soft QCD cloud cannot prevent two heavy particles from simply passing by each other without forming a bound object. The argument, however, is that the reaction goes into the excited level of the two $\chi$ system
of the size of order $\sim\Lambda_{QCD}^{-1}$. At low enough temperatures the angular momentum
of such a state  is close to the typical angular momentum of two colliding hadrons, $L_i\sim (m\chi T)^{1/2}\Lambda_{QCD}^{-1}$ so that one may satisfy the angular momentum conservation law by emission of a few pions.
Assuming that the reaction to such an excited level is exothermic the geometrical cross-section appears
to be a reasonable estimate.
After the excited state with two $\chi$'s forms it decays to the ground level and  $\chi$'s annihilate.
As a result the relic abundance can be reduced to the values below $\Omega_\chi h^2\sim  10^{-6}$
for a TeV mass particles, where they don't affect the Lithium abundance.

Definitely, many of the details of this story are rather uncertain at the quantitative level and
this conclusion has to be taken with a grain of salt. We discuss some of the involved uncertainties in more detail
in section~\ref{split}.
%
Interestingly, in the models we are discussing here, one may avoid  going into the detailed discussion of this complicated process and be rather confident that the residual abundance of the colored particles is close to that given by the perturbative calculation. The reason is that in order for the above mechanism to operate the two
$\chi$ particles in the bound state should be able to annihilate with each other. This is the case for some candidate
long-lived colored particles, such as gluino in the split SUSY scenario or stop NLSP decaying into gravitino, but not always true.

For instance, for antisymmetric  $\chi$ annihilation is possible in some of the bound states (e.g., $\chi_u\chi_{\bar u}$), but not in the others (e.g., $\chi_u\chi_{\bar Q}$).
Once formed, these bound states go to the Coulombic ground state, which is compact and doesn't get converted into other
mesons any longer. Through weak decays such a meson decays to the energetically preffered neutral
ground state, that survives until individual $\chi$ particles decay. In fact, it is likely that an original $\chi$-hadron
is a baryon, given that the reaction converting $\chi$-meson and ordinary proton into $\chi$-baryon and pion is exothermic.
This doesn't change the story much; after two transitions one obtains in this case baryons containing three $\chi$'s,
such as $\chi_u\chi_Q\chi_Q$. Again, such a baryon will decay to the stable state and will survive till individual $\chi$ particles decay.
Similarly, for fundamental $\chi$ an order one fraction of them ends up being in the compact baryonic state
$\chi_d\chi_d\chi_d$ which is safe with respect to annihilations.

We don't attempt here to analyze the above processes at the precision level, but this discussion implies that
the resulting abundance of colored $\chi$'s while being somewhat
reduced from its perturbative value is still high enough to solve
 $\LiSix$ problem, or even both Lithium problems.

\subsection{Supersymmetric Axion}
\label{Axinos}
We already mentioned dimension 5 decays involving the gravitino as one of the solution of the Lithium problems, that does not involve new particles beyond those present in the MSSM. There is another well motivated particle in the MSSM  that may have similar effects -- the axino (see, \cite{Covi:2001nw} for a detailed discussion of axino properties and cosmology).

The axion is a well-motivated new pseudo-scalar particle. It solves the strong CP-problem, and may constitute a fraction, or all, of the dark matter. Axion-like particles are also generic in string models.
In supersymmetric models, the axion
is a part of a chiral supermultiplet $S$, so it comes together with the scalar (saxino) and the fermionic (axino, $\tilde a$) superpartners. Being a
(pseudo)Goldstone boson, the axion supermultiplet couples to the MSSM fields suppressed by the PQ breaking scale $f_a$. The leading interactions are with the gauge fields
\be
\label{axino}
{\cal L}=\int d^2\theta{S\over f_a}\sum_{i=1}^3{C_i\alpha_i\over 4\pi}W_{(i)}^2+h.c.\;,
\ee
where  $C_i$ are model-dependent coefficients of order one.
These interactions are often generated at the one-loop level, for instance,
by integrating out  vector-like fields acquiring the mass from the vev of $S$ due to interactions like
$S\bar Q Q$ (KSVZ model \cite{Kim:1979if,Shifman:1979if}).
As a result, as compared to the  general model-independent analysis above
these dimension 5 operators contain extra one-loop suppression factors. Consequently, to be relevant for the Lithium problem the high-energy scale $f_a$ entering here has to be somewhat lower than the GUT scale, $f_a\sim 10^{14}$~GeV. Still, this scale is intriguingly close to the GUT scale.

The relevance of axino for the Lithium problems crucially depends on its mass. We will focus our attention on the gravity mediated SUSY breaking, so that there is no light gravitino. Then the saxion receives a mass of the same order as other soft masses,
$\sim F_{susy}/M_{Pl}$. On the other hand, the mass of axino is a hard SUSY breaking term and its value is highly model-dependent. If it is generated
at the loop level, it is suppressed by at least one extra loop factor, and varies between $\sim $GeV down to the keV range. It may also be generated at the tree level
after SUSY is broken, if SUSY breaking triggers some singlet fields to develop vev's giving rise to the axino mass.
In this way the axino acquires mass of the same order as other soft masses.

In all other respects, the axino is a particular example of adding an MSSM singlet. If the axino is lighter than the LSP, and the MSSM LSP is bino-like, it will decay to axino and  photon.
The corresponding lifetime is
\[
\tau\sim10^3\mbox{sec}\left({f_a\over 10^{14}\mbox{GeV}}\right)^2 \left({1\mbox{TeV}\over \Delta m_{\chi a}}\right)^3 \;,
\]
where $\Delta m_{\chi a}$ is the axino-LSP mass difference.
The hadronic branching fraction in this case is due to decays with virtual photon producing quarks and is at the few percent level, see Eqn. \eqref{Eqn: off-shell photon to fermions}.
If the resulting axino is the only cold dark matter component now these decays may solve the $\LiSeven$ problem, but
not the $\LiSix$ problem. If the dominant component of the cold dark matter is the axion, and the LSP is light, so that its thermal abundance is low, the LSP decay to the axino may solve both Lithium problems.

\subsection{Split SUSY}
\label{split}

\begin{figure}[htbp]
\begin{center}
\includegraphics[width=3.0in]{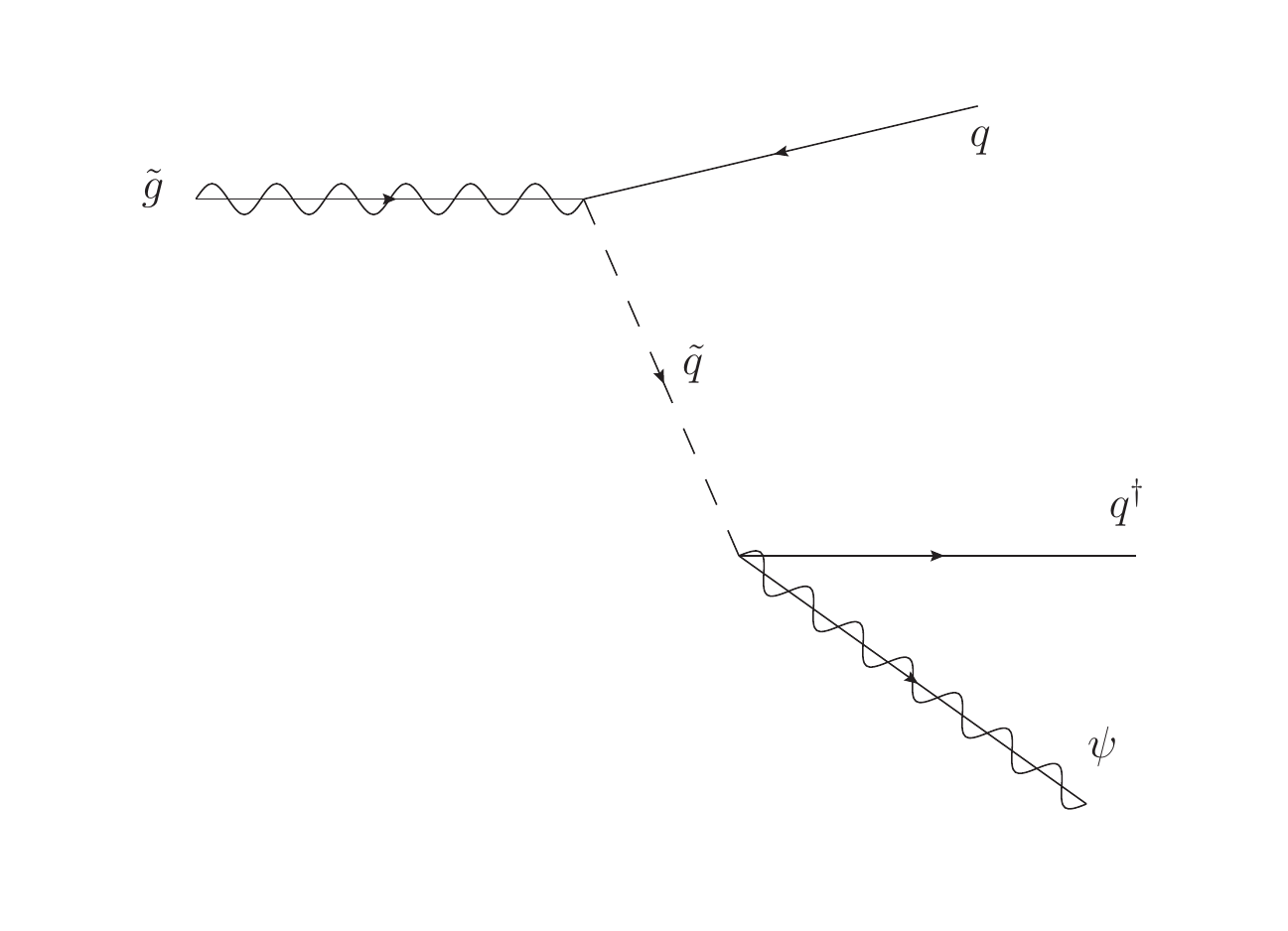}
\caption[Gluino decay in Split SUSY]{Gluino decay through an off-shell squark}
\label{gdecay}
\end{center}
\end{figure}

In the Split SUSY framework \cite{ArkaniHamed:2004fb}, the SUSY breaking scale is not the TeV but an intermediate scale up to $10^{12}$ GeV. All the scalars are at that scale except one Higgs that is tuned to be light. The gauginos and Higgsinos are protected by chiral symmetries and they are at the TeV scale, giving thermal dark matter and allowing for the gauge couplings to still unify at the GUT scale. The gluino of split SUSY can only decay through an off-shell squark (see Fig. \ref{gdecay}) and its lifetime is set by the SUSY breaking scale:
\begin{eqnarray}
\Gamma_{\tilde{g}}\sim &&\frac{1}{128\pi^3}\frac{m_{\text{\tiny{gluino}}}^5}{m_{SUSY}^4}\nonumber \\
\sim && \left(100 ~\text{sec}\right)^{-1} \left( \frac{m_{\text{\tiny{gluino}}}}{1~\text{TeV}}\right)^5 \left(\frac{2 \times 10^{9}~\text{GeV}}{m_{SUSY}}\right)^4
\end{eqnarray}
When the SUSY breaking scale is $10^9-10^{10}$ GeV,  and the gluino lifetime is $100-1000$ sec, it becomes an excellent candidate for solving the Lithium problem(see Fig. \ref{Fig:LithiumRelicAbundance}), if its abundance is between $\Omega_{\tilde{g}} h^2 \sim 10^{-4}-10^{-1}$.

Even though the gluino lifetime is well determined by the squark masses, its abundance is difficult to know because it involves strong dynamics \cite{ColoredLuty, GluinoCosmology}. At a temperature of $\sim \frac{m_{\text{\tiny{gluino}}}}{20}$ the gluino will thermally freeze-out. Its cross-section is calculable, since QCD is still perturbative at those times, and its abundance is roughly given by $\Omega h^2 \sim 10^{-3}\left( \frac{m_{\text{\tiny{gluino}}}}{1~\text{TeV}} \right)^2$. After the QCD phase transition, the gluino gets dressed into colored singlet states, R-hadrons, which have a radius of $\sim \Lambda_{QCD}^{-1}$. The question is if these extended states can bind to form gluinonium states, $\tilde{g}-\tilde{g}$, with a geometric cross-section, $\pi \Lambda_{QCD}^{-2} \sim 30~\text{mbarn}$, that will lead to a second round of gluino annihilation and suppress its abundance by several orders of magnitude.

Whether this second round of annihilations happens depends on the details of R-hadron spectroscopy. 

For example, even if the lightest state is an R-$\pi$, the reaction:
\begin{eqnarray}
\text{R-}\pi+\text{baryon} \rightarrow \text{R-baryon} + \pi
\end{eqnarray}
may well be exothermic due to the lightness of $\pi$ and, if it has a geometric cross-section, most of the gluina will end up in R-baryon states. Gluino annihilation now depends on the reaction:
\begin{eqnarray}
\text{R-baryon}+\text{R-baryon} \rightarrow \tilde{g}\text{-}\tilde{g} + 2~\text{baryons}.
\end{eqnarray}
If this reaction is exothermic, it may suppress the gluino abundance by many orders of magnitude. If this reaction is endothermic for high angular momentum gluinonium states, it may leave a significant number of gluinos in R-hadrons and there is no significant reduction in the gluino abundance. These uncertainties render the final gluino abundance incalculable.

However, the case where there is no second round of annihilations after the QCD phase transition provides an interesting scenario for the LHC. The gluino mass range that gives the measured $\LiSeven$ abundance is $300-7000$ GeV \cite{GluinoCosmology}. If the gluino is lighter than $\sim2$ TeV it will be produced at the LHC, stop inside the detector and then decay, allowing its discovery, the measurement of its lifetime, and the determination of the SUSY breaking scale \cite{StoppingGluinos}. As the gluino is the heaviest of the gauginos, this also suggests that all low scale Split particles have a good chance of being discovered at the LHC. For a 300 GeV gluino of a 1000 s lifetime, its decay can solve both the $\LiSeven$ and $\LiSix$ problems (see Fig. \ref{Fig:LithiumRelicAbundance}) and it will be abundantly produced at the LHC. The potential discovery of a gluino with a lifetime $100-1000$ sec at the LHC will solidify the primordial origin of the discrepancy between the measured $Li$ abundances and sBBN. 


\section{Models for Lithium and Decaying Dark Matter}
\label{LithiumPAMELA}

The primordial lithium abundance discrepancies and the observations of PAMELA/ATIC can be explained by the decays of  a TeV mass particle through dimension 5 and 6 GUT suppressed decays (see sections \ref{Sec: Dim 6 decays} and \ref{LithiumIntroduction}). The dimension 6 operators

\begin{equation}
\frac{S_m^{\dagger}S_m 10_f^{\dagger} 10_f}{\Mgut^2} \text{, } \frac{S_m^{\dagger}S_m\FiveBar_f^{\dagger} \FiveBar_f}{\Mgut^2} \text{, } \frac{S_m^{\dagger}S_m H_{u\l d\r}^{\dagger} H_{u\l d\r}}{\Mgut^2} \text{ and } \frac{S_m^2 \mathcal{W}_{\alpha} \mathcal{W}^{\alpha}}{\Mgut^2}
\end{equation}
in section \ref{Sec: Dim 6 decays} allow decays between a singlet sector $S_m$ and the MSSM with lifetimes long enough to explain the signals of PAMELA/ATIC. A large class of dimension 5 operators were discussed in section \ref{LithiumIntroduction}. In particular, operators of the form  
\begin{equation}
\frac{10_m S_m \FiveBar_f \FiveBar_f}{\Mgut} \text{ and } \frac{S_m \mathcal{W}_{\alpha} \mathcal{W}^{\alpha}}{\Mgut} 
\end{equation}
allow for the population of the singlet $S_m$ sector through dimension 5 decays from the MSSM or a new TeV scale $SU(5)$ vector-like sector, for example $10_m$.  In this section, we will show that both of these decays can be naturally embedded into SUSY GUT models. These models can naturally solve both the primordial lithium abundance problem and the observations of PAMELA/ATIC. We also consider models yielding mono-energetic photons that give qualitatively new signals for Fermi. 

\subsection{$SO(10)$ Model}

In $SO(10)$, the MSSM superpotential is: 

\begin{equation}
W_{MSSM} = \lambda_f 16_f 16_f 10_h + \mu 10_h 10_h
\end{equation}
where $16_f$ and $10_h$ are family and higgs multiplets. We add TeV scale multiplets  $\l16_m, \bar{16}_m\r$ and a GUT scale $10_{\gut}$ along with the following interactions: 

\begin{equation}
\label{SO10GUT}
W^{\prime} = \lambda 16_m 16_f 10_{\gut} + m 16_m \bar{16}_m + \Mgut 10_{\gut} 10_{\gut}
\end{equation}

These interaction terms allow for R-parity assignments -1 for $10_{\gut}$ and +1 for $16_m$ and preserves a $m$ parity under which $16_m$ and $10_{\gut}$ have odd parity.  Integrating out the $10_{\gut}$ field and the $SO(10)$ gauge bosons, we generate the dimension 5 and 6 operators: 

\begin{equation}
\label{SO10Dim5and6}
\int d^2 \theta \l \frac{16_m 16_m 16_f 16_f}{\Mgut}\r \text{, } \int d^4 \theta \l\frac{1}{16 \pi^2}\r \l\frac{16_m 16_m 10^{\dagger}_h}{\Mgut}\r \text{ and } \int d^4 \theta \l\frac{16^{\dagger}_m 16_m 16^{\dagger}_f 16_f}{M_{B-L}^2}\r
\end{equation}
where $M_{B-L}$ is the vev  that breaks the $SO(10)$  $U(1)_{B-L}$ gauge symmetry. The R and $m$ parity assignments forbid dangerous, lower dimensional Kahler operators like $10_{\gut}^{\dagger} 10_h$ and $16_m^{\dagger} 16_f$. The dimension 5 operators in (\ref{SO10Dim5and6}) connect the components of the $16_m$ multiplet that are charged under the standard model to the singlet component $S_m$ of the $16_m$ and the MSSM. However, the only operator in (\ref{SO10Dim5and6}) that allows for two singlet fields $S_m$ to be extracted from $16_m$ is the dimension 6 operator  $\l\frac{16^{\dagger}_m 16_m 16^{\dagger}_f 16_f}{M_{B-L}^2}\r$ which yields  $\l\frac{S^{\dagger}_m S_m 16^{\dagger}_f 16_f}{M_{B-L}^2}\r$. The phenomenology of this model is identical to that of the $SU(5) \times U(1)_{B-L}$ model discussed below. The decays of the standard model multiplets in $16_m$ to the singlets $S_m$ and the MSSM fields at $\sim$ 1000 s can solve the primordial lithium abundance problems while the decays between the singlets and the MSSM at $\sim 10^{26}$ s can reproduce the observations of PAMELA/ATIC. 

\subsection{$SU(5) \times U(1)_{B-L}$ Model}
\label{SU5U1B-L}

We  consider a $SU(5) \times U(1)_{B-L}$ model, with $U(1)_{B-L}$ broken at the scale $M_{B-L}$ near the GUT scale. The MSSM is represented in this model by the superpotential:

\begin{equation}
W_{MSSM} = \lambda^u_f 10_f 10_f H_u + \lambda^d_f 10_f \FiveBar_f H_d + \mu H_u H_d
\end{equation}
where $10_f$ and $\FiveBar_f$ are the standard model generations, $\lambda^u_f$ and $\lambda^d_f$ are the yukawa matrices and $H_u$ and $H_d$ are the higgs fields. We add a GUT scale $\l 10_{\gut}, \TenBar_{\gut} \r$, a TeV scale $\l 10_m, \TenBar_m \r$ and a singlet $S_m$ to this theory with the following additional terms $W^{\prime}$ in the superpotential: 

\begin{equation}
\label{Eqn:Dim5and6Interactions}
W^{\prime} = \lambda_1 10_m S_m \TenBar_{\gut} + \lambda_2 10_{\gut} \FiveBar_f \FiveBar_f +  \Mgut 10_{\gut} \TenBar_{\gut} + m 10_m \TenBar_m + m_s S_m S_m 
\end{equation}

The mass terms $m$ and $m_s$ are at the TeV scale and  $\Mgut$ is at the GUT scale. These interactions allow R parity assignments  +1 for $10_m$, $S_m$ and $10_{\gut}$. The superpotential also conserves a $m$ parity under which $S_m$ and $\l 10_m, \TenBar_m \r$ have parities -1.  Soft SUSY breaking will contribute to the scalar masses and lead to mass splittings between the fermion and scalar components. In particular, the singlet fermion mass $m^f_s$ will be different from the singlet scalar mass $m_{\tilde{s}}$. Integrating out the GUT scale field $10_{\gut}$ and the broken $U(1)_{B-L}$ gauge sector, we get the dimension 5 and 6 operators:

\begin{equation}
\label{Eqn:Dim5and6EffectiveOperators}
\int d^2\theta \left(\frac{10_m S_m \FiveBar_f \FiveBar_f}{\Mgut} \right)  \text{, } \int d^4 \theta \left(\frac{S_m^{\dagger}S_m Y^{\dagger}Y}{M_{B-L}^2}\right) \text{ and }  \int d^4 \theta \l\frac{1}{16 \pi^2}\r \left(\frac{S_m^{\dagger}S_m Y^{\dagger}Y}{\Mgut^2}\right)
\end{equation}
Here the $Y$ represent the other chiral multiplets $10_m$, $10_f$, $\FiveBar_f$, $H_u$ and $H_d$ in the model. The gauge symmetries of the standard model, supersymmetry and R and $m$ parities ensure that the operators in (\ref{Eqn:Dim5and6EffectiveOperators}) are the lowest dimension operators that connect particles carrying $m$ parity ({\it i.e.} $S_m$ and $10_m$) and the MSSM. 


\begin{figure}
\begin{center}
\includegraphics[scale=1.0]{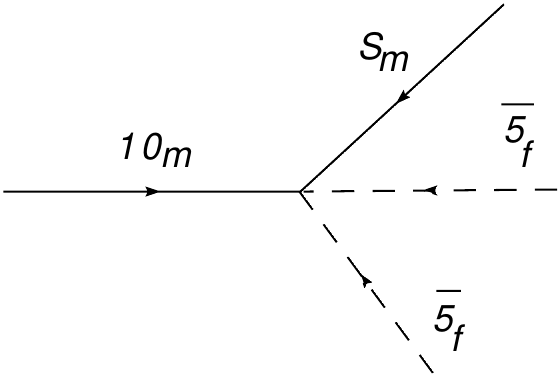}
\caption{ \label{Fig:Dim5Decay} $10_m$ decaying to $S_m$ and sleptons.  }
\end{center}
\end{figure}

Consider the phenomenology of this theory when the mass $m$ of the $10_m$ particles are greater than the singlet masses $m_s$ and $m_{\tilde{s}}$. The $10_m$ are produced with a thermal abundance due to standard model gauge interactions. They will decay to the singlets $S_m$ and the MSSM particles through the dimension 5 operator in (\ref{Eqn:Dim5and6EffectiveOperators}) with lifetime $\tau \sim 1000 \text{ s} \l\frac{1 \text{ TeV}}{\Delta m }\r^3 \l\frac{M_H}{10^{16} \text{ GeV}}\r^2$ where $\Delta m$ is the mass difference between the $10_m$ and the singlet  (see figure \ref{Fig:Dim5Decay}). Following the discussion in section \ref{LithiumIntroduction}, the decays of the electroweak and colored multiplets in the $10_m$ can solve the primordial $\LiSeven$ and $\LiSix$ abundance problems. The decays of the $10_m$ generates a relic abundance of the singlets $s_m$ and $\tilde{s}_m$.  Since R and m parities are conserved in this model, decays between the singlets and the MSSM have to involve the dimension 6 operators in (\ref{Eqn:Dim5and6EffectiveOperators}) when the $10_m$ fields are heavier than the singlets. These operators are identical to the R-parity conserving operators discussed in section \ref{Sec: Dim 6 decays}. Following the discussion in that section, the decays mediated by these operators can explain the observations of PAMELA/ATIC. 

A decaying particle can explain the observations of ATIC if its mass is $\sim 1.2$ TeV (see section \ref{Sec: Astro signals} and \cite{ATIC}). We first analyze the case where one component of the singlet has a mass $\sim$ 1.2 TeV and this component decays to its superpartner and the MSSM. Without loss of generality, we assume that the scalar singlet $\tilde{s}_m$ is the decaying particle with mass $m_{\tilde{s}} \sim 1.2 \text{ TeV}$. The singlet fermion $s_m$ with a mass $m_s$ and the MSSM LSP with mass $M_{LSP}$ are light. A relic abundance of $\tilde{s}_m$ is generated in the early universe from the decays of the thermally produced $10_m$. The relic energy density of the $10_m$ is $\sim 0.2 \l \frac{m}{\text{1.5 TeV}}\r^2$.  $10_m$ decays to both  $s_m$ and $\tilde{s}_m$. However, since $s_m$ is lighter than  $\tilde{s}_m$, the branching fraction for decays to $s_m$ is higher due to the larger phase space available for the decay.  When $m_s \ll m_{\tilde{s}} $, the branching fraction for decays to $\tilde{s}_m$ is $\sim \l 1 - \frac{m_{\tilde{s}}}{m}\r^3$. The relic energy density $\Omega_{\tilde{s}_m} h^2$ of $\tilde{s}_m$ is $\sim  0.2 \l \frac{m}{\text{1.5 TeV}}\r^2 \l \frac{m_{\tilde{s}}}{m}\r  \l 1 - \frac{m_{\tilde{s}}}{m}\r^3 $. For $m \sim 1.5 \text{ TeV}$ and $m_{\tilde{s}} \sim 1.2 \text{ TeV}$, we get  $\Omega_{\tilde{s}_m} h^2 \sim 10^{-3}$. The decays of $\tilde{s}_m$ can explain the observations of PAMELA/ATIC if the lifetime $\sim 10^{24} \text{ s}$. This lifetime can be obtained from the dimension 6 GUT suppressed operators in (\ref{Eqn:Dim5and6EffectiveOperators}) if the $U(1)_{B-L}$ symmetry is broken slightly below the GUT scale  with a vev $M_{B-L} \sim 3 \times 10^{15} \text{ GeV}$. 

The other possible decay topology is for the MSSM LSP to decay to the singlets, with the LSP mass $M_{LSP} \sim 1.2 \text{ TeV}$. Stringent limits from dark matter direct detection experiments \cite{DirectDetection} and heavy element searches \cite{HeavyElement} require the thermally generated $10_m$ abundance to decay to the MSSM and the singlets.  The $10_m$ can decay to the MSSM and the singlets $s_m$ through the dimension 5 operator in (\ref{Eqn:Dim5and6EffectiveOperators}) if the spectrum permits the decay. Since the $10_m$ has R parity +1, its fermionic components  $10_m$ can decay only if their mass is greater than the LSP mass $\sim 1.2 \text{ TeV}$. The scalars $\tilde{10}_m$ can decay to the standard model and the singlets as long as the scalars are heavier than the singlets. However, if the scalars $\tilde{10}_m$ and the singlets  are too light, the MSSM LSP will decay to $\tilde{10}_m$ and the singlet fermion through the dimension 5 operator in  (\ref{Eqn:Dim5and6EffectiveOperators}). This decay occurs with a lifetime $\sim 1000 \text{ s}$ and will not explain the PAMELA observations. This decay mode must be shut off in order for the MSSM LSP to decay through the dimension 6 operator in (\ref{Eqn:Dim5and6EffectiveOperators}). This is achieved if the sum of the $\tilde{10}_m$ and singlet masses are larger than the LSP mass.


\subsection{Supersymmetric Axions}

\begin{figure}[t]
\begin{center}
\includegraphics[width=3.5 in]{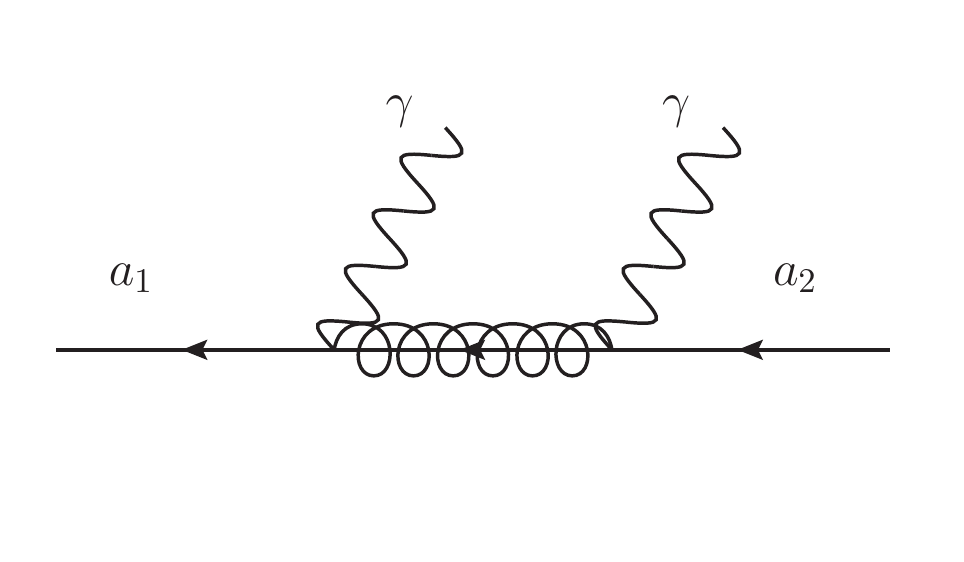}
\caption[Axino decay]{ \label{Fig:axinodecay} Axino decay to the lighter axino and two photons.  }
\end{center}
\end{figure}
The interesting example of the setup that provides both dimension 5 and dimension 6 mediated decays is a
mild generalization of the axino model for Lithium.
Namely, let's assume that there are two axion-like particles corresponding to different PQ symmetries, so that they are coupled
to MSSM through different combinations of operators as in (\ref{axino}) (in other words, coefficients $C_i$ are different in the two cases).

In this case there are two axinos and after the LSP decays the dark matter will be a mixture of the two. The heavier of the axinos is unstable, but its decays involve insertion of two dimension five operators, so effectively
it has dimension 6 suppression.
Typically, the fastest decay channel is $\tilde a_1\to\tilde a_2+2\gamma$ (see Fig.~\ref{Fig:axinodecay}),
giving
\[
\tau\sim10^{33}\mbox{sec}\left({f_a\over 10^{14}\mbox{GeV}}\right)^4 \left({m_{\chi}\over\mbox{TeV}}\right)^2 \left({\mbox{TeV}\over \Delta m_{12}}\right)^7\;,
\]
where $m_{12}$ is the axino mass difference.
The decays involving other gauge bosons are suppressed by the
higher mass of the intermediate gaugino in the diagram in Fig.~\ref{Fig:axinodecay}.
As discussed in section \ref{Axinos}, axino masses are highly model dependent
and may be as high as of order $m_{SUSY}$. If that is the case for one of the axinos, its decays will produce monoenergetic photons
potentially observable by Fermi.

\begin{figure}[t]
\begin{center}
\includegraphics[width=3.5 in]{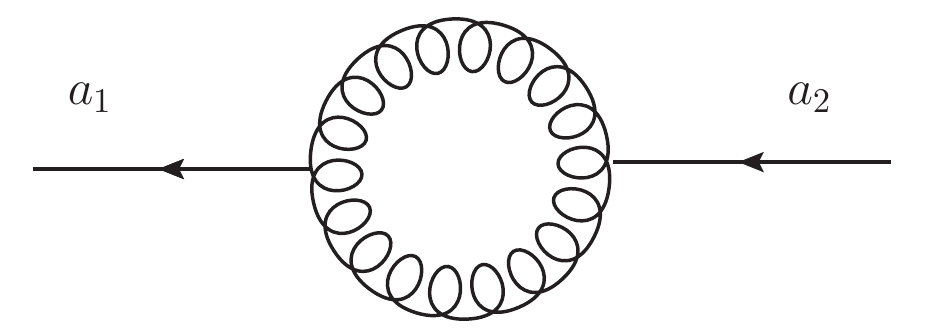}
\caption[Axino mixing]{ \label{Fig:mixing} Mixing between two axinos induced by the gauge loop.  }
\end{center}
\end{figure}

Note that at the one loop level the two axions (and axinos) are mixed by the gauge loops (see Fig.~\ref{Fig:mixing}). This mixing is generated at the high scale,
so there is no reason for it to be small. However, this does not change the conclusion that the decay of one axino to the other is
effectively dimension 6. Indeed, after one diagonalizes axinos kinetic terms and mass matrices, at the dimension 5 level the resulting eigenstates  have only couplings of the form (\ref{axino}). An intuitive way to understand this is to note that axion is a (pseudo)Goldstone boson, so each axion (and axino) carries a factor of $f_a^{-1}$ with it. Consequently processes involving two of them have dimension 6 suppression.

In principle, one may generalize this story to the broader class of models explaining Lithium problems by dimension 5
decays. Namely, one may
 consider  two  singlet fields (not necessarily axinos),  that do not couple directly to each other and
decay through the dimension 5 operators listed in the Table~\ref{Tab:LithiumOperators}.
However, generically, there is a danger that the dimension 5 operator involving two singlets can be generated and as a result the transition of one singlet to another will be rapid. One can avoid this problem for some choices of operators by imposing additional symmetries, and for  other choices of operators one may check that there are UV divergent diagrams already at the level of effective theory that makes this proposal invalid.  The advantage of the axino scenario, is that the absence of the dangerous dimension 5 operators is built in automatically.

\section{Astrophysical Signals}
\label{Sec: Astro signals}

In this section we consider the astrophysical signals of dark matter decaying through dimension 6 operators at current and upcoming experiments including PAMELA, ATIC, and Fermi.  The main astrophysical signals of long-lived particles decaying at around 1000 s from dimension 5 operators are changes to the light element abundances as produced during BBN.  As well as solving the Lithium problems, such a decay could make a prediction for the abundance of $^9\text{Be}$ \cite{Jedamzik2008}.  These decays may also generate a significant component of the current dark matter abundance of the Universe (see section \ref {LithiumPAMELA}), resulting in the production of naturally warm dark matter that may contribute to observed erasure of small structure \cite{WarmDarkMatter}.

\subsection{Electrons and Positrons and Signals of SUSY}

The PAMELA satellite has reported~\cite{Adriani:2008zr} a rise in the observed positron ratio above the expected background starting at about 10 GeV and continuing to 100 GeV where the reported data ends. The ATIC balloon~\cite{ATIC} experiment has observed an interesting feature in the combined flux of electrons and positrons  which begins to deviate from a simple power law around 100 GeV, peaks at around 650 GeV, and descends steeply thereafter. The HESS experiment~\cite{HESS} has added new data above 700 GeV, clearly observing the steep decline of the spectrum. It is attractive to explain both the PAMELA and the ATIC signals as coming from a new population of electrons and positrons which are the products of dimension six decay of dark matter suppressed by the GUT scale. As we have noted in section~\ref{sec:explaining-pamela}, such a GUT suppressed decay agrees with the ATIC and PAMELA rates, both of order $10^{26}$ seconds. In this section we will show that the spectral shape observed by both experiments can be explained as well. We will find that decays of dark matter allow for a much better fit to ATIC data, as compared with models considered thus far, when dark matter is allowed to decay to both standard model particles and superpartners. 

Beyond the rough agreement in rates, additional important information may be gained by studying the spectral shape of the ATIC feature as well as the PAMELA rise. 
These shapes will become even clearer as upcoming experiments will add new data.
The PAMELA experiment is expected to release positron data up to $\sim270$ GeV as well as measure the combined electron positron flux up to 2 TeV. The Fermi satellite, though optimized for gamma rays, has a large acceptance and is also capable of measuring the combined flux.  HESS is also capable of measuring the combined flux and may be able to add to their already published by measuring the flux at energies down to $\sim 200$ GeV, checking the ATIC bump.
As has been shown for annihilations~\cite{NealNew}, assuming a new smooth component of positrons that would explain the PAMELA rise accompanied by an equal spectrum of new electrons and extrapolating to higher energies, gives rough qualitative agreement with the height and slope of the ATIC feature. Here we would like to stress that analysis of the precise shape of electron-plus-positron flux, particularly as data improves, may allow for surprising discoveries. This may already be demonstrated by examining ATIC data closely and considering models that fit its shape.

In what follows, we will take the published ATIC data with only statistical errors as an example, and assume for now that there are no large systematics.  Thus, we will discuss in detail how the spectrum may be fit, including its various features.  Such an assumption may well be wrong.  ATIC may have many systematic effects in the data, rendering small features meaningless.  We will take it seriously merely to give an example of all the information that may be extracted from a full spectrum, including small features on top of the overall bump.  From this we learn an interesting new qualitative feature, that dark matter decays can generically produce several features in the spectrum and not merely a smooth rise and fall around the dark matter mass.  However, we do not believe that the ATIC data are conclusive yet.  Future data with better systematics, including from the Fermi satellite, are necessary before we can definitively conclude anything from the details of the electron spectrum.

\begin{figure}
\begin{center}
\includegraphics[width=3.3 in]{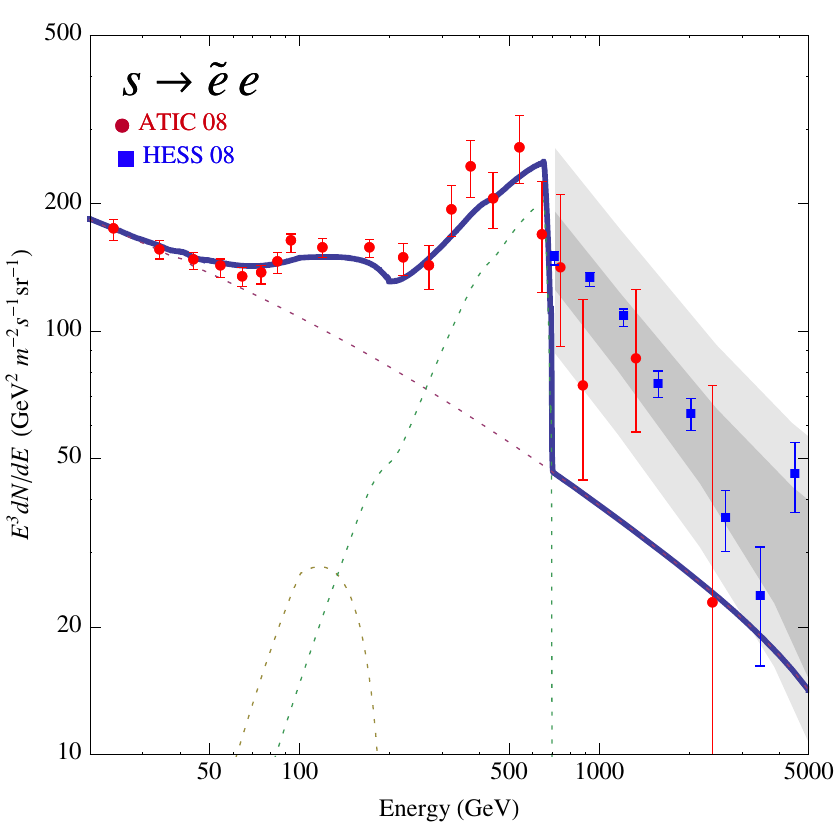}\qquad
\includegraphics[width=3.3 in]{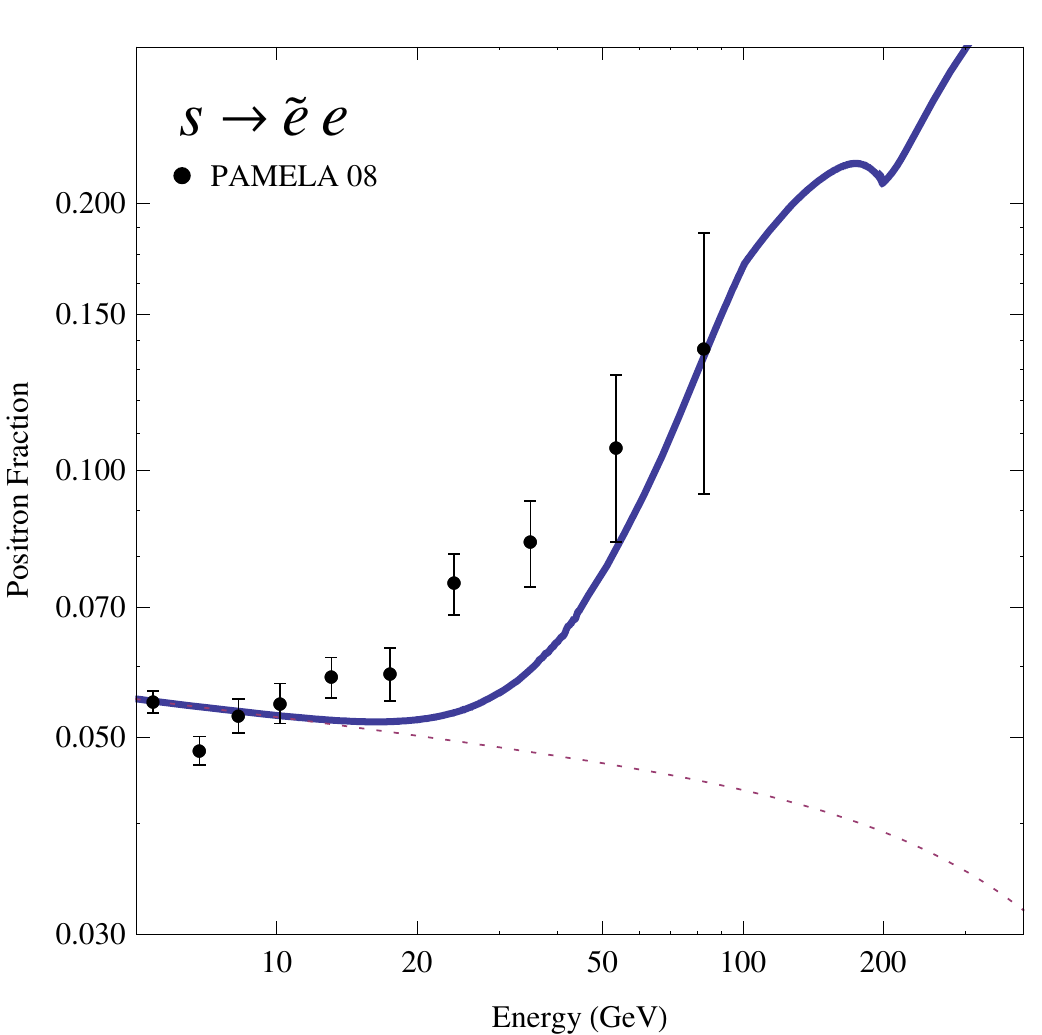}
\caption[ATIC and PAMELA spectra for $\mathrm{DM}\to e\tilde e$]{ \label{fig-ATIC1} (Color online) Left: The combined electron-positron flux for a heavy dark matter candidate decaying to an electron and a selectron. The selectron subsequently decays to a neutralino and a positron giving a two features in the observed flux. The dotted curves are the various components of the flux - electron, selectron and background.
ATIC (red circles) and HESS (blue squares) data are shown. The darker grey band is the HESS systematic error and the lighter grey is the area covered by this error after  including the uncertainty in the energy scale shift in both directions. Right: The positron fraction for the same decay and PAMELA data. The dotted line is the expected background.}
\end{center}
\end{figure}

The ATIC spectrum, shown in figure~\ref{fig-ATIC1}, may in fact contain two features -- a soft feature around 100-300 GeV and a hard feature at 300-800 GeV. The hard feature may extend to yet higher energies if HESS data is considered. 
The collaboration itself and the following literature has generally highlighted the hard and more pronounced feature, however it should be noted that the statistical errors in the 100-300 GeV range are quite small, and when compared with a power law background extrapolated from low energies, the soft feature may well be more significant statistically. The soft feature also has a peculiar shape; a sharp rise just below 100 GeV, followed by a nearly flat, or slightly descending spectrum up to 300 GeV. 
The shape of the hard feature is less clear, and may be interpreted either as ending sharply at 800 GeV or ending more smoothly around 2 TeV if HESS data is added.
It is tempting to fit the two ATIC features by two different products of the same decay which would give these shapes.

\begin{figure}
\begin{center}
\includegraphics[width=3.5 in]{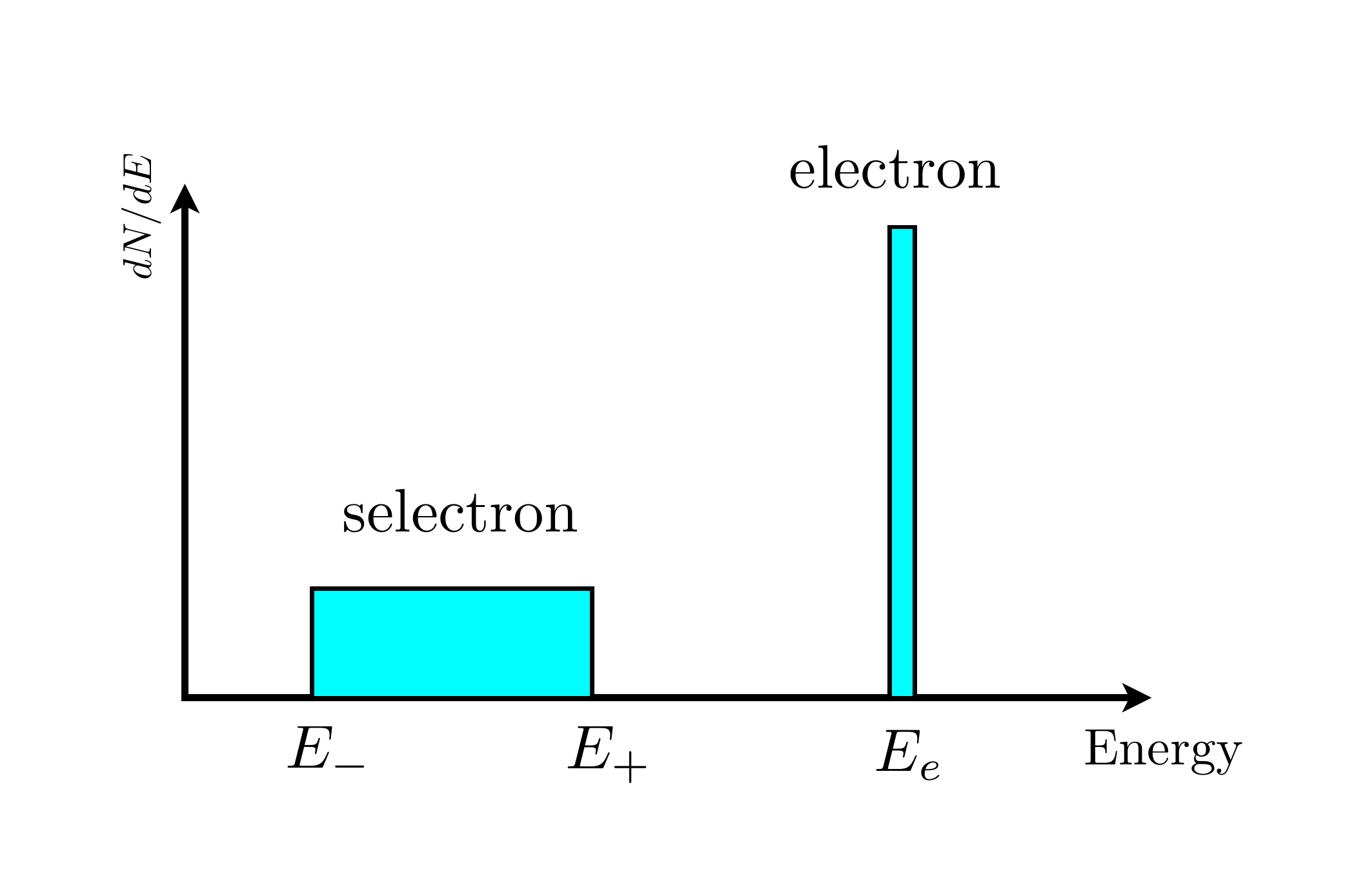}
\caption[Schematic injection spectrum for the decay $\mathrm{DM}\to e\tilde e$]{ \label{fig-box} A schematic plot of a spectrum of the electron-positron pair emitted when a singlet dark matter particle decays to an electron and a selectron. Such a spectrum provides an explanation of the two features in the ATIC flux.}
\end{center}
\end{figure}

If dark matter is a singlet which is heavier than the MSSM LSP, the products of its decay may be superpartners. If this is the case,  the decay of dark matter may potentially allow us to discover supersymmetry and measure its spectrum. Consider for example the simple dimension six operator of equation~(\ref{SdS10d10}), which allows the singlet fermion to decay to an electron and a selectron with a lifetime given by
\begin{equation}
\tau_{s\to\tilde e e} \sim 2 \times 10^{26} 
\left(\frac{1\mbox{ TeV}}{\Delta m}\right)^3
\left(\frac{1\mbox{ TeV}}{\langle \tilde s \rangle}\right)^2 
\left(\frac{M_{GUT}}{10^{16}\mbox{ GeV}}\right)^4 
\mbox{ sec}\,.
\label{rate-electron-selectron}
\end{equation}
This is a two body decay and both particles are mono-energetic. In particular the electron (or positron) energy will be
\begin{equation}
E_e=\frac{m_s^2-m_{\tilde e}^2}{2 m_s}
\label{electron-line}
\end{equation}
The selectron, being a scalar will then decay isotropically in its rest frame. Assuming the decay is directly to an electron and a neutralino LSP the isotropic decay will give a flat energy distribution in the ``lab'' frame between two edges
\begin{equation} 
E_+=\frac{m_s(m_{\tilde e}^2-m_{LSP}^2)}{2m_{\tilde e}^2} \qquad
E_-=\frac{m_{\tilde e}^2-m_{LSP}^2}{2m_s}
\label{selectron-edges}
\end{equation} 
The combined spectrum of the electron-positron pair emitted in the decay is shown schematically in figure~\ref{fig-box}.
Note that if one were to measure the three energies $E_e$, $E_-$ and $E_+$, the mass of the dark matter, the selectron and the LSP may be solved for without ambiguity.
We further notice that the flat spectrum that the selectron produces is reminiscent of the plateau above 100 GeV in the ATIC flux. The hard monochromatic electron may produce the hard ATIC feature. 

In the top left panel of figure~\ref{fig-ATIC1} we show that such a decay can indeed fit the ATIC data remarkably well. 
We have used GALPROP (v5.01)~\cite{galprop} to generate electron and positron background and also to inject and propagate the signal, modifying the dark matter annihilation package to one for decays. 
In order to reproduce the ATIC shape, including its sharp rises in flux, we used a propagation model with a relatively thin diffusion region,  ``model B'' of~\cite{Cholis:2008hb} ($L=1$ kpc) which was found to agree with cosmic ray data. In addition, to the propagation model the ATIC flux depends on the electron background spectrum which is still highly uncertain~\cite{Harnik:2008uu} (a slope of $\sim-3.15\pm0.35$), because it can only be constrained by the electron data itself as well as by photons to some extent\footnote{One could expect that low energy ATIC data itself would constrain the background slope. However this region may be somewhat contaminated by signal, loosening the constraint. In fact, extrapolating a power law to high energies and taking the steep decline seen by HESS into account, may imply that the background is on the steeper end of the uncertainty.}.   This is in contrast with the positron background which may be indirectly linked to spectra of nuclei.  In this case we have taken a background electron flux with a slope of $-3.3$ at 20 GeV. 

Considering ATIC data alone (we will discuss HESS later), the spectrum that gave a good fit (by eye) was $(E_e, E_-,E_+)=(700, 100,200)$ GeV
which may be inverted to produce a heavy spectrum of 
\begin{equation}
m_s\sim2800\mbox{ GeV}\qquad
m_{\tilde e}\sim2000\mbox{ GeV}\qquad
m_{LSP}\sim1800\mbox{ GeV}\,.
\end{equation}
It is remarkable that such a good fit to both ATIC features may be achieved by assuming the same rate of injection for both the hard electron and the selectron, strengthening the case for them to come from the same decay. Assuming the singlet $s$ makes up all of dark matter we can deduce its lifetime from the rate that fits th ATIC spectrum to be $1.5\times 10^{26}$ seconds. Using equation~(\ref{rate-electron-selectron}), this decay is probing a scale of $0.8\times 10^{16}$, remarkably close to the GUT scale.

The heavy selectron is required so that the selectron will be mildly boosted producing the feature at relatively low energies. 
The superpartner spectrum is quite sensitive to the values of $(E_e, E_-,E_+)$ and thus to the shape of the combined electron-positron flux. The uncertainties on these masses are thus large, being affected by  propagation uncertainties as well as the statistical and systematic errors of ATIC, HESS and future data. It will be interesting to perform a more systematic analysis as data improves. 

It is also interesting to consider other ways to get a similar spectrum. For example, if dark matter is a scalar singlet, it may be decaying both to an $e^+e^-$ pair and to a $\tilde e^+ \tilde e^-$ pair.  This would be given for example by the operator $S^\dagger\, 10_f H_u^\dagger \bar 5_f$ from Section~\ref{Sec: S-number violating} where the Higgs gets a vev.  This operator does not have a helicity suppression because the scalar $\tilde{s}$ decays to two left-handed spinors.  In order to  reproduce the hard ATIC feature  at 700 GeV and the soft feature between 100 and 200 GeV, the spectrum is lighter
\begin{equation}
m_{\tilde s}\sim1400\mbox{ GeV}\qquad
m_{\tilde e}\sim660\mbox{ GeV}\qquad
m_{LSP}\sim500\mbox{ GeV}\,.
\end{equation}
However, if the operators that lead to decays to electrons and to selectrons are suppressed by the same high scale, as would be expected in a supersymmetric theory, one would expect that the decay to electrons would be more rapid due to phase space. The ATIC spectrum on the other hand requires the two rates to be equal, which would imply the two rates must be tuned independently somehow.

On the top right panel of figure~\ref{fig-ATIC1} we show the positron fraction this decay, $s\to\tilde e^\pm e^\mp$, produces. Qualitatively the positron fraction agrees with that seen by PAMELA, a monotonic rise. However, the positron fraction in this case stays flat until roughly 40 GeV before climbing rather abruptly whereas the PAMELA data goes up more smoothly. Though the fit is not perfect, the quantitative disagreement is not very significant. At yet higher energies, beyond the range of current data, the positron fraction undergoes a wiggle, transitioning from the soft to the hard components of the spectrum. This feature is an interesting prediction for future PAMELA data or other experiments such as AMS2.

The mild disagreement between the PAMELA data and the spectrum that fits ATIC exemplifies that there is a mild tension between the detailed shape of the spectra observed by the two experiments. This tension may be phrased model independently by noticing that given a smooth power law for the electron background, say the dotted background line on the right panel of figure~\ref{fig-ATIC1}, the difference between it and the ATIC spectrum may be interpreted as signal, half of which is electrons and half of which is positrons. This new component of positrons can be combined with the positron background and the total flux to give a positron fraction which has the peculiar shape seen in the right panel of figure~\ref{fig-ATIC1}. The mild tension is thus between the two data sets, regardless of whether the signal originates from annihilation or decays. 

\begin{figure}
\begin{center}
\includegraphics[width=3.3 in]{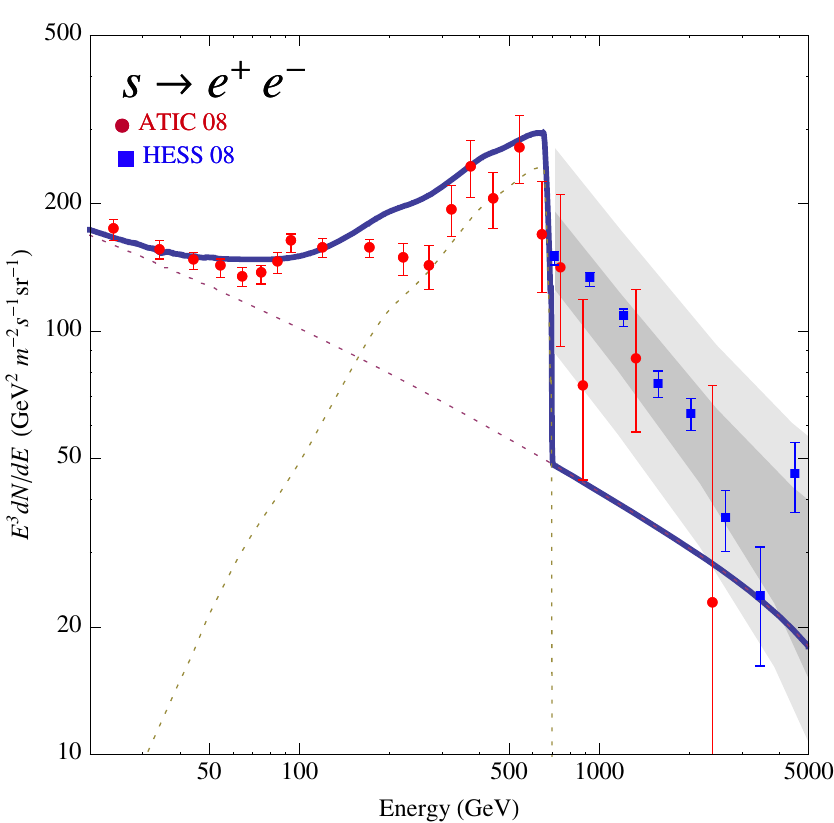}\qquad
\includegraphics[width=3.3 in]{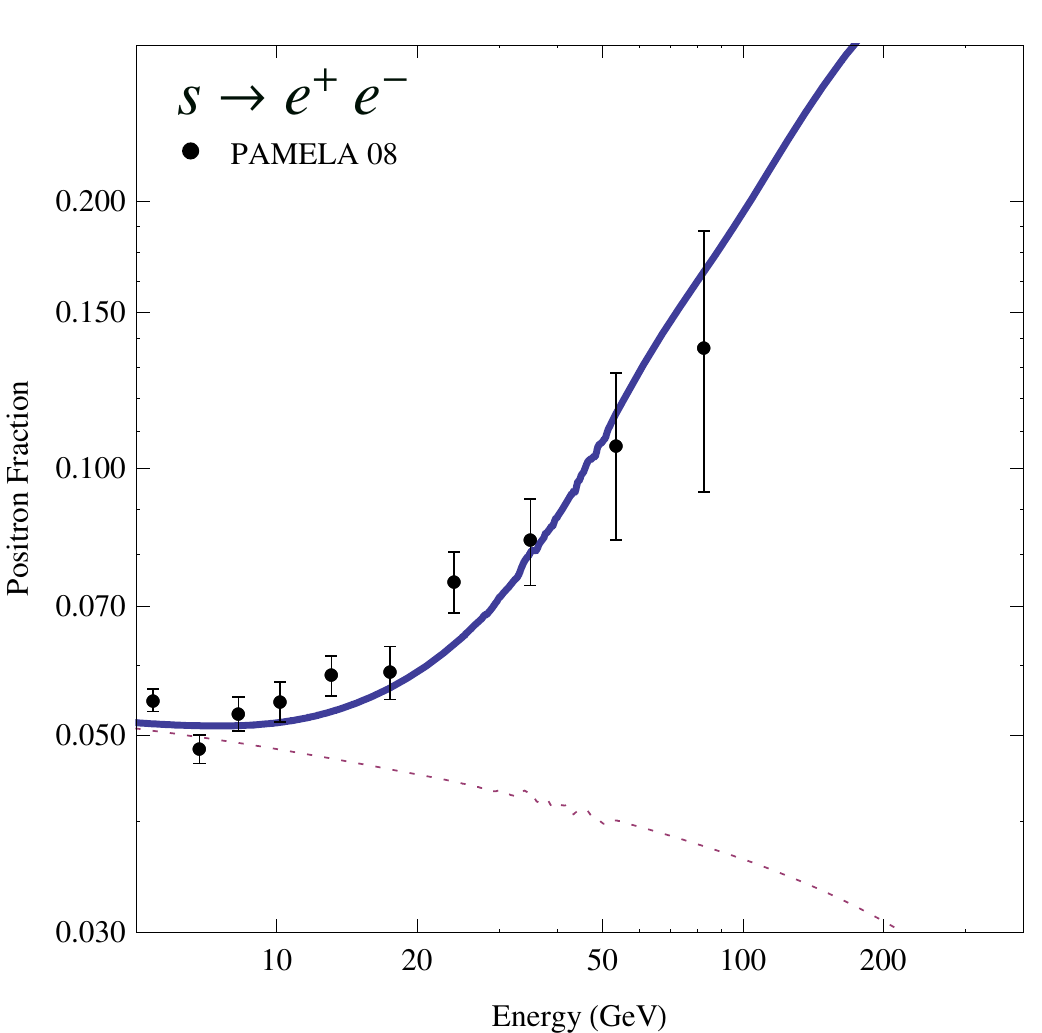}
\caption[ATIC and PAMELA spectra for $\mathrm{DM}\to e e$]{ \label{fig-ATIC2} (Color online) Left: The combined electron-positron flux for a heavy dark matter candidate decaying to an electron-positron pair. The dotted curves are the various components of the flux - signal and background. ATIC (red circles) and HESS (blue squares) data are shown. The darker grey band is the HESS systematic error and the lighter grey is the area covered by this error after  including the uncertainty in the energy scale shift in both directions. Right: The positron fraction for the same decay and PAMELA data. The dotted line is the expected background. In this decay channel the fit to ATIC data is inferior to that in the previous case, but the fit to PAMELA data is slightly improved}
\end{center}
\end{figure}

An alternative approach is to begin with a smoother signal that reproduces the PAMELA shape very well, but to give up on the various ATIC features and treat ATIC as a single smoother bump. This is shown in figure~\ref{fig-ATIC2} in which both the combined flux and the positron fraction are shown for a dark matter with a mass of 1.4 TeV that is decaying to an electron-positron pair.
The lifetime of dark matter in this case is $3\times 10^{26}$ seconds which may result by a dimension six operator suppressed by a scale of $1.5\times10^{16}$ GeV. This spectrum does not require sharp features and a more standard propagation model was used, (the diffusive convective model 
of~\cite{Lionetto:2005jd}).

\begin{figure}
\begin{center}
\includegraphics[width=3.3 in]{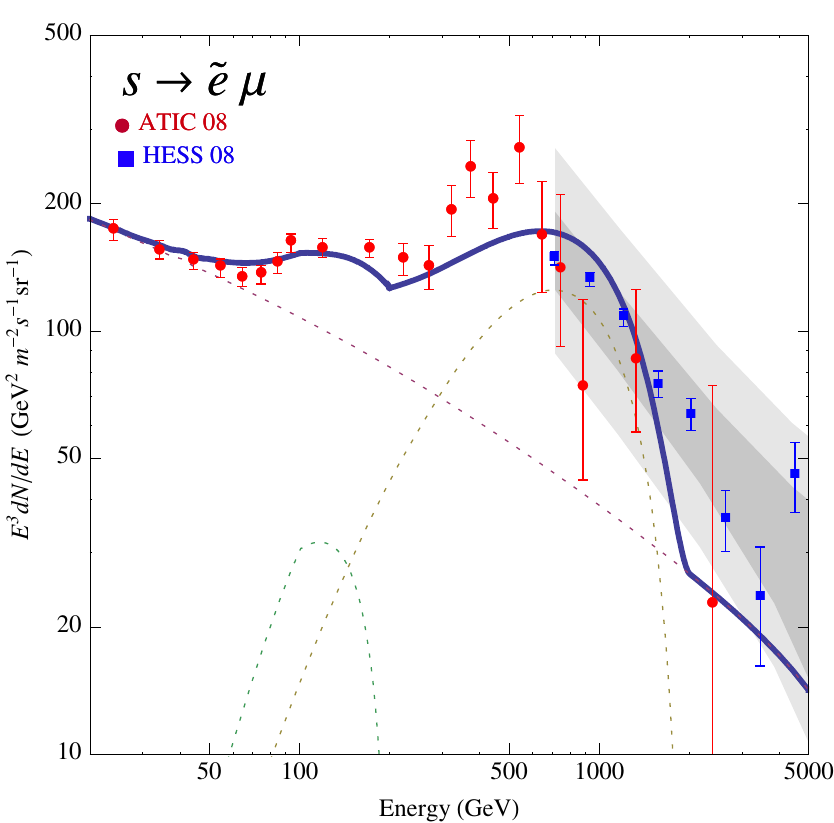}\qquad
\includegraphics[width=3.3 in]{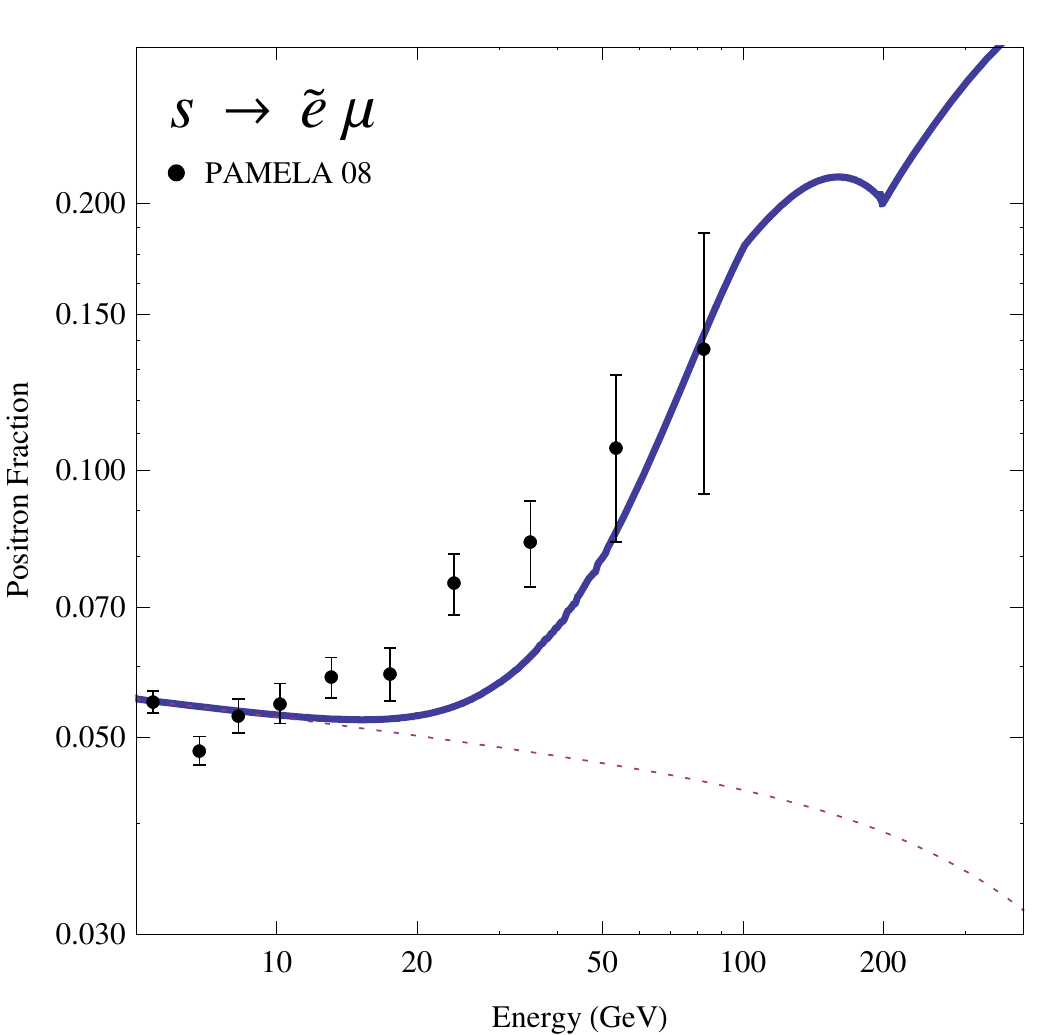}
\caption[ATIC and PAMELA spectra for $\mathrm{DM}\to \mu\tilde e$]{ \label{fig-ATIC3} (Color online) Left: The combined electron-positron flux for a heavy dark matter candidate decaying to a hard muon (2 TeV) and a selectron. The selectron subsequently decays to a neutralino and a positron giving a two features in the observed flux. The dotted curves are the various components of the flux - muon, selectron and background. The more rounded spectrum of the hard muon is in qualitative agreement with HESS data. ATIC (red circles) and HESS (blue squares) data are shown. The darker grey band is the HESS systematic error and the lighter grey is the area covered by this error after  including the uncertainty in the energy scale shift in both directions. Right: The positron fraction for the same decay and PAMELA data. The dotted line is the expected background.}
\end{center}
\end{figure}

Though the hard ATIC peak by itself fits a sharp electron positron peak at around 700 GeV, when combined with recent HESS data the hard feature may be interpreted as a broader bump, extending to a couple of TeV. The ambiguity in interpretation arises from the uncertainty of whether HESS is seeing the background beyond the ATIC bump, or the decline of the bump itself to a lower and steeper background. Though the HESS statistical errors are small, these should be added to a larger systematic error (shown as a grey band in the figures), and to an overall energy scale uncertainty (whose effect is roughly shown in the figure as a wider and lighter grey band). Nonetheless, the HESS decline motivates fitting the harder ATIC feature with a smoother spectrum. This may be done without losing the good fit to the soft feature given by a mono-energetic selectron. In figure~\ref{fig-ATIC3} we show the combined flux and positron fraction from a heavy dark matter singlet decaying to a mono-energetic muon, with an energy of 2 TeV, and a slow slepton, that produces a box-like spectrum as in figure~\ref{fig-box}.
The shape of  the hard ATIC peak including the HESS decline is qualitatively reproduced, within the propagation uncertainties. This particular spectrum may be achieved if a very heavy dark matter candidate (of order 8 TeV) is decaying to a heavy slepton (at 5.6 TeV) which decays to a slightly lighter neutralino (5.5 TeV). Though this spectrum is very heavy, is remarkable that a good fit to both features may be gotten from a single decay.

\subsection{Gamma-Rays}

Gamma rays can provide the best evidence that dark matter is the explanation for the observed electron/positron excesses.  Gamma rays produced in our galaxy at energies of hundreds of GeV are not bent or scattered and so provide clean directional and spectral information.  Such high energy gamma-rays have been searched for with Atmospheric Cherenkov Telescopes (ACTs) such as HESS and will be searched for in the upcoming Fermi telescope (GLAST).  Gamma-rays from dark matter decays or annihilations will appear as a diffuse background with greater intensity in directions closer to the galactic center.  There is also an expected astrophysical diffuse background and so to distinguish the dark matter signal may require spectral information as well.  The classic signals of dark matter are lines, edges, or bumps in the gamma-ray spectrum.

Dark matter annihilations do not generically produce gamma rays as a primary annihilation mode.  By contrast, dark matter decays may directly produce photons as a primary decay mode, as discussed in Section \ref{Sec: Dim 6 decays}.  Operators such as $S^2 \mathcal{W}_\alpha \mathcal{W^\alpha}$ or $\frac{m_{SUSY}}{M_{GUT}^2}H_d W  \partial\hspace{-2mm}/\hspace{.3mm} \bar 5_f^\dagger $ cause decays of $\psi \to \gamma \gamma$ or $\psi \to \gamma \nu$ which produce monoenergetic photons and neutrinos.  This will appear as a line in the gamma-ray spectrum, easily distinguishable from backgrounds if the rate is fast enough.  This would also be a line in the diffuse neutrino spectrum, perhaps detectable at upcoming experiments such as IceCube, but it would likely be seen first in gamma rays.  The exact reach of upcoming experiments depends on the energy of the line as well as the halo profile of the dark matter.  However, it is clear from the conservative limits in Table \ref{Tab: astro limits} that HESS observations are already in the range to detect such decays.  Fermi and future ACT observations will significantly extend this reach, probing GUT scale suppressed dimension 6 operators.

\begin{figure}
\begin{center}
\includegraphics[width=6.0 in]{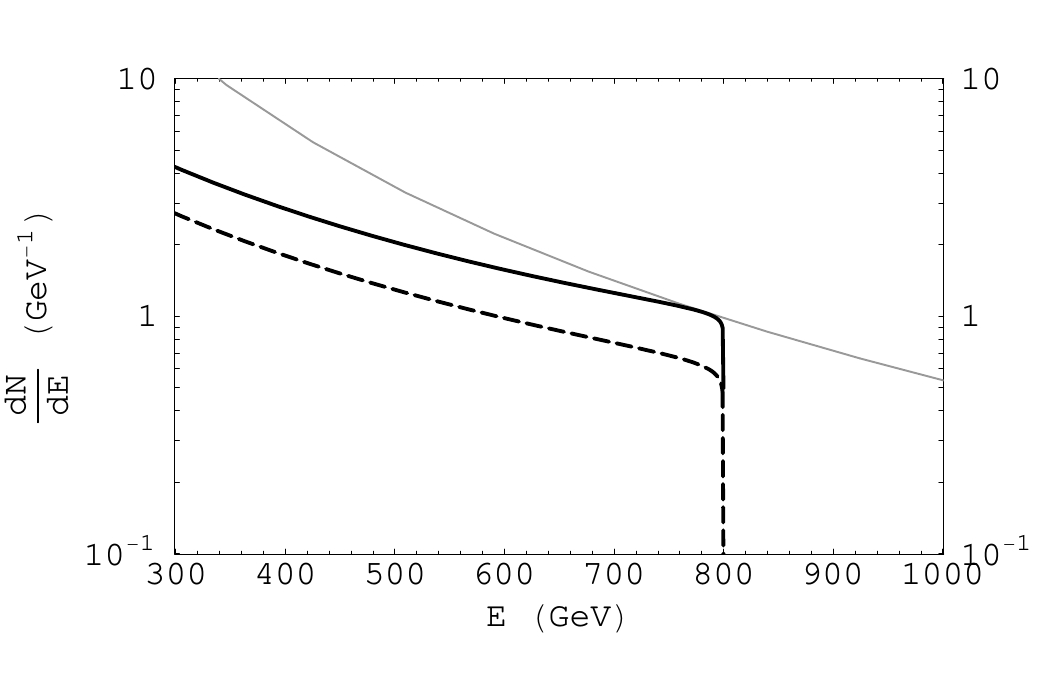}
\caption[Final State Radiation Spectrum]{ \label{Fig: FSR spectrum} The spectrum of final state radiation from a $m_\psi = 1600~\GeV$ dark matter particle decaying in the galaxy with lifetime $\tau = 10^{26} ~\s$ as would be seen at Fermi with a field of view = 1 sr near the galactic center.  The black curve is for the decay channel $\psi \to e^+e^-$, the dashed is for decays to $\mu^+\mu^-$.  Gray is the expected background.}
\end{center}
\end{figure}

Even if the primary decay mode does not include photons, they will necessarily be produced by final state radiation (FSR), i.e. internal bremsstrahlung, from charged particles that are directly produced.  In particular, if dark matter decay explains the electron/positron excesses then the decay must produce a large number of high energy electrons and positrons.  These decays will then also produce high energy gamma-rays through FSR with a spectrum given in Eqn. \eqref{Eqn: FSR spectrum}.  Such decays could come, for example, from the operators  from Section \ref{Sec: Dim 6 decays}.  Figure \ref{Fig: FSR spectrum} shows the spectrum of final state radiation expected from decays of dark matter in the galaxy to $e^+e^-$ and $\mu^+\mu^-$ as would be seen by Fermi.  We take the effective area $ = 8000 ~\cm^2$, the field of view = 1 sr annulus around the galactic center but at least than $10^\circ$ away from it, and the observing time = 3 yr.  This spectrum exhibits an `edge' at half the dark matter mass which could allow these gamma-rays to be distinguished spectroscopically from the diffuse astrophysical background.

The third general mechanism which produces photons from dark matter decays (or annihilations) is hadronic decay (e.g. to $q\overline{q}$) producing $\pi^0$'s which then decay to photons.  These generally yield a soft spectrum, similar to the expected diffuse astrophysical backgrounds.  Thus, we will not consider these signals, as a definitive detection could be challenging.  However it is worth noting that these signals can be quite large because they are not suppressed by the factor of $\sim 10^{-2}$ that suppresses final state radiation.

\begin{figure}
\begin{center}
\includegraphics[width=6.0 in]{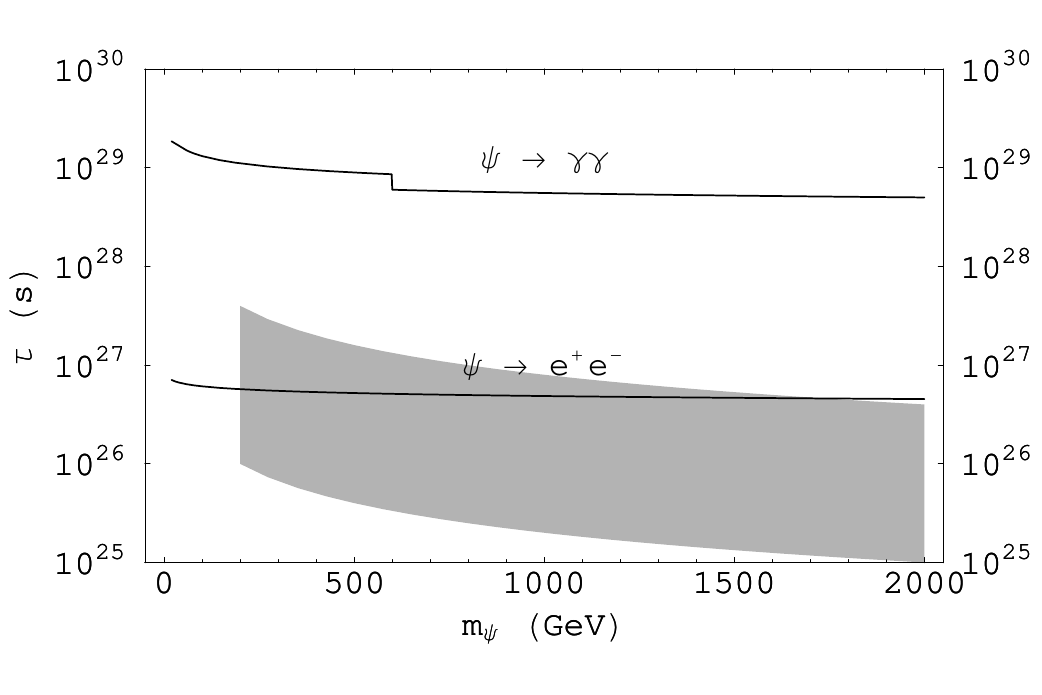}
\caption[Estimated Fermi Reach]{ \label{Fig: Fermi reach} A guess for the sensitivity of Fermi.  The solid lines are the expected reach of the Fermi telescope in the lifetime to decay into the modes $\psi \to \gamma\gamma$ and $\psi \to e^+e^-$ as a function of the mass of the decaying dark matter.  The signal from the second mode comes from internal bremsstrahlung gamma-rays from the electrons (but the plotted reach is for the primary decay mode into just $e^+e^-$).  The sensitivities are conservatively estimated using the Burkert (isothermal) profile, though the NFW profile gives essentially the same result.  The grey band is the region of lifetimes to decay to $e^+e^-$ that would explain the PAMELA positron excess from Eqn. \eqref{Eqn: pamela lifetimes}.  The ATIC excess would be explained in this same band for masses near $m_\psi \approx 1500 ~\GeV$.}
\end{center}
\end{figure}

Fig. \ref{Fig: Fermi reach} shows an estimate for the sensitivity of the Fermi telescope to dark matter that decays either to $\gamma\gamma$ or $e^+e^-$.  The sensitivity to decays to $e^+e^-$ comes observing the edge in the gamma-ray spectrum from final state radiation, as in Fig. \ref{Fig: FSR spectrum}.  Note though, the plot shows the reach of Fermi in the lifetime of the primary decay $\psi \to e^+e^-$ and not the FSR decay $\psi \to e^+e^-\gamma$.  Roughly we see that the reach for $e^+e^-$ is a factor of $10^2$ worse than for $\gamma\gamma$, which is the expected probability to emit final state radiation.  The sensitivity to the gamma ray line signal ($\psi \to \gamma\gamma$) was found by demanding that the signal (which is entirely within one energy bin) be larger than 3 times the square root of the expected diffuse astrophysical background in that bin.  The signals were calculated from Eqns. \eqref{Eqn: Decay flux} and \eqref{Eqn: FSR spectrum}.  The expected background was estimated as \cite{Birkedal:2005ep, Bergstrom:1997fj, Hunter:1997we}
\begin{equation}
\label{Eqn: diffuse gamma background}
\frac{d^2 \Phi}{dE d\Omega} = 3.6 \times 10^{-10} \left( \frac{100 ~\GeV}{E} \right)^{2.7} ~\cm^{-2} ~\s^{-1} ~\sr^{-1} ~\GeV^{-1}.
\end{equation}
The energy resolution of Fermi was estimated from \cite{Fermi webpage} as the given function below $300~\GeV$ and 30\% between 300 GeV and 1 TeV.  The sensitivity to the edge feature was estimated (similarly to \cite{Birkedal:2005ep}) by requiring that the number of signal photons within 50\% of the energy of the edge be larger than 5 times the square root of the number of background photons within that same energy band.  This sensitivity increases as the size of the energy bin used is increased and does not depend on the actual energy resolution of the detector (except for the assumption that it is smaller than the energy bin used).  This is clearly only a crude approximation of the actual statistical techniques which would be used to search for an edge.

We exclude a $10^\circ$ half-angle cone around the galactic center from our analysis because the diffuse gamma ray background is expected to be much lower off the galactic plane.  The annulus we consider, though it overlaps the galactic plane, is a good approximation to the dark matter signal from off the galactic plane.  Ignoring the galactic center reduces the signal.  To be conservative we use Eqn. \eqref{Eqn: diffuse gamma background} for the background everywhere, though it is probably an overestimate for the background off the galactic plane.  Fermi may also be able to resolve point sources of gamma rays, further reducing the astrophysical diffuse background.

From Fig. \ref{Fig: Fermi reach} we see that the most of the range of lifetimes which could explain the PAMELA and ATIC excesses as a decay to $e^+e^-$ should be observable at Fermi from the final state radiation produced.  The same conclusion would hold for decays to $\mu^+\mu^-$.  However, decays to three body final states would soften the produced electron and positron spectrum, thus softening the spectrum of FSR gamma-rays.  This signal could be more difficult to observe with Fermi, though the injected electrons and positrons must always be rather high energy in order to explain ATIC so there is a limit to how soft the gamma-ray spectrum could be for any model which explains ATIC.

\begin{figure}
\begin{center}
\includegraphics[width=6.5 in]{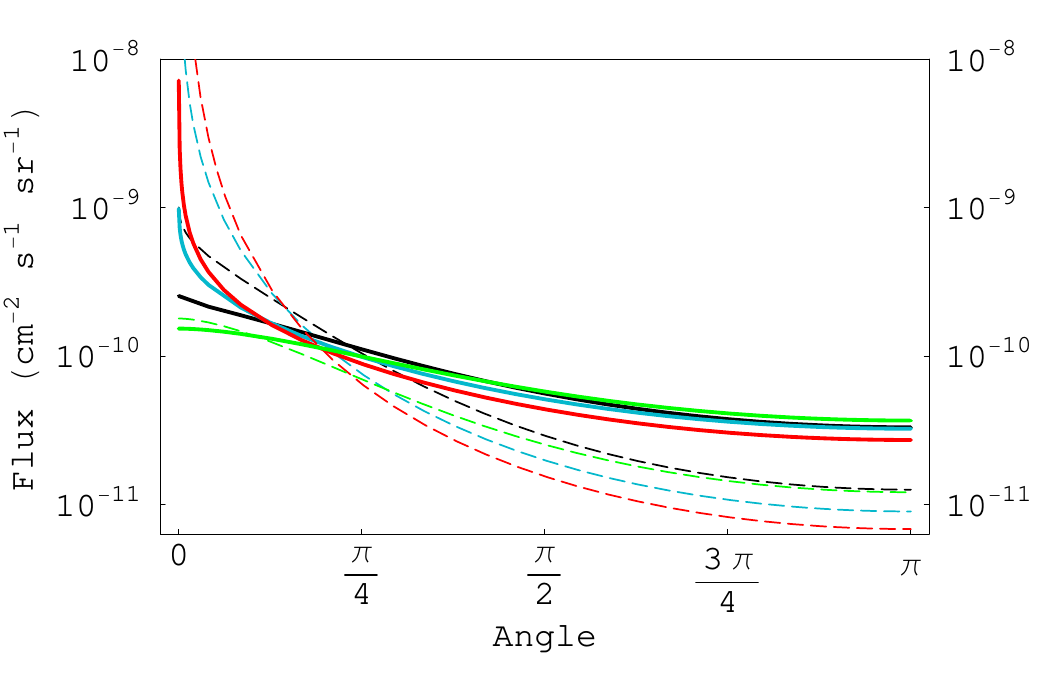}
\caption[Gamma Ray Flux vs Angle]{ \label{Fig: photon flux vs angle} (Color online) The flux of galactic gamma rays versus the angle from the galactic center for dark matter decay to 2 photons with $m_\psi = 1~\TeV$ and $\tau = 4 \times 10^{28} \, \s$ versus dark matter annihilation to 2 photons with $m_\psi = 500 ~\GeV$ and $\sigma v = 3 \times 10^{-26} \frac{\cm^3}{\s}$.  Note that the spike at 0 angle is cut off by a finite angular resolution taken to be $3 \times 10^{-6}$ sr.  Solid lines are the fluxes from decays, dashed lines are from annihilations.  Listed in order from top to bottom on the left edge of the plot: Moore profile (red), NFW profile (blue), Kravtsov profile (black), Burkert or isothermal profile (green).}
\end{center}
\end{figure}

It may be possible to distinguish dark matter decays from annihilations, so long as the dark matter-produced component of the total diffuse gamma ray background can be distinguished from the spectrum.  The intensity of dark matter produced gamma rays can then be measured as a function of the observing angle.  Because the decay rate scales as $n$ while the annihilation rate scales as $n^2$, the dependence of the gamma ray intensity on the angle from the galactic center is different for decays and annihilations for a given halo profile.  Of course, uncertainty in the halo profile could make it difficult to distinguish between the two.  In Fig. \ref{Fig: photon flux vs angle} we have plotted the galactic gamma ray flux as a function of the angle from the galactic center at which observations are made (these halo profiles are spherically symmetric around the galactic center so only the angle from the center matters).  The intensity is shown for both decays and annihilations to 2 photons from a variety of halo profiles.  We have chosen to scale the rates to the standard annihilation cross section ($\sigma v = 3 \times 10^{-26} \frac{\cm^3}{\s}$) and the decay rate that corresponds to this from Eqn. \eqref{Eqn: annihilations to decays} ($\tau = 4 \times 10^{28} ~\s$).  The shape of these curves is independent of the overall normalization.  The decay curves in Fig. \ref{Fig: photon flux vs angle} exhibit a universal behavior for large angles $\gtrsim \frac{\pi}{8}$, independent of the halo profile.  This stems from the fact that the integral of $n$ is essentially just the total amount of dark matter and is relatively insensitive to the distribution.  Note that this universal shape of the decay curve is significantly different from the shape of the annihilation curves.  With enough statistics, this difference would be readily apparent.  The annihilation curve from the Burkert profile is most similar to the decay curve and so would be the most difficult to distinguish.

A telescope such as Fermi is ideally suited to the task of measuring the angular distribution of the gamma-rays because of its large, $\OO (1~\sr)$, field of view and coverage of the entire sky.  Further, astrophysical backgrounds are significantly lessened away from the galactic ridge, giving an advantage to Fermi over the existing HESS observations which are mostly dominated by backgrounds.  Fermi has an acceptance of $\sim 3 \times 10^{11} ~\cm^2 ~\s ~\sr$.  Thus we can see from the scale in Fig. \ref{Fig: photon flux vs angle} that at least with a relatively strong decay mode into photons, Fermi could have enough statistics to differentiate the decay signal from an annihilation signal.

If the WMAP haze is due to dark matter decays producing electrons and positrons in the galactic center it may be possible to detect inverse Compton scattered photons with Fermi.  A study has been done for annihilating dark matter \cite{Hooper:2007gi}.  It may be interesting to check both the possibility of explaining the original WMAP haze and of observing the corresponding inverse Compton gamma rays at Fermi, for decaying dark matter as well.

\section{LHC Signals}
\label{Sec: LHC signals}

In this section, we study the LHC signals of the dimension 5 and 6 GUT suppressed operators considered in this paper. The dimension 5 operators lead to particle decays with lifetimes $\sim 100-1000 \text{ s}$ and these can have striking LHC signatures, particularly when the decaying particle is colored or has electric charge. The dimension 6 operators lead to particle decays with lifetimes $\sim 10^{26} \text{ s}$ and these decays will not be visible at the LHC. However, the requirement that these operators explain the observations of PAMELA/ATIC leads to constraints on the low energy SUSY spectrum and these constraints can be probed at the LHC. In the following subsections, we analyze the signals of these two kinds of operators. 

\subsection{Signals of Dimension 5 Operators}
\label{Dim5LHCSignals}

The dimension 5 operators introduced in section  \ref{LithiumIntroduction} mediate decays between the MSSM and a new TeV scale $\chi$. The $\chi$ is in a vector-like representation of $SU(5)$. In the following subsections, we analyze the cases when $\chi$ is a $\left(10, \TenBar\right)$, a $\left(5, \FiveBar\right)$ or a singlet  of $SU(5)$ and summarize our results in Table \ref{Tab:LithiumSignatures}. 

\subsubsection{Decouplet Relic}
\label{TenLHCSignals}

 The most striking collider signals arise in the case when $\chi$ is a $\left(10, \TenBar\right)$. In this case, all the particles in the multiplet are either colored or carry electric charge. When these particles are produced at a collider, they will leave charged tracks as they barrel out of the detector. This signal should be easily visible over backgrounds at the LHC \cite{KraanSkands}.  A significant fraction of these particles will stop in the calorimeters and their late time ($\sim 100 - 1000 \text{ s}$) decays will lead to energy deposition in the calorimeters with no activity in either the tracker or the muon chamber further enhancing the visibility of this signal  \cite{StoppingGluinos}. The LHC reach for such colored or electrically charged stable particles was analyzed in \cite{KraanSkands} and found to be $\sim$ 1 TeV for a right handed positron and at least 2 TeV for colored particles. As discussed in sub section \ref{ChiDecayToMSSM}, the primordial $\LiSix$ problem can be solved through the decays of a electroweak multiplet only if its mass $M_{E_\chi} \lessapprox 1 \text{ TeV}$. In this scenario, the colored multiplets in the $\chi$ could also  be light enough to be produced at the LHC. 

The production of these colored multiplets allows for more dramatic signatures at the LHC. Let us first consider the case when the fermionic components of the colored  $SU(2)$ doublets $\left(U, D\right)$ in $\left(10, \TenBar\right)$ are lighter than the scalar components, so that the scalar components, if produced, decay rapidly to the fermionic components. The masses of the components of the colored $SU(2)$ doublet fermions $\left(U^f, D^f\right)$ in $\left(10, \TenBar\right)$ are split by  $\sim \alpha M_Z \sim 350 \text{ MeV}$ due to electroweak symmetry breaking \cite{MinimalDarkMatter, ScottWells}. When the heavier component of the doublet is produced at the LHC, it will decay to the lighter one with a lifetime $\sim \frac{1}{G_F^2 \left(350 \text{ MeV}\right)^5} \sim 4 \times 10^{-11} \text{ s} = 1.2 \text{ cm}$, producing a displaced vertex that releases an energy $ \sim 350 \text{ MeV}$ resulting in the production of soft pions, muons and electrons. The lighter component  will barrel through the detector where it will leave a charged track and may eventually stop and decay after $100 - 1000 \text{ s}$. While it is difficult to trigger on the soft particles produced from the displaced vertex \cite{MinimalDarkMatter, ScottWells}, the detector will trigger on the charged track produced from the decay and this track could potentially be used to identify the displaced vertex and the soft particles produced with it. The observation of the displaced vertex, while experimentally difficult, will unveil the presence of the $SU(2)$ colored multiplet. 

It is also possible that the scalar components of the colored doublets are lighter than their fermionic partners. In this case, the fermionic components will decay rapidly to the scalar components. The contribution from electroweak symmetry breaking to the  mass splitting between the components of the colored $SU(2)$ scalar doublets is $\sim \cos \l 2 \beta \r M_W^2$ \cite{SUSYPrimer}. The decays of the heavier scalar component to the lighter component will be rapid, producing the lighter component and hard jets or leptons. The lighter component will again barrel through the detector producing a charged track and a late decay. This will also be a striking signal at the LHC. 

Since the $\chi$ are always pair produced and decay at similar times,  it may be possible to determine their lifetime by correlated measurements of double decay events in the calorimeter  \cite{StoppingGluinos}. If all the particles in the multiplet are light enough to be produced at the LHC, a measurement of their mass and lifetime will be a direct probe of the GUT scale. The decay rate $\Gamma$ of a particle of mass $M$ through a dimension 5 operator scales as $\Gamma \sim M^3$. A measurement of the masses and decay rates of the colored and electrically charged multiplets can then be used to determine if $\Gamma$ and $M$ scale as expected for a dimension 5 decay operator. Upon establishing this, a measurement of the scale $\Lambda$ mediating the decay can be inferred from the relation $\Gamma \sim \frac{M^3}{\Lambda^2}$. 

\subsubsection{Fiveplet Relic}

The colored particles in $\left(5, \FiveBar\right)$ will give rise to charged tracks and late time decays similar to the decays of the colored multiplets in $\left(10, \TenBar\right)$ as discussed earlier in subsection \ref{TenLHCSignals}. Electroweak symmetry breaking causes mass splittings between the electrically charged component $l^{+}_\chi$  and the neutral $l^{0}_\chi$. The collider phenomenology of the lepton doublet $l_\chi$ depends upon the spectrum of the theory. We first consider the case when the scalar components $\tilde{l}_\chi$ are lighter than their fermionic partners $l^f_\chi$, so that the $l^f_\chi$, if produced, decay rapidly to the $\tilde{l}_\chi$. Upon electroweak symmetry breaking, the masses of the scalar components $\tilde{l}^{+}_\chi$ and $\tilde{l}^{0}_\chi$ are split by $\sim \cos\l 2\beta\r M_W^2$ \cite{SUSYPrimer}.  Depending upon the value of $\cos \l2 \beta\r$, the $\tilde{l}^{+}_\chi$  can be lighter than  $\tilde{l}^{0}_\chi$. In this case, the $\tilde{l}^{+}_\chi$ will produce charged tracks as it traverses through the detector. Some of the  $\tilde{l}^{+}_\chi$s will stop in the detector and result in late time decays. The  $\tilde{l}^{+}_\chi$  will rapidly decay to the $\tilde{l}^{0}_\chi$ when  the $\tilde{l}^{0}_\chi$ is lighter than the $\tilde{l}^{+}_\chi$. This decay produces hard jets and leptons along with missing energy from the $\tilde{l}^{0}_\chi$ that leave the detector and decay outside it. 

We now consider the case when the fermionic components $l^f_\chi$ are lighter than their scalar partners $\tilde{l}_\chi$. In this case, the scalars $\tilde{l}_\chi$, if produced, will rapidly decay to their fermionic partners $l^f_\chi$. Electroweak symmetry breaking makes the electrically charged fermionic component $l^{f^{+}}_\chi$ heavier than the neutral component $l^{f^{0}}_\chi$ by $\sim \alpha M_Z \sim 350 \text{ MeV}$ \cite{MinimalDarkMatter, ScottWells}. The  $l^{f^{+}}_\chi$ decays  to the $l^{f^{0}}_\chi$  with a lifetime $\sim \frac{1}{G_F^2 \left(350 \text{ MeV}\right)^5} \sim 4 \times 10^{-11} \text{ s} = 1.2 \text{ cm}$ yielding a displaced vertex. This displaced vertex produces soft pions, muons and electrons with energies $\sim 350 \text{ MeV}$. The softness of these particles makes it difficult to trigger on them unless the production of these particles is accompanied by other hard signals like initial or final state radiation of photons \cite{ScottWells}. One possible source for this hard signal is the decay of the scalars $\tilde{l}_\chi$. If the $\tilde{l}_\chi$ are produced, they will rapidly decay to their fermionic counterparts. The LSP will also be produced in this process resulting in the activation of missing energy triggers. 

Another possible source for this hard signal are the colored particles in the multiplet. In the model discussed in sub section \ref{SU5U1B-L}, the colored particles in the $SO(10)$ multiplet $16_\chi$  can decay to the lepton doublets in the multiplet through dimension 5 operators. In this case, the late time decays of the colored multiplet can produce $l^{f^{+}}_\chi$. This decay will be accompanied by a release of hadronic energy at the location of the colored particle and a charged track that follows the trajectory of the produced  $l^{f{+}}_\chi$. This track ends in a displaced vertex when the  $l^{f{+}}_\chi$ decays to the $l^{f{0}}_\chi$. The late time decay followed by the charged track can be used to trigger on this event. The discovery of soft particles at the location of the displaced vertex, while difficult experimentally,  will unveil the presence of the lepton doublet.

\subsubsection{Singlet Relic}

The LHC signals for decays involving operators with singlet $\chi$s is dependent on the MSSM spectrum. If the MSSM LSP is heavier than $\chi$, then the MSSM LSP will decay to $\chi$ at $\sim 1000$ s. Cosmologically, this decay could solve the Lithium abundance problems and give rise to a dark matter abundance of $\chi$. In this scenario, since the MSSM LSP is no longer the dark matter, the MSSM LSP could be charged.  The SUSY particles produced at the LHC will rapidly decay to the MSSM LSP and if this MSSM LSP is charged or colored, it will give rise to charged tracks and late decays in the detector. The signals of SUSY in this scenario are significantly altered since the decays of SUSY particles are no longer associated with missing energy signals as the charged MSSM LSP can be detected. However, if the MSSM LSP is neutral, it will decay to the $\chi$ outside the detector and the LHC signals of this scenario will be identical to that of conventional MSSM models.

The cosmological lithium problem could also be solved by decays of singlet $\chi$s to the MSSM if the $\chi$ is heavier than the MSSM LSP. In this case, an abundance of $\chi$ must be generated through non standard model processes. This abundance can be generated thermally through new interactions like a low energy $U(1)_{B-L}$ or through the decays of heavier, thermally produced standard model multiplets to $\chi$. In this scenario, these new particles could be discovered at the LHC, for example through $Z^{\prime}$ gauge boson searches. It is also possible that the initial abundance of $\chi$ was generated through a tuning of the reheat temperature of the Universe. This scenario is devoid of new signals at the LHC. 

\begin{table}[ t]
\centering
\begin{tabular}{|l||l||l||}
\hline
$\chi \; SU(5)$ Rep. & Spectrum & Prominent Signals and Features\\
\hline
$ \left(10, \TenBar\right)$& $M_\chi > M_{LSP}$& Charged tracks from colored and electrically charged particles.\\
& & Measurement of mass, lifetime from stopped colored and charged particles.  \\
& & Potentially visible displaced vertex from colored doublet decays. \\
& & Infer dimensionality and scale of decay from mass, lifetime measurements. \\
\hline
$\left(5, \FiveBar\right)$ & $M_\chi > M_{LSP}$ & Charged tracks from colored particles. \\
& & Measurement of mass, lifetime from stopped colored particles.  \\
& & Charged track from scalar lepton doublet. \\
& & Potentially visible displaced vertex from lepton doublet decays. \\
\hline
Singlet & $M_{LSP} > M_\chi$ & Possibility of charged or colored LSP giving rise to charged tracks. \\
& & Charged LSP detection enables better measurement of SUSY spectrum.\\ 
& & \\   
           & $M_\chi > M_{LSP}$  & Discover new particles ({\it e.g.} TeV scale $U(1)_{B-L}$) for thermally produced $\chi$.  \\
\hline
\end{tabular}
\caption[LHC Signals of Dimension 5 Operators]{\label{Tab:LithiumSignatures} Prominent signals of the dimension 5 decay operators considered in section \ref{LithiumIntroduction}. The signals are classified on the basis of the $SU(5)$ representation of the new particle $\chi$ and its mass $M_\chi$. When $M_\chi > M_{LSP}$, the primordial lithium problem is solved by the decays of $\chi$ to the MSSM and vice-versa when $M_{LSP}  >  M_\chi$. Bounds on dark matter direct detection \cite{DirectDetection} force $\chi$ to decay to the MSSM LSP when $\chi$ ({\it e.g} $\left(5, \FiveBar\right)$,  $\left(10, \TenBar\right)$) has dirac couplings to the $Z$.}
\end{table}

\subsection{Signals of Dimension 6 Operators}
\label{Dim6LHCSignals}

The dimension 6 GUT suppressed operators discussed in this paper mediate decays between singlet $S$s and the MSSM when the operators conserve R-parity or cause decays of the LSP to the standard model when the operators violate R-parity (see section \ref{Sec: Dim 6 decays}).  The decay rate for these processes is $\Gamma \sim 10^{-27} \text{ s}^{-1}$. The largest production channel for these particles at the LHC would be through the decays of colored particles. For example, the LSP would be produced in the decays of gluinos. The production cross-section for a 200 GeV gluino at the LHC is $\sim 10^7 \text{ fb}$\cite{StoppingGluinos}. With a integrated luminosity of 100 $\text{fb}^{-1}$, the LHC will produce $\sim 10^9$ LSPs through the decays of the gluino. This number is far too small to allow for the observation of the decay. Similarly, the number of singlet $S$s that can be produced either through a new low energy gauge symmetry ({\it e.g.} $U(1)_{B-L}$) or through the decays of other particles that carry standard model quantum numbers is also too small to allow for the observation of the decay of the $S$. 

The most promising signatures of these dimension 6 GUT suppressed operators are the astrophysical signatures of their decay.  In section \ref{Sec: Dim 6 decays}, we show that these dimension 6 GUT suppressed decays  can explain the observations of PAMELA/ATIC. PAMELA has observed an excess in the positron channel while it does not see an excess in the anti proton channel. The hadronic branching fraction of these decays must be smaller than 10 percent in order to account for the PAMELA signal (see section \ref{Sec: Dim 6 decays}). The need to suppress the hadronic branching fraction of these decays can be used to arrive at constraints on the superpartner spectrum. 

For concreteness, we consider the model in section \ref{RParityConserving} where the scalar component $\tilde{s}$ of the $S$ gets a TeV scale vev $\langle \tilde{s} \rangle$. This leads to the operators $\frac{\langle \tilde{s} \rangle s \tilde{l}l}{M_{GUT}^2}$ and $\frac{\langle \tilde{s} \rangle s \tilde{q}q}{M_{GUT}^2}$ where $s$ denotes the fermionic component of $S$, $\l \tilde{l}, l \r$ represent slepton and lepton fields and $\l \tilde{q}, q \r$ represent squark and quark fields. This operator will mediate the decay of $s$  to the MSSM LSP if the mass $M_{s}$ of $s$ is greater than the LSP mass $M_{LSP}$. The hadronic branching fraction of this decay depends upon the mass splittings $\Delta M_{s \tilde{q}}$ and  $\Delta M_{s \tilde{l}}$ between $s$ and the squarks and sleptons respectively. The following options emerge: 

\begin{itemize}
\item $M_{s} > M_{\tilde{l}}$ and $M_{s} > M_{\tilde{q}}$:  This decay produces   on shell sleptons and squarks.  The hadronic branching fraction is $\sim \l\frac{\Delta M_{s \tilde{q}}}{\Delta M_{s \tilde{l}}}\r^3$ and this fraction is smaller than 10 percent if $\Delta M_{s \tilde{q}} < 0.45 \; \Delta M_{s \tilde{l}}$.

\item $M_{s} < M_{\tilde{l}}$ and  $M_{s} < M_{\tilde{q}}$: This decay produces electrons and quarks through off-shell sleptons and squarks respectively. The hadronic branching fraction of this decay mode is $\l \frac{M_{\tilde{l}}}{M_{\tilde{q}}} \r ^4$ where $M_{\tilde{l}}$ and $M_{\tilde{q}}$ are slepton and squark masses respectively. This branching fraction is smaller than 10 percent if $M_{\tilde{l}} <  1.8 \; M_{\tilde{q}}$.

\item $M_{s} > M_{\tilde{l}}$ and  $M_{s} < M_{\tilde{q}}$: This decay produces sleptons on shell. The dominant hadronic branching fraction of this decay mode arises from final state radiation of $W$ and $Z$ bosons yielding a branching fraction $\sim 10^{-2}$ for $M_{s} > 1$ TeV. The hadronic branching fraction produced as a result of decays through off-shell squarks is smaller than $10^{-2}$ since it is suppressed by additional phase space factors and couplings. 
\end{itemize}

The cases where the slepton is light enough to be produced on shell  ({\it e.g.} $M_{s} > M_{\tilde{l}}$) allow for the possibility of another correlated measurement at the LHC and astrophysical experiments like PAMELA/ATIC. The decay of the $s$ to a on-shell lepton and slepton will generate a primary source of hot electrons from the leptons produced in this primary decay.  These electrons will produce a "bump" in the spectrum that will be cut off at roughly $\sim \frac{M_{s}}{2}$. The sleptons produced in this process are also unstable and will rapidly decay to a lepton and the LSP. This process produces a secondary lepton production channel that will also lead to a spectral feature cut off at $M_{\tilde{l}} - M_{LSP}$. This astrophysical measurement can be correlated with measurements of this mass difference at the LHC and these independent measurements can serve as a cross-check for this scenario. 

The other possible decay in this model is the decay of the LSP to the $s$ which will happen if $M_{LSP} > M_{s}$. In this case, leptons and quarks are produced through off-shell sleptons and squarks respectively (see figure \ref{fig-DMdecay}). The hadronic branching fraction of this decay is given by  $\l \frac{M_{\tilde{l}}}{M_{\tilde{q}}} \r ^4$. This branching fraction is smaller than 10 percent if $M_{\tilde{l}} <  1.8 \; M_{\tilde{q}}$. 
We note that the model discussed in this section naturally allows for small hadronic branching fractions. The hadronic branching fraction is determined by slepton and squark masses and since squarks are generically expected to be heavier than sleptons due to RG running, the hadronic branching fraction of these models is generically small. These decays have a hadronic branching fraction that is at least $\sim 10^{-2}$ for $M_{s} \sim 1$ TeV due to final state radiation of $Z$ and $W$ bosons. A measurement of anti-protons at PAMELA that indicates a hadronic branching fraction smaller than $\sim 10^{-2}$ will rule out this model. A hadronic branching fraction that is larger than $10^{-2}$ will however constrain the squark and slepton masses as discussed earlier in this section.


\section{Conclusions}

In broad classes of theories, global symmetries that appear in nature, such as baryon number, are accidents of the low energy theory and are violated by GUT scale physics.  If the symmetry stabilizing a particle, such as the proton, is broken by short-distance physics then it will decay with a long lifetime.  In a sense, a particle is continually probing the physics that allows it to decay.  Thus, though the effects of GUT scale physics are suppressed at low energies, this suppression is compensated for by the long time scales involved.  Intriguingly, short-distance physics can be unveiled by experiments sensitive to long lifetimes.

Astrophysics and cosmology provide natural probes of long lifetimes.  The dark matter particle appears stable, but dimension 6 GUT suppressed operators can cause it to decay with a long lifetime $\sim 10^{26} ~\s$.  This can lead to observable signals in a variety of current experiments including PAMELA, ATIC, HESS, and Fermi.  Such observations can also probe the TeV scale physics associated with dark matter annihilations.  However, annihilations cannot proceed through GUT scale particles since such a cross section would be far too small.  Thus, the observation of annihilations can probe TeV physics but does not probe GUT physics directly.

This leads to interesting differences between the expected signals from dark matter decays and annihilations \cite{MinimalDarkMatter, NimaDougTracyWeiner}.  Decaying dark matter can directly produce photons with an $\OO(1)$ branching ratio, as seen in Section \ref{Sec: Dim 6 decays}.  It can also produce light leptons without helicity or p-wave suppression and give signals in the range necessary to explain the PAMELA/ATIC excess without boost factors \cite{Nardi:2008ix, MaximPospelov, YanagidaOne, YanagidaTwo, YanagidaThree}.  Additionally, in our framework the hadronic branching fractions can be naturally suppressed since sleptons tend to be lighter than squarks, perhaps explaining the lack of an antiproton signal at PAMELA.  
We have also found that the signals produced by decays may show many interesting features beyond just a bump in the spectrum of electrons and positrons.
For example, the spectrum from decays may accommodate the secondary feature visible in the ATIC data at energies between 100 - 300 GeV.  The SUSY spectrum can allow for the direct production of a slepton, lepton pair from the decaying dark matter.  In this case, the subsequent decay of the slepton provides a secondary hot lepton that can give rise to the observed secondary ATIC feature.  It is also common in our framework to have two dark matter particles, scalar and fermion superpartners, that both decay at late times producing interesting features in the observed spectra of electrons or photons.

Such models will be tested at several upcoming experiments.  The spectrum of electrons and positrons will be measured to increasing accuracy by Fermi, HESS, and PAMELA, testing the observed excesses.  These measurements could potentially reveal not just the nature of dark matter, but also the mass spectrum of supersymmetry and GUT scale physics.  Excitingly, Fermi and ground-based telescopes such as HESS or MAGIC will be providing measurements of the gamma-ray spectrum in the near future.  These measurements have the potential to test many of our models, and in particular may be able to observe final state radiation from models that explain the electron/positron excess.  Further, they could distinguish decaying and annihilating dark matter scenarios by the different angular dependence of the gamma ray signals.

The supersymmetric GUT theories in which the dark matter decays through dimension 6 operators often have TeV mass particles that decay via dimension 5 operators with a lifetime of $\sim$ 1000 sec.  This results in relic decays during BBN which could explain the observed primordial Lithium abundances.  This TeV mass particle can be produced at the LHC and if charged or colored will stop and decay inside the detector within $\sim$ 1000 sec.  This will allow the measurement of its properties and directly make the connection with primordial nucleosynthesis.

\section*{Acknowledgments}
We would like to thank Nima Arkani-Hamed, Douglas Finkbeiner,  Raphael Flauger, Stefan Funk, Lawrence Hall, David Jackson, Karsten Jedamzik, Graham Kribs, John March-Russell, Igor Moskalenko, Peter Michelson, Hitoshi Murayama, Michele Papucci, Stuart Raby, Graham Ross, Martin Schmaltz, Philip Schuster, Natalia Toro, Jay Wacker, Robert Wagoner, and Neal Weiner for valuable discussions.  PWG acknowledges the hospitality of the Institute for Advanced Study and was partially supported by NSF grant PHY-0503584.

\end{document}